\documentstyle[aps,prb,eqsecnum,tighten,preprint]{revtex}
\begin{document}
\title{Theory of two-dimensional quantum Heisenberg antiferromagnets with a 
nearly-critical ground-state}
\author{Andrey V. Chubukov${}^{1,2}$, Subir Sachdev${}^{1}$, and 
Jinwu Ye${}^{1}$}
\address{
${}^{1}$Departments of Physics and Applied Physics, P.O. Box 208120,\\ 
Yale University, New Haven, CT 06520-8120,\\
and ${}^{2}$P.L. Kapitza Institute for Physical Problems, Moscow, Russia}
\date{April 16, 1993; revised July 21, 1993}
\maketitle
\begin{abstract} 
We present the general theory of clean, two-dimensional, quantum Heisenberg
antiferromagnets which are close to the zero-temperature quantum transition
between ground states with and without long-range N\'{e}el order.
While some of our discussion is more general, the bulk of our theory
will be restricted to antiferromagnets in which the N\'{e}el order
is described by a $3$-vector order parameter.
For N\'{e}el-ordered states, `nearly-critical' means that the
ground state spin-stiffness, $\rho_s$, satisfies $\rho_s \ll J$, where
$J$ is the
nearest-neighbor exchange constant, while `nearly-critical' quantum-disordered
ground states have a energy-gap, $\Delta$, towards excitations with 
spin-1, which satisfies $\Delta \ll J$.
The allowed temperatures, $T$, are also smaller
than $J$, but {\em no} 
restrictions are placed on the values of $k_B T /\rho_s$ or 
$k_B T/\Delta$.
Under these circumstances, we show that the 
wavevector/frequency-dependent uniform and staggered spin 
susceptibilities, and the specific heat, are completely universal functions
of just three thermodynamic parameters. On the ordered side, these
three parameters are $\rho_s$, the $T=0$ spin-wave velocity
$c$, and the ground state staggered moment $N_0$;
previous works have noted the universal dependence of the susceptibilities
on these three parameters 
only in the more restricted regime of $k_B T \ll \rho_s$.
On the disordered
side the three thermodynamic parameters 
are $\Delta$, $c$, and the spin-1 quasiparticle residue ${\cal A}$.
Explicit 
results for the universal scaling functions are obtained 
by a $1/N$ expansion on the $O(N)$ quantum non-linear sigma model,
and by
Monte Carlo
simulations. These calculations lead to
a variety of testable predictions for neutron scattering,
NMR, and magnetization measurements. Our results are in good 
agreement with a number of numerical simulations and 
experiments on undoped and lightly-doped
$La_{2-\delta} Sr_{\delta}
Cu O_4$.
\end{abstract}
\narrowtext
\section{INTRODUCTION}
\label{intro}

The subject of two-dimensional quantum antiferromagnetism has witnessed a 
remarkable revival in recent years\cite{theory,numerics,trieste}. 
This is largely due to the intense interest
in understanding the properties of the $CuO_2$ layers in the high temperature 
superconductors. However, the experimental motivation is not limited
to these cuprate compounds; recent 
investigations\cite{ramirez,broholm,otherafm,elser,clarke} 
have refocussed interest
on a number of other layered insulating compounds which are rather well
described as Heisenberg antiferromagnets at low temperatures~\cite{harrison}.

The bulk of the existing theoretical work on two dimensional antiferromagnets 
can be divided into two broad classes:

First, there are the studies of the low
temperature properties of antiferromagnets with well-established long-range
N\'{e}el order in their ground state\cite{theory,numerics,Pokr,Andreev}. 
Definitive results have been obtained
for these systems by Chakravarty {\em et.al.\/}\cite{CHN,Tyc}. 
They showed that the
long-wavelength, low energy properties were well described by a mapping to a
{\em classical\/} two-dimensional Heisenberg magnet. All effects of quantum 
fluctuations could be absorbed into almost innocuous renormalizations of the
coupling constants. Good agreement with neutron scattering experiments
on $La_2 Cu O_4$ was obtained.

Second, there have been numerous investigations on spin-fluid, 
or quantum disordered,
ground states~\cite{Kalm_Laugh,sun,spn,CCL,Affl_Marst,Wieg-Khv}.
 These are states in which quantum-fluctuations 
have removed all vestiges of the N\'{e}el
order. An entirely new physical picture is necessary for visualizing the ground
states: it is phrased most often in terms of resonating singlet valence-bonds
between pairs of quantum spins~\cite{Anders}. Many interesting questions on the
presence of alternative symmetry breaking in the ground state
have been addressed.
The nature of the excitations above the ground state is also of some interest.
The two experimentally distinguishable possibilities are 
({\em i\/}) the low-lying states correspond
to those associated with weakly-interacting spin-1/2 quanta (`spinons'), or
({\em ii\/}) the spinons are confined in pairs, leading to integer-spin
excitations with infinite lifetimes.
The reader is referred to a recent
review\cite{trieste} 
where many of these questions are addressed in greater
detail.

In this paper, we present a detailed theory of quantum 
antiferromagnets which fall
{\em in between\/} 
the above two classes. These nearly-critical antiferromagnets are 
neither strongly N\'{e}el ordered nor fully quantum disordered in the ground
state. The ordered N\'{e}el moment, if it exists, is much smaller than
the ordering moment of the corresponding classical antiferromagnet.
If the ordering moment is absent, the resulting 
quantum-disordered ground state
has an energy-gap towards excitations with non-zero spin which is 
much smaller than all microscopic energy scales in the problem. 
Such nearly critical antiferromagnets were considered briefly by
Chakravarty {\em et. al.\/}~\cite{CHN}; 
they identified three different regimes
of behavior (See Fig.~\ref{phasediag}) which we now outline.
At short length and energy scales 
the spin correlations in these
antiferromagnets are essentially critical.
The spins fluctuate strongly between
ordered and non-ordered configurations. At very low temperatures,
the quantum-fluctuating 
system only makes up its mind at a fairly large scale, 
and crosses over to behavior characteristic
of either a N\'{e}el-ordered ground state 
(this is the `renormalized-classical'
region of Fig.~\ref{phasediag})
or a
quantum-disordered ground state (the `quantum-disordered' region of 
Fig.~\ref{phasediag}; see also Fig.~\ref{komdep}). At larger $T$
there is another, very interesting possibility:
the critical quantum fluctuations may be quenched by thermal effects
{\em before} the system has had a chance to undergo the above
crossover to one of its two ground states (the `quantum-critical'
region of Figs.~\ref{phasediag} and~\ref{komdep}).
The system then does not display the properties of
either 
N\'{e}el-order or quantum-disorder at {\em any length scale\/};
instead it crosses over from critical spin fluctuations 
to a thermally-induced, quantum-relaxational regime which will be 
described for the first time
in this paper. 

The above three crossover occur at a large length-scale in
nearly-critical antiferromagnets, suggesting that their properties
should be universal.
The bulk of this paper is devoted to making this statement more precise.
We will indeed find that the 
long-wavelength, low-energy uniform and staggered
spin susceptibilities are completely characterized 
by universal scaling functions.  The contribution of the spins to the 
specific heat will also be found to be similarly universal.
The only required inputs are three thermodynamic parameters, which will be
described more precisely below. Some of the results reported here have been
discussed briefly in recent reports\cite{jinwu,andrey}.
Some other recent work has also discussed quantum criticality
near related magnetic phase transitions~\cite{tsvelik,millis}.

Our motivation in examining nearly-critical antiferromagnets
comes primarily from numerous recent experiments on
undoped and weakly doped $La_{2-\delta} Sr_{\delta} Cu O_{4}$.

We consider first undoped $La_2 Cu O_4$.
It is by now well established that $La_2 Cu O_4$ is described
extremely well as a spin-1/2 square lattice Heisenberg antiferromagnet
with nearest-neighbor interactions\cite{theory}. 
Almost all theoretical studies\cite{CHN,hasen1,hasen2} of
this system have focussed primarily on its 
properties at very low temperatures,
where a description in terms of classical fluctuations of the 
N\'{e}el order parameter is appropriate; this range of temperatures
was referred to as the `renormalized-classical' region~\cite{CHN}.
There is good accord between theory\cite{theory,CHN,Chak-Orbach} and 
experiments\cite{Yamada,gabe,Slichter} at these low
temperatures ($T$):
the correlation length $\xi(T)$ increases exponentially with
falling $T$,
the equal-time structure
factor, $S(k)$, at zero momentum behaves as 
$S(0) \propto T^2 \xi^2$, and the $^{63}Cu$ 
spin-lattice relaxation rate $1/T_{1}$ decreases rapidly with the 
increasing $T$, all of which is  
in good agreement with the renormalized classical theory. 
However, recent experiments\cite{Slichter} have shown that 
at intermediate temperatures ($T \geq 0.4J$ where $J$ is the nearest-neighbor
exchange constant),
$1/T_{1}$ becomes nearly independent of $T$. 
The crossover to this behavior occurs at a $T$ which is small compared
to $J$, so one can hope that a low-energy theory 
of the expected crossover to the quantum-critical region 
might still be appropriate; the possibility of such a crossover
was already noted earlier~\cite{Chak-Orbach}.
We will show in this paper that this new behavior is  well described 
quantitatively by a theory of the quantum-critical spin fluctuations.
The correlation length $\xi (T)$ is also expected to display a crossover
at these temperatures~\cite{CHN};
unfortunately, there are no  experimental data for
$\xi$ for $T \geq 0.4J$.
However, strong support for our interpretation 
comes from the experimental~\cite{Johnson} and
numerical~\cite{Singh-Gelfand,Makivic,Sokol} measurements of the
 uniform static spin susceptibility $\chi_{u}^{{\rm st}}(T)$. 
We will show that its temperature-dependent slope above $T \sim 0.35 J$
is in excellent
accord with the predictions of the quantum-critical theory, and in
clear disagreement (by a factor of three)
with the result deduced from
a renormalized-classical theory.
Taken together, we will argue that the above results imply the
following: 
the square lattice spin-1/2 Heisenberg antiferromagnet
with nearest-neighbor exchange has long-range N\'{e}el order in its
ground state, and is well described by a renormalized classical theory
at low temperatures; however it is apparently
close enough to a quantum phase transition
to a quantum disordered phase to display quantum-critical spin
fluctuations  over an appreciable range of
intermediate temperatures.

Consider next weakly doped $La_{2-\delta} Sr_{\delta} Cu O_4$.
Nonzero doping ($\delta \geq 0$) has two important consequences:
({\em i\/}) the bare value of the spin-stiffness $\rho_s$ takes
a smaller value, pushing the antiferromagnet closer to the quantum phase
transition, and ({\em ii\/}) the antiferromagnet is perturbed 
by the random potential of the dopant ions.
Randomness is a relevant perturbation at the $T=0$ quantum phase 
transition\cite{jinwu} and must be included in any theory of the 
low $T$ properties. 
Dynamic neutron scattering measurements\cite{birg,keimer2} near 
$\delta = 0.04$ show
that the spin spectrum can be collapsed in a scaling plot in which the
measurement frequency $\omega$ is scaled by $T$; a related dependence
of the susceptibility on $\omega$ and $T$ was also discussed in 
the phenomenological theory of the marginal Fermi liquid~\cite{varma}.
We have argued that this
$\omega / T$ scaling is in fact a rather general property of quantum-critical
spin fluctuations, {\em even in the presence of randomness and
doping\/}~\cite{jinwu}. 
Furthermore, the pre-factor of the  scaling function at low
$\omega$ and $T$ shows clear evidence of the importance of 
randomness~\cite{jinwu}. However 
many other experiments\cite{Slichter,keimer2} 
on doped 
$La_{2-\delta} Sr_{\delta} Cu O_4$ have been performed at relatively
large temperatures.
For these $T$ we assume that the antiferromagnet is insensitive to
the weak randomness and is still in the vicinity of the pure fixed
point. The primary effect of non-zero doping will then be
to reduce the value of $\rho_s$ - an immediate consequence is
that the NMR relaxation $1/T_1$ should be nearly $T$ independent
(the quantum-critical behavior) over a  $T$ range 
which
increases with doping.  
This is indeed what is observed. 
We will discuss this and
other comparisons with experiments on the doped cuprates in
more detail later in Section~\ref{comp_exp}. 
With a single adjustable parameter (the doping
dependent $\rho_s$) we will find good agreement between 
our theory and the experiments. 

We turn now to a more complete description of our results.
Consider the antiferromagnet described by the following Hamiltonian
\begin{equation}
{\cal H} = \sum_{i<j} J_{ij} {\bf S}_i \cdot {\bf S}_j ,
\label{calH}
\end{equation}
where $i,j$ extend over the sites of two-dimensional lattice, 
${\bf S}_i$ are on-site spin operators acting on states with spin $S$
on the site $i$, and the $J_{ij}$ are exchange constants which fall off
rapidly with the separation between $i$ and $j$. The $J_{ij}$ are assumed
to be invariant under the  translation symmetry of the underlying lattice.
The energy scale $J$ will be used to denote the largest
of the $J_{ij}$.
The $J_{ij}$ are predominantly
antiferromagnetic, so that the classical ground state has no average
uniform magnetization. We will be interested primarily in antiferromagnets
which undergo a zero-temperature quantum phase transition from a 
N\'{e}el-ordered to a quantum disordered 
state as the ratios of the $J_{ij}$ are
varied, and the strength of the quantum fluctuation increases. 
Let us represent this strength by an all-purpose coupling constant
$g$; the system is assumed to be N\'{e}el-ordered for $g$ smaller than 
a critical coupling $g_c$ and quantum-disordered for $g > g_c$. We will assume
further that the critical ground state at $g=g_c$ is described 
by a continuum field theory of excitations which propagate with a non-singular
spin-wave velocity $c$ - a number of explicit mean-field solutions of such
transitions have this property\cite{sun,spn}, although 
other possibilities have also been 
discussed~\cite{CCL,Ioffe_Lar,Chub1}.
It should be possible to extend our results to 
antiferromagnets
which have different velocities for different spin-wave 
polarizations\cite{Andreev,halpsas}, but
we will not consider this complication here. 
In Section~\ref{comp_exp}
 we will argue that the following scaling results also apply
unchanged to the corresponding quantum phase transitions in 
lightly-doped antiferromagnets.
  
The nature of the classical ordering helps us identify the proper continuum
fields necessary for a hydrodynamic theory of the 
quantum phase transition. 
The first of these is
of course the N\'{e}el order parameter. We will restrict the analysis
in this paper to antiferromagnets with an ordinary vector order parameter. This
type of order parameter is associated with ground states with collinear 
spin-ordering {\em i.e.\/} the on-site spin condensates are either parallel or
anti-parallel to each other.
More complicated order parameters can arise in systems with stronger
frustration
{\em e.g.\/} the triangular and kagom\'{e} lattices which have co-planar
spins and matrix order
parameters\cite{dombre,delamotte,kagtotal,kagome};
we will not discuss these complications here.
Returning to the vector order-parameter case,  we assume 
that the condensate is oriented in the
$\pm z$ direction, and define a continuum
quantum field ${\bf n} ( {\bf r} )$, which will be used as the hydrodynamic 
order-parameter variable, as:
${\bf n} ( {\bf r}_i ) = {\bf S}_i$ on sites where the condensate points up,
and $n^{z} ( {\bf r}_i ) = - S^{z}_i; ~ n^{\pm} ( {\bf r}_i ) =  S^{\mp}_i$
on sites where the condensate points down.
The staggered magnetization $N_0$ is then
\begin{equation}
N_0 = \langle n_z ( {\bf r} ) \rangle_{T=0}.
\end{equation}
Upon approaching the critical point at $T=0$, this 
staggered magnetization will vanish as\cite{CHN,Ma}
\begin{equation}
N_0 \sim (g_c - g)^{\beta},
\label{n0beta}
\end{equation}
where $\beta$ is a universal critical exponent.
We have $N_0 = 0$ for $g > g_c$. Furthermore, 
$\langle n_z ( {\bf r} ) \rangle = 0$ at all finite $T$
because it is not possible to break a continuous non-abelian symmetry in a 
two-dimensional system. At the critical point, 
$g=g_c$, equal-time ${\bf n} ( {\bf r})$ correlations
will decay with an anomalous power law\cite{Ma}
\begin{equation}
\langle {\bf n} ( {\bf r} ) 
\cdot {\bf n} (0) \rangle \sim \frac{1}{r^{D-2+\eta}},
\end{equation}
where $D=3$ is the dimension of space-time.

The second important hydrodynamic 
variable is the magnetization density quantum field ${\bf M} ( {\bf r})$
\begin{equation}
{\bf M} ( {\bf r}_i ) = \frac{g \mu_B}{a^2} {\bf S}_i ,
\end{equation}
where $a^2$ is the volume per spin, and $g \mu_B$ is the
gyromagnetic ratio of each spin.
Although the ordering has no net magnetization, magnetization fluctuations
decay slowly due to the conservation law on the total magnetization.

Finally, the Hamiltonian ${\cal H}$ is itself associated with a conserved
total energy. The contribution of the spins to the specific heat
per unit volume, $C_V$,
is the appropriate experimental observable, sensitive to this
hydrodynamic quantity.

The hydrodynamic properties of the order parameter and magnetization
fluctuations
can be determined from the following
two retarded response functions:
\begin{eqnarray}
\chi_s ( k , \omega ) \delta_{\ell,m} 
= &&-\frac{i}{\hbar} \int d^2 r \int_0^{\infty} dt \nonumber \\
&&~~~~
\langle [ n_{\ell} ( {\bf r}, t) , n_m (0,0) ] \rangle e^{-i ({\bf k} \cdot 
{\bf r} - \omega t)}, \nonumber \\
\chi_u ( k , \omega ) \delta_{\ell m}
 = && -\frac{i}{\hbar} \int d^2 r \int_0^{\infty} dt \nonumber \\
&&~~~~
\langle [ M_{\ell} ( {\bf r}, t) , M_m (0,0) ] \rangle e^{-i ({\bf k} \cdot 
{\bf r} - \omega t)}, 
\end{eqnarray}
where all fields have now acquired a Heisenberg-picture time ($t$) dependence,
the indices $\ell , m$ extend over the three spin directions $x,y,z$,
and the average is with respect to a thermal Gibbs ensemble at a temperature $T$.
These correlation function are the dynamic staggered and 
uniform spin-susceptibilities respectively; their values predict the result
of 
essentially all the experiments that have been performed on antiferromagnets.
An important exception 
is the Raman-scattering cross-section~\cite{Park} - we will not discuss its properties
here.

We now present the scaling forms satisfied by $\chi_s$, $\chi_u$, and 
$C_V$ in the vicinity
of the quantum phase transition at $g=g_c$. The temperature, $T$, is taken to 
be non-zero, but must satisfy
\begin{equation}
k_B T \ll J.
\end{equation}
A non-zero $T$ implies the absence of a spin-condensate, and the response functions
are therefore rotationally invariant.
It is useful to describe separately the scaling properties of magnets
with and without N\'{e}el order in their ground state. This will be followed
by a discussion of the relationship between the two cases.

\subsection{N\'{e}el ordered ground state:}
\label{introneel}
The scaling properties should clearly depend upon a variable which measures
the distance of the ground state from criticality. The most convenient
choice
is the ground state 
spin-stiffness\cite{helicity} $\rho_s$. Its value can be easily
determined by experiments and by various numerical 
analyses on model Hamiltonians, with no arbitrary overall scale factors.
In two-dimensions, $\rho_s$ has the dimensions of energy, and the requirement
that the magnet is not too far from criticality is
\begin{equation}
\rho_s \ll J .
\end{equation}
Upon approaching $g_c$, $\rho_s$ obeys Josephson scaling\cite{josephson}
\begin{equation}
\rho_s \sim  (g_c - g)^{(D-2)\nu} ,
\label{rhosdm2n}
\end{equation}
where $\nu$ is the usual correlation length exponent.
We can now state one of the central results of this paper:
For $g \leq g_c$, and under the conditions on $T$ and $\rho_s$ noted above,
the values of $\chi_s$, $\chi_u$ and $C_V$ satisfy the following scaling
forms
\widetext
\begin{mathletters}
\begin{eqnarray}
\chi_s ( k , \omega ) &=& \frac{N_0^2}{\rho_s} 
\left( \frac{\hbar c}{k_B T} \right)^2
\left( \frac{N k_B T}{2 \pi \rho_s } \right)^{\eta}
\Phi_{1s} \left( \frac{\hbar c k}{k_B T} , \frac{\hbar \omega}{k_B T}
, \frac{N k_B T}{2 \pi \rho_s} \right), 
\label{phi1s}\\ 
\chi_u ( k , \omega ) &=&  
\left( \frac{g \mu_B}{\hbar c} \right)^2 k_B T~
\Phi_{1u} \left( \frac{\hbar c k}{k_B T} , \frac{\hbar \omega}{k_B T}
, \frac{N k_B T}{2 \pi \rho_s} \right),
\label{phi1u} \\
C_V &=& \frac{3 \zeta (3)}{\pi} 
k_B \left( \frac{k_B T}{\hbar c} \right)^2 \Psi_1 \left(
\frac{N k_B T}{2 \pi \rho_s} \right),
\label{psi1}
\end{eqnarray}
\end{mathletters}
\narrowtext
where $N$ is the number of components of the order-parameter,
$\zeta$ is the Reimann zeta function,
and $\Phi_{1s}$, $\Phi_{1u}$ and $\Psi_1$ 
are completely universal, dimensionless, functions ($\Phi_{1u}$ and $\Phi_{1s}$
are complex while $\Psi_1$ is real)
defined such that they remain finite as $T/ \rho_s \rightarrow \infty$;
this will also be true for other scaling functions introduced below. 
For simplicity, 
we have explicitly specialized to antiferromagnets with spacetime dimension
$D=2+1$, although analogous results for 
general $D$ are not difficult to write down.
Particularly striking is the absence of any non-universal scale
factors (in either the arguments or the prefactors of the scaling
functions) in all scaling forms. Everything is fully determined by 
the values of $\rho_s$, $c$ and $N_0$ and there is no further dependence
on lattice scale physics.
The universal dependence of the spin susceptibilities
on $\rho_s$, $N_0$ and $c$ was implicit in the analysis of 
Chakravarty {\em et.al.\/}\cite{CHN} (see also the recent
work of Hasenfratz and Niedermayer\cite{hasen1}) for the low $T$ regime
$T \ll \rho_s$; our results are however valid for {\em all\/} values
of $T /\rho_s$. Also Castro Neto and Fradkin\cite{eduardo} have recently
discussed closely related scaling forms for $C_V$ near general quantum
phase transitions in $2+1$ dimensions. The 
coefficient of $\Psi_1$ in (\ref{psi1}) has been chosen to be 
the specific heat of a
single gapless bose degree of freedom with dispersion $\omega = c k$ in
2 dimensions. The number $\Psi_1 (T \rightarrow 0)$ is thus a measure of the 
effective number of such modes in the ground state.
This number is given by $\Psi_1 ( 0 )$ for the ordered N\'{e}el phase,
and by $\Psi_1 ( \infty )$ for the quantum critical state at $g=g_c$.
 
We note further that all scaling forms
continue to be valid even at $g=g_c$: the prefactor of $\Phi_{1s}$ in
(\ref{phi1s}) remains non-singular at $g=g_c$ because of the results
(\ref{n0beta}), (\ref{rhosdm2n}) and the exponent identity\cite{Ma}
\begin{equation}
2 \beta = (D - 2 + \eta) \nu.
\label{betaeta}
\end{equation}

The arguments of the scaling functions will occur frequently in this paper.
We therefore introduce the dimensionless variables
\begin{equation}
\overline{k} = \frac{\hbar c k}{k_B T} ~~~;~~~ \overline{\omega} 
= \frac{\hbar \omega}{k_B T},
\label{barkdef}
\end{equation}
which represent momentum and frequency measured in units of a scale set
by the absolute temperature $T$. The third argument
\begin{equation}
x_1 = \frac{N k_B T}{2 \pi \rho_s},
\label{x1def}
\end{equation}
determines whether the antiferromagnet is better described at the
longest distances as a quantum-critical or a renormalized-classical model
(see Fig.~\ref{phasediag}). 
The factor of $N$ in the definition of $x_1$ is to facilitate the large $N$ limit
in which $\rho_s \sim N$; the variable $x_1$ will therefore remain of order unity.
The factor of $1/(2 \pi)$ is purely for future notational convenience.
For large $x_1$, the energy scale $k_B T$ is the largest energy which
first cuts-off the critical spin fluctuations,
and the system never fully realize that its coupling $g$ is in fact
different from  $g_c$ and that the ground state is ordered: the spin-fluctuations
are quantum-critical at the shortest scales, and are eventually quenched
in a universal way by the temperature. For small $x_1$ the antiferromagnet is in the
renormalized-classical region. There is a large
intermediate scale over which the antiferromagnet behaves as if it has
long-range N\'{e}el order; eventually, strong two-dimensional 
classical thermal fluctuations of the
order-parameter destroy the N\'{e}el order. 
 
\subsection{Quantum-disordered ground state:}
We now consider the case $g \geq g_c$. 
We will assume that the quantum-disordered state has a gap towards
all excitations. This has certainly been satisfied by all explicit large $N$
constructions of such states in frustrated antiferromagnets 
in the vicinity of the transition to long-range
N\'{e}el order\cite{sun,spn}. We will assume further that there are no
deconfined spin-1/2 excitations above the ground state: this is expected to 
be true in systems with a collinear N\'{e}el order parameter\cite{spn}. 
Antiferromagnets on the  triangular
or kagome lattices with coplanar spin correlations are expected to possess
deconfined spin-1/2 spinon excitations\cite{spn,kagome} - their critical properties
will therefore not be described by the present theory.
For the case of confined spinons which is under consideration here, the lowest
excitation with non-zero spin will carry spin 1. Further, at $T=0$,
this excitation should
have an infinite lifetime at small enough $k$.
The distance from criticality
is conveniently specified by the gap, $\Delta$, to this spin 1 excitation.
The equal-time order parameter correlation function will decay exponentially
on a scale $\xi$ which is inversely proportional to $\Delta$. Therefore
$\Delta$ will vanish as 
\begin{equation}
\Delta \sim (g - g_c )^{\nu}
\label{deltan}
\end{equation}
upon approaching criticality. One is not too far from criticality provided
\begin{equation}
\Delta \ll J.
\label{deltasmJ}
\end{equation}
Further, we need an observable which sets the scale for order-parameter
fluctuations. On the ordered side this was done by $N_0$. A convenient
choice on the disordered side is to use an amplitude of the local,
on-site, dynamic
susceptibility, $\chi_L$. This susceptibility is defined by
\begin{eqnarray}
\chi_L ( \omega ) \delta_{\ell m} &=& - \frac{i}{\hbar} \int_{0}^{\infty}
dt \langle [ S_{i\ell} ( t) , S_{i m} ( 0) ] \rangle \nonumber \\  
&\approx& \delta_{\ell m} \int \frac{d^2 k}{4 \pi^2} \chi_s ( k , \omega ).
\label{childef}
\end{eqnarray}
In principle, $\chi_u$ also contributes to $\chi_L$, but when (\ref{deltasmJ}) is
satisfied its contribution
is subdominant to that from $\chi_s$, and can therefore be 
neglected.
For $g > g_c$, it can be shown that at $T=0$, $\chi_L$ has the following
imaginary part for small $\omega$ close enough to 
the threshold $\Delta$:
\begin{equation}
\left. \mbox{Im} \chi_L ( \omega ) \right|_{T=0} 
= \frac{{\cal A}}{4} ~\mbox{sgn}( \omega )  
\theta ( \hbar |\omega| - \Delta ),
\end{equation}
where $\theta$ is the unit step function, and ${\cal A}/4$ is an amplitude with
the dimensions of inverse-energy. 
We will show later that the discontinuity $i{\cal A}/4$ in the local
dynamic susceptibility is precisely a quarter of the
 quasiparticle residue ${\cal A}$
of
the low-lying spin-1 excitation.
As $g$ approaches $g_c$, ${\cal A}$ vanishes as
\begin{equation}
{\cal A} \sim (g - g_c)^{\eta\nu} .
\label{aen}
\end{equation}
We have now assembled all the variables necessary for obtaining the scaling
forms for $\chi_s$ and $\chi_u$ for $g \geq g_c$. The relations analogous
to (\ref{phi1s}), (\ref{phi1u}) and (\ref{psi1}) are
\widetext
\begin{mathletters}
\begin{eqnarray}
\chi_s ( k , \omega ) &=& {\cal A} 
\left( \frac{\hbar c}{k_B T} \right)^2
\left( \frac{k_B T}{\Delta} \right)^{\eta}
\Phi_{2s} \left( \frac{\hbar c k}{k_B T} , \frac{\hbar \omega}{k_B T}
, \frac{k_B T}{\Delta} \right), 
\label{phi2s}\\ 
\chi_u ( k , \omega ) &=&  
\left( \frac{g \mu_B}{\hbar c} \right)^2 k_B T~
\Phi_{2u} \left( \frac{\hbar c k}{k_B T} , \frac{\hbar \omega}{k_B T}
, \frac{k_B T}{\Delta} \right),
\label{phi2u}\\
C_V &=& \frac{3 \zeta (3)}{\pi } 
k_B \left( \frac{k_B T}{\hbar c} \right)^2 \Psi_2 \left(
\frac{k_B T}{\Delta} \right),
\label{psi2}
\end{eqnarray}
\end{mathletters}
\narrowtext
where $\Phi_{2s}$, $\Phi_{2u}$, and $\Psi_2$ are completely universal functions,
which are finite in the limit $T/ \Delta \rightarrow \infty$.
The physical response functions are again completely determined
by three thermodynamic parameters: $\Delta$, $c$, and ${\cal A}$,
with no further sensitivity to lattice scale physics.
As in Sec~\ref{introneel}, all scaling forms
continue to be valid even at $g=g_c$: the prefactor of $\Phi_{2s}$ in
(\ref{phi2s}) remains non-singular at $g=g_c$ because of the scaling results
(\ref{deltan}), (\ref{aen}).

We also introduce, for future convenience, the variable
\begin{equation}
x_2 = \frac{k_B T}{\Delta},
\end{equation}
which determines whether the antiferromagnet is in the quantum-critical or
quantum disordered regions (see Fig~\ref{phasediag}). 
For large $x_2$, the temperature $T$ predominates
the small zero-temperature gap, $\Delta$, and the system may as well be
at $g=g_c$. For small $x_2$, the ground-state gap $\Delta$ quenches
the spin fluctuations, putting the system in the quantum-disordered region.
The thermodynamics is well described in terms
of a dilute gas of activated excitations.

\subsection{Critical point:}
We have obtained above two separate universal scaling forms at the critical
coupling $g=g_c$, but $T$ finite, by taking the limits $g \nearrow g_c$ and 
$g \searrow g_c$. It follows therefore that the two results must be simply related:
\begin{eqnarray}
\Phi_{2s} ( \overline{k} , \overline{\omega} , x_1 = \infty ) &=& 
Z_Q \Phi_{1s} ( \overline{k} , \overline{\omega} , x_2 = \infty ), \nonumber \\
\Phi_{2u} ( \overline{k} , \overline{\omega} , x_1 = \infty ) &=& 
 \Phi_{1u} ( \overline{k} , \overline{\omega} , x_2 = \infty ), \nonumber \\
\Psi_2 ( \infty ) &=& \Psi_1 ( \infty ),
\label{zqdef} 
\end{eqnarray}
where $Z_Q$ is a universal number. 
Recall that the universal functions have been chosen to have a 
a finite limit as $x_{1,2} \rightarrow
\infty$. 

Note that there is no rescaling factor
for the uniform susceptibility and the specific heat. 
This is because their overall scale is
universal and was not set by some thermodynamic observable, as was the case
for the staggered susceptibility. This universality in scale is related to the
fact that ${\bf M}$ and the energy are conserved quantities: 
this will discussed further
in Sec~\ref{phenom}. 
For the staggered susceptibility, we have performed a $1/N$ expansion 
of the scale factor on
antiferromagnets with an $N$-component order-parameter and found
\begin{equation}
Z_Q = 1 - \frac{0.229191243}{N} .
\label{zqval}
\end{equation}

Actually it is not just the values, but the entire asymptotic
expansions of the universal functions which have matching conditions
at $x_{1,2} = \infty$. As the antiferromagnet has no phase transition
at finite temperature, all its measurable properties should be smooth
functions of the bare coupling constant $g-g_c$ provided $T \neq 0$. 
This fact, combined with $x_1 \sim (g_c - g)^{-\nu}$, $x_2 \sim (g -
g_c)^{-\nu}$, can be used to easily deduce the constraints on the
asymptotic expansions of $\Phi_{1s}$, $\Phi_{1u}$, $\Psi_1$,
$\Phi_{2s}$, $\Phi_{2u}$, $\Psi_2$ about $x_{1,2} = \infty$.

\subsection{Experimental Observables}
 
The main purpose of the rest of the paper is to describe the universal
functions $\Phi_{1s}$, $\Phi_{1u}$, $\Psi_1$, $\Phi_{2s}$, $\Phi_{2u}$,
and $\Psi_2 $ as completely 
as possible. An large amount of information is contained in them;
in particular, as we shall see later, in a suitable limit they
contain the complete static and dynamic scaling functions of Chakravarty
{\em et. al.\/}\cite{CHN} and Tyc {\em et. al.\/}\cite{Tyc}.
A large number of experimentally testable quantities can be
obtained from these functions; now we highlight some of the most important
by endowing them with their own scaling functions.

We begin with the measurements related to $\chi_u$.
\subsubsection{Static, uniform spin susceptibility and spin diffusivity}
The conservation of ${\bf M}$ makes the small $k$ and $\omega$ dependence of
$\chi_u$ rather simple. 
In the hydrodynamic limit, $\omega \tau_{\ell} \ll 1$, where $\tau_{\ell}$ 
is a typical lifetime of excitations, 
the magnetization fluctuations must obey a 
diffusion equation; we find for $g \leq g_c$ that
\begin{equation}
\Phi_{1u} ( \overline{k} , \overline{\omega} , x_1 ) = \Omega_{1} ( x_1 ) 
\frac{ D_{1} (x_1) \overline{k}^2}{-i \overline{\omega} + D_{1} ( x_1 )
 \overline{k}^2}; ~~~\overline{k} , \overline{\omega} \ll 1 ,
\end{equation}
and an analogous expression for $g \geq g_c$ with $1 \rightarrow 2$.
The functions $\Omega_{1} (x_1)$, $\Omega_{2 } (x_2 ) $, 
$D_{1 } ( x_1) $, $D_{2 } ( x_2 )$ are all universal. From (\ref{phi1u}), 
(\ref{barkdef}),
(\ref{phi2u}) we see that they are related to the static, uniform spin
susceptibility $\chi_u^{{\rm st}}$, and spin diffusion constant $D_S$. 
We have for $g \leq g_c$
\begin{eqnarray}
\chi_u^{{\rm st}} ( T ) &=& \left( \frac{g \mu_B }{\hbar c} \right)^2 k_B T ~
\Omega_{1} ( x_1 ), \nonumber \\
D_S ( T) & =& \frac{\hbar c^2}{k_B T} D_{1} ( x_1 ),
\label{chistdef}
\end{eqnarray}
and analogous expressions for $g \geq g_c$ with $1 \rightarrow 2$.

\subsubsection{Wilson Ratio}
The Wilson ratio is defined by
\begin{equation}
W = \frac{k^{2}_B T \chi^{st}_u (T)}{(g\mu_B )^2 C_V (T)}~.
\label{Wilson}
\end{equation}
Its properties are therefore easily obtainable from our scaling results
for $\chi^{st}_u$ and $C_V$. It follows from  (\ref{psi1}), (\ref{psi2}) and
(\ref{chistdef}) that $W$ is a completely universal function of $x_1 ~(x_2)$
for $g \leq g_c ~(g \geq g_c )$. We have for $g \leq g_c$
\begin{equation}
W = \frac{\pi}{3 \zeta (3)}~~\frac{\Omega_1 (x_1)}{\Psi_1 (x_1)} .
\label{Wilson2}
\end{equation}
and analogous expression for $g \geq g_c$ with $1 \rightarrow 2$.

We turn next to experiments sensitive to $\chi_s$. 
\subsubsection{Structure factor}
The equal-time spin structure
factor, $S(k)$, is related to $\chi_s ( k , \omega )$ by
\begin{eqnarray}
S ( k ) \delta_{\ell m} &=& \int d^2 r \langle n_{\ell} ( {\bf r} , 0)
n_{m} ( 0 , 0) \rangle e^{- i {\bf k} \cdot {\bf r} } \nonumber \\
&=& \hbar \int_{-\infty}^{\infty} \frac{d \omega}{\pi} 
\frac{\delta_{\ell m}}{1 - e^{-\hbar\omega/(k_B T)}} 
\mbox{Im} \chi_s ( k , \omega ).
\label{flucdiss}
\end{eqnarray}
{}From (\ref{phi1s}), we deduce that for $g \leq g_c$ 
it satisfies the scaling form
\begin{equation}
S ( k ) = 
\frac{N_0^2}{\rho_s} 
\frac{(\hbar c)^2}{k_B T} 
x_1^{\eta}~
\Xi_{1} ( \overline{k} , x_1 ),
\label{sksig1}
\end{equation}
where the universal function $\Xi_1$ is given by
\begin{equation}
\Xi_1 ( \overline{k} , x_1 ) = \int_{-\infty}^{\infty} \frac{d \overline{\omega}}{\pi}
\frac{1}{1 - e^{-\overline{\omega}}} \mbox{Im} \Phi_{1s} 
( \overline{k} , \overline{\omega}, x_1 ).
\label{xi1phi1s}
\end{equation} 
Similarly for $g \geq g_c$, we can relate $S ( k)$ to a universal function
$\Xi_2 ( \overline{k} , x_2 )$ with the prefactor $N_0^2 / \rho_s$ replaced
by ${\cal A}$ and the subscript $1 \rightarrow 2$.

\subsubsection{Antiferromagnetic correlation length}
We define the correlation length $\xi$ from the long-distance,
$e^{-r/\xi}$, decay of the equal time ${\bf n}$ - ${\bf n}$
correlation function. This correlation function will have such an
exponential decay for all $g$ provided $T \neq 0$ (the actual
asymptotic form also has powers of $r$ as a prefactor). 
Equivalently, one can define $\xi$
as $\kappa^{-1}$, where $i \kappa $ is the location of the pole of
$S(k)$ closest to the real $k$ axis. 
The scaling function for $\xi$ for $g \leq g_c$ is
\begin{equation}
\xi^{-1} = \frac{k_B T}{\hbar c} X_1 ( x_1 ).
\label{scalexi}
\end{equation}
For $g \geq g_c$ we have an identical form with $1 \rightarrow 2$.
Since the correlation lengths must match at $g=g_c$, we clearly have
$X_1 ( \infty ) = X_2 ( \infty )$. The universal 
linear $T$ dependence of $\xi^{-1}$ at $g=g_c$ ($x_1 = \infty$) 
was noted by
Chakravarty {\em et. al.\/}~\cite{CHN}.
\subsubsection{Local susceptibility}
\label{secdefchil}
{}From its definition (\ref{childef}), and the scaling form (\ref{phi1s}),
we can deduce the following for $g \leq g_c$:
\begin{equation}
\mbox{Im} \chi_{L}( \omega ) = \frac{N_0^2}{\rho_s} x_1^{\eta}
| \overline{\omega} |^{\eta} 
F_1 ( \overline{\omega} , x_1 ),
\label{chilscale}
\end{equation}
where the universal function $F_1$ is 
\begin{equation}
F_1 ( \overline{\omega} , x_1 ) = \frac{1}{\overline{\omega}^\eta} \int \frac{d^2 \overline{k}}{4 \pi^2}
\mbox{Im} \Phi_{1s} ( \overline{k} , \overline{\omega} , x_1 ).
\label{f1def}
\end{equation}
On general grounds we expect $\mbox{Im} \chi_L ( \omega ) \sim \omega$
for small $\omega$ and $T$ non-zero; this implies that 
\begin{equation}
F_1 \sim \mbox{sgn} ( \overline{\omega} ) 
|\overline{\omega}|^{1 - \eta}~~~~\mbox{for small $\overline{\omega}$}.
\end{equation}
In principle, the real part of $\chi_L$ also has a singular piece
which satisfies a scaling form analogous to (\ref{chilscale}); however 
the momentum integral in (\ref{f1def}) is divergent at the upper cutoff
(provided $\eta > 0$).
Thus $\mbox{Re} \chi_{L}$ has a leading contribution which is
non-universal and dominated by lattice scale physics. There is no such problem
for $\mbox{Im} \chi_{L}$ however - in this case the momentum integral 
sums over intermediate states which are on-shell and only long-wavelengths
contribute. Finally we note that a similar scaling form 
for $\mbox{Im} \chi_{L}$ for $g \geq g_c$ can be obtained by
replacing $N_0^2 / \rho_s$ by ${\cal A}$ 
and the substitution $1 \rightarrow 2$.

\subsubsection{NMR relaxation rate} 
We consider the relaxation of nuclear spins
coupled to electronic spins of the antiferromagnet ({\em e.g.\/}
$Cu$ spins in $La_2 Cu O_4$). We assume that the relaxation is dominated by
contributions near the antiferromagnetic ordering wavevector. After suitably
accounting for lattice-scale form factors and integrating out high-energy lattice
excitations, a coupling $A_{\pi}$ between the nuclear spins
and the antiferromagnetic order parameter ${\bf n}$ can be obtained.
The typical frequencies in NMR experiments are much smaller than the temperature
and the  
relaxation rate, $1/T_1 $ of the nuclear spins is given by
\begin{equation}
\frac{1}{T_1} = \lim_{\omega \rightarrow 0} 2 A_{\pi}^2 \frac{k_B T}{\hbar^2
\omega} \int \frac{d^2 k}{4 \pi^2} \mbox{Im} \chi_s ( k , \omega ) .
\label{1T1}
\end{equation}
{}From (\ref{chilscale}) we deduce the following result for $1/T_1$ for
$g \leq g_c$
\begin{equation}
\frac{1}{T_1} = \frac{2 A_{\pi}^2 N_0^2}{\hbar\rho_s} x_1^{\eta}
R_{1} ( x_1 ) ,
\label{nmrscale}
\end{equation}
where $R_{1} (x_1 ) $ is a universal function given by
\begin{equation}
R_{1} ( x_1 ) = \lim_{\overline{\omega} \searrow 0} 
\frac{F_1 ( \overline{\omega} , x_1 )}
{\overline{\omega}^{1-\eta}} .
\label{r1qdef}
\end{equation}
As before, the scaling form for $1/T_1$ for $g \geq g_c$ involves replacing
$N_0^2 / \rho_s$ by ${\cal A}$ and replacing $1 \rightarrow 2$ on the 
right-hand-side.

We will discuss the general form of our results for the scaling functions for
the different regions of the phase diagram in turn
(Fig.~\ref{phasediag}).  We will
consider first the quantum-critical region ($x_1 \gg 1$, $x_2 \gg 1$),
followed by the renormalized-classical ($x_1 \ll 1$) and the
quantum-disordered ($x_2 \ll 1$) regions.
Precise numerical results will not be presented here: the reader is
referred to Sections~\ref{NLsmodel}-\ref{secqd} for precise results
obtained in the $1/N$ expansion of the $O(N)$ non-linear sigma model.

\subsection{Quantum-critical region ($x_1 \gg 1$ or $x_2 \gg 1$)}
At short length/time scales the spin fluctuations in any nearly-critical
antiferromagnet should be indistinguishable from those at the critical point.
The special property of the quantum-critical region is that the deviations
from criticality at longer length/time scales arise primarily from the 
presence of a finite $T$. The fact that the ground state of the system
is not exactly at the critical point is never terribly important, and the
system does not display behavior characteristic of either ground state
at any length/energy scale. The critical
spin fluctuations are instead quenched in a universal way by thermal
relaxational effects. 

It should therefore be evident that there are two distinct types of
spin fluctuations
in wavevector/frequency space (Fig.~\ref{komdep}). 
With either $\hbar c k$ or $\hbar \omega$
significantly larger than $k_B T$, the spin-dynamics is that of
the critical 2+1 dimensional field theory of the critical point $g=g_c$.
Otherwise, damping from thermally-excited, critical spin-waves
produces a regime of quantum-relaxational dynamics. 

The crossover from the 2+1 critical to quantum-relaxational behavior
is clearly evident in the dynamic staggered 
susceptibility (See Fig.~\ref{figphi1s}).
In the 2+1 dimensional critical
region ($\overline{k} \gg 1$ or $\overline{\omega} \gg 1$) we have
\begin{equation}
\Phi_{1s} ( \overline{k}, \overline{\omega} , \infty ) = 
\frac{A_{Q}} {(\overline{k}^2 - \overline{\omega}^2 )^{1-\eta/2}}~;~~
\mbox{$\overline{k} \gg 1$ or $\overline{\omega} \gg 1$} ,
\label{phi1sqc}
\end{equation}
where $A_{Q}$ is a universal number,
and the exponent $\eta = 8/(3 \pi^2 N)$ at order $1/N$
but is known\cite{abe} to order $1/N^3$
For $N=3$, precision Monte Carlo simulations\cite{holm} place the value
of $\eta$ around $\eta \approx 0.028$; this extremely small, but positive,
value of $\eta$ will have important consequences for experiments.
Note that in this regime $\mbox{Im} \Phi_{1s}$ is non-zero
only for $\overline{\omega} > \overline{k}$ where one obtains a broadband
spectrum of critical spin-waves. The small value of $\eta$ implies however
that the damping is small and the spectrum is almost a delta-function.
 In the quantum-relaxational regime ($\overline{k} \ll 1$ and $\overline{\omega}
\ll 1$), there is strong damping due to thermally-excited, critical spin-waves
and excitations are not well-defined. However the spectrum of overdamped
spin-waves remains universal and is shown in Fig.~\ref{figphi1s} 

The same crossover is also present in the structure factor. Its
universal scaling function $\Xi_1 ( \overline{k}, \infty) = Z_Q^{-1}
\Xi_2 ( \overline{k} , \infty)$ has the 2+1 critical form
$\Xi_1 ( \overline{k}, \infty ) \sim \overline{k}^{-1+\eta}$ at large
$\overline{k}$, and a Lorentzian form at small $\overline{k}$. 

The various static observables have a value set by the absolute
temperature in a universal way. Their $T$ dependence can be
deduced easily from the scaling forms with the knowledge that
all scaling functions
were chosen to have a finite limit as $x_{1,2} \rightarrow \infty$.
The first corrections awa4y from $x_{1,2} = \infty$ are 
given by~\cite{erratum}
\begin{equation}
U(x_{1,2}) = U_{\infty} + U_{1,2} /x_{1,2}^{1/\nu} +
\ldots~~~~\mbox{$x_{1,2} \gg 1$}
\end{equation}
where $U$ represents any of the scaling functions $X$, $\Omega$,
$\Psi$ for the correlation length, uniform static susceptibility,
and specific heat respectively. The form of the subleading term
above follows from the requirement that the physics at finite
temperature is a smooth function of the bare coupling $g-g_c$.
Chakravarty {\em et. al.\/}~\cite{CHN} have noted the $x_1 = \infty$
result for the
case of the correlation length.

\subsection{Renormalized classical region ($x_1 \ll 1$)}
\label{intro_ren_cl}
We now describe our results at low temperatures on the ordered side,
$g < g_c$, $x_1 \ll 1$. Extensive results on a closely related regime
have been obtained by Chakravarty {\em et.al.\/}\cite{CHN} and 
Tyc {\em et.al.\/}\cite{Tyc}.       
The relationship between our and their results is discussed
below. We will find that, in the appropriate limit, our scaling
functions reduce exactly to theirs.

We begin with a qualitative discussion of the nature of the spin
correlations in the renormalized-classical region. The spin-fluctuations now 
fall naturally into {\em three} different regimes in wavevector-frequency
space (see Fig.~\ref{komdep}). 
At the largest $\overline{k}, \overline{\omega}$ 
we have 2+1
dimensional critical spin fluctuations which are essentially identical to those
in the quantum critical region and are therefore described by a
staggered spin susceptibility $\Phi_{1s}$ similar to that in (\ref{phi1sqc}).
Upon moving to longer distances/times, the first crossover occurs at
length (time) scales of order $\xi_J$ ($\xi_J / c$) to a `Goldstone' regime
where the spin dynamics is well described by rotationally averaged  spin-wave
fluctuations about a N\'{e}el ordered ground state. The scale $\xi_J$, controlling
the critical to Goldstone crossover,
is the Josephson correlation length~\cite{josephson}, 
and determines the vicinity of the
ground state of the antiferromagnet to the quantum phase transition. 
Near $g_c$, $\xi_J$ diverges as
\begin{equation}
\xi_J \sim (g_c - g )^{-\nu} .
\end{equation}
The second crossover in the renormalized classical region (Fig.~\ref{komdep})
occurs at the length scale $\xi$, which is  the actual correlation
length.  At this scale, strong, 
classical, two-dimensional, thermal fluctuations
of locally N\'{e}el ordered regions destroy the long range N\'{e}el order, so
that at
 scales larger than $\xi$, the antiferromagnet again appears
disordered, with all equal-time spin correlations decaying exponentially
in space. 
The scale $\xi$
is roughly given by
\begin{equation}
\xi \sim \xi_J \exp\left( \frac{N}{(N-2)x_1} \right) ,
\label{xit1}
\end{equation}
where we have omitted pre-exponential power-law factors of $x_1$.
For small $x_1$, $\xi$ is clearly much larger than $\xi_J$; the three
regimes in Fig.~\ref{komdep} are therefore well separated.

Our explicit results for the scaling functions will be
restricted to the vicinity
of the second crossover described above: thus they are valid when
$x_1 \ll 1$,
$k \xi_J \ll 1$, and $\omega \xi_J/c \ll 1$. In this regime, all of
our scaling functions, which in general depend upon three arguments
$\overline{k}$, $\overline{\omega}$, and $x_1$, collapse into
reduced scaling function depending only on 
two arguments measuring momentum and frequency,
$k\xi$ and $\omega \xi / c$, 
Further, the reduced scaling functions turn out to be exactly those
obtained by Chakravarty {\em et. al.\/}\cite{CHN} Tyc {\em et. al.\/}\cite{Tyc}
and Hasenfratz {\em et. al.\/}\cite{hasen1,hasen2}.
At first sight this result may
seem a bit surprising. Our results were obtained for antiferromagnets
with a small stiffness, {\em i.e.} $\rho_s \ll J$, while in other 
work\cite{CHN,Tyc,hasen1,hasen2}, no restrictions was 
 placed  on the value of $\rho_s$. 
The equivalence to order $1/N$ between the two theories is a consequence of
the fact
that the low $T$ results in the renormalized classical region 
contain {\em no corrections of order $\rho_s / J$} to order $1/N$. 
In a recent analysis
Hasenfratz {\em et. al.\/}\cite{hasen1,hasen2}. 
have suggested that such corrections are in fact 
absent at low $T$ at all orders
in $1/N$.

We show the nature of the collapse in the scaling functions
explicitly for the structure factor.
Chakravarty {\em et. al.\/}~\cite{CHN} proposed the scaling form
\begin{equation}
S(k) = S(0) ~f(k\xi) ,
\label{RD1}
\end{equation}
where $f$ is a universal function. We find that $f$ is related to 
the scaling function $\Xi_1$ by
\begin{eqnarray}
f(y) = \lambda_f \left(\frac{N}{N-2} \right)^{1/(N-2)} 
&& ~\lim_{x_1 \rightarrow 0} \left\{ 
 (x_1 )^{\eta - 1/(N-2)}  X_1^2 (x_1) \right. \nonumber \\
 \times \Xi_1 && \left. [ y X_1 (x_1) ,
x_1 ] \right\},
\label{fxi}
\end{eqnarray}
where the universal number $\lambda_f$ was found to be 
$\lambda_f = 1 - 0.188/N + {\cal O}(1/N^2)$. 

In Section~\ref{secrc} we will present the details of our
computations of the reduced scaling functions of the
renormalized-classical regime in a $1/N$ expansion. Our results agree
with those of Chakravarty {\em et. al.\/}~\cite{CHN}; however we are
also able to obtain a number of universal amplitudes which were
previously only determined by numerical simulations.

Another of our new results here is the low $T$ dependence ($x_1 \ll 1$) 
of the scaling
functions for the uniform
susceptibility and specific heat: 
\begin{eqnarray}
\Omega_1 ( x_1) &=& \frac{1}{\pi x_1} + \frac{N-2}{N\pi} \nonumber \\
\Psi_1 ( x_1 ) &=& N-1
\end{eqnarray}
These results have also been obtained independently by Hasenfratz
{\em et. al.\/}~\cite{hasen1}
The first of these results will be important to us later in the
comparison with experiments. The second result simply implies that
the number of low energy degrees of freedom are the $N-1$ spin waves.
The two results together also imply that the Wilson ration $W \sim
1/T$ at low temperatures provided $g < g_c$.
 
\subsection{Quantum disordered region ($x_2 \ll 1$)}
At $T=0$ ($x_2 = 0$), the ground state of the antiferromagnet has a
gap towards all excitations in this region. Unlike both previous
regions, therefore, finite $T$ is almost always a
weak perturbation on the $T=0$ results; all finite temperature
corrections are accompanied by factors of $\exp(-\Delta/(k_B T)) =
\exp (-1/x_2) \ll 1$. Furthermore, the ground state is rather well
described by the $N=\infty$ theory. 
We will therefore refrain, here, from giving complete
expressions for all the observables; we refer the reader to 
Section~\ref{Ninftyggtgc} for the exact results at
$N=\infty$. 

However, thermal effects and $1/N$ corrections are important at
measurement frequencies smaller than $\Delta$; in this region
the dynamics is controlled by a dilute concentration of thermally
excited quasiparticles. 
The dissipative effects of such quasiparticles will be discussed
in Section~\ref{secqc}.

We now discuss the plan of the remainder of the paper. 
In Section~\ref{phenom} we present a phenomenological derivation of the 
scaling forms used above. 
Section~\ref{NLsmodel} introduces the quantum $O(N)$ non-linear sigma model
and presents the complete solution for all the scaling functions
at $N=\infty$. 
The formal structure of the 
model at order $1/N$ is also discussed. The remainder of the paper contains the details
of the calculations. The calculations are discussed for the quantum-critical, 
renormalized-classical, and quantum-disordered regions in turn in 
Sections~\ref{secqc}, \ref{secrc}, \ref{secqd}.
Finally in Section~\ref{comp_exp} the comparison with experimental
results is presented. The appendixes contain 
a discussion of the effects of disorder and Berry phases, results of
a Monte-Carlo simulation, and some technical details.

\section{PHENOMENOLOGICAL DERIVATION OF SCALING FORMS}
\label{phenom}

In this section we will present a phenomenological derivation of the scaling
forms (\ref{phi1s}), (\ref{phi2s}) for the order-parameter dynamic 
susceptibility,
the scaling forms (\ref{phi1u}), (\ref{phi2u}) for the uniform spin 
susceptibility, and the scaling forms (\ref{psi1}), (\ref{psi2}) for the
specific heat. These
are valid in the
vicinity of a quantum phase transition in a two-dimensional quantum Heisenberg
antiferromagnet from a state with long-range N\'{e}el order to
a spin-fluid state. 
We will only consider the case in which the N\'{e}el order parameter is a
$3$-vector.
The following discussion does not explicitly refer
to the quantum non-linear sigma model. It should instead be regarded
more generally 
as a study of the consequences of the scaling hypothesis on 
a quantum phase transition in a Heisenberg antiferromagnet. 
The non-linear sigma model provides a 
realization
and verification of these hypothesis for a particular field theory.

Let us first present the precise ingredients from which our results
follow.
\begin{enumerate}
\item
The spin-wave velocity, $c$, should be non-singular at the $T=0$, quantum
fixed-point separating the two phases. 
We will also assume, for simplicity, that there is no spatial anisotropy;
this assumption is not crucial and our results can be easily extended to 
include quantum transitions in anisotropic systems like 
spin chains coupled in a plane. These systems will of course have two spin-wave
velocities, whose effects can be absorbed into a rescaling of lengths.
In the presence of spatial isotropy, and for the case of the
vector order-parameter, the non-singularity of the spin-wave velocity
implies that the critical field-theory has the Lorentz invariance of
2+1 dimensions.
\item
The antiferromagnet in the vicinity of the critical point
satisfies `hyperscaling'~\cite{Ma} hypothesis. For 
a quantum transition in $D=2+1$ dimensions, this hypothesis implies
that the singular part of the free energy density (${\cal F}_s$)
at $T=0$ has the form
\begin{equation}
{\cal F}_s = \hbar c \tilde{\Upsilon} \xi^{-D}
\label{twoscalef}
\end{equation}
where $\xi$ is the correlation length which diverges at the transition
(of course, on the ordered side $\xi = \xi_J$). The number $\tilde{\Upsilon}$
is dimensionless, and like all such numbers at the critical point,
it is expected to be universal. The statement of the universality
of $\tilde{\Upsilon}$ is known in the literature as
the hypothesis of `two-scale factor'
universality\cite{stauffer,two-scale}.
\item
Turning on a finite temperature places the critical field theory in a 
slab geometry which is infinite in the two spatial directions, but of 
finite length,  
\begin{equation}
L_{\tau} = \frac{\hbar c}{k_B T} ,
\label{ltdef}
\end{equation}
in the imaginary time ($\tau$) direction. The consequences of a finite $T$
can therefore be deduced by the principles of finite-size scaling\cite{binder}.
\end{enumerate}

In the following we measure temperature, time, and length in units
which make $k_B$, $\hbar$, and $c$ equal to unity.

\subsection{Staggered susceptibility, $g< g_c$}

Consider first the application of the scaling hypothesis to 
the staggered spin susceptibility for $g < g_c$. A straightforward application
of finite-size scaling yields
\begin{eqnarray}
\chi_s ( k , \omega ) &=& A L_{\tau}^{2-\eta} \tilde{\Phi} \left( k L_{\tau} , 
\omega L_{\tau} , \frac{\xi_J}{L\tau} \right) \nonumber \\
&=& \frac{A}{T^{2-\eta}} \tilde{\Phi} \left( \overline{k} , 
\overline{\omega} , \frac{\xi_J}{L\tau} \right) ,
\label{finsize}
\end{eqnarray}
where $A$ is a non-universal amplitude, and the scaling function
$\tilde{\Phi}$ is universal upto an overall pre-factor. In particular there
are no non-universal metric factors\cite{privman} in any 
of the three arguments of $\tilde{\Phi}$.
We now wish to eliminate the dependence of this result on $\xi_J$.
It has been argued recently~\cite{matt,helicity} that the result (\ref{twoscalef})
can be extended to deduce a simple, universal
relationship between $\rho_s$ and $\xi_J$, valid in the limit
$\xi_J \rightarrow \infty$:
\begin{equation}
\rho_s = \frac{\hbar c \Upsilon}{\xi_J}
\label{twoscale}
\end{equation}
where $\Upsilon$ is another universal number. Now,
from (\ref{x1def}), (\ref{twoscale}) and (\ref{ltdef}) we see that that the
third argument
\begin{equation}
\frac{\xi_J}{L_{\tau}} = \frac{\Upsilon T}{\rho_s} = \frac{N
\Upsilon}{2 \pi} x_1 .
\end{equation}
We therefore define a new universal function $\Phi_{1s}$
\begin{equation}
\Phi_{1s} ( \overline{k} , \overline{\omega} , x_1 ) \equiv
\tilde{\Phi} \left( \overline{k}, 
\overline{\omega}, \frac{N \Upsilon}{2 \pi} x_1 \right) .
\label{phi1sphit}
\end{equation}
We shall soon see that $\Phi_{1s}$ is the same function as in (\ref{phi1s}).

To make $\Phi_{1s}$ completely 
universal, we have to fix its overall scale, which we now do. 
First, we notice that in the renormalized
classical region ($x_1 \ll 1$), there is a Goldstone 
regime ($k \xi_J \ll 1$, $k \xi 
\gg 1$, see Fig.~\ref{komdep} 
and Section~\ref{intro_ren_cl}) of non-interacting spin-waves.
In this regime, the hydrodynamics predicts that the
static staggered susceptibility has a simple form:
\begin{eqnarray}
\chi_s ( k , \omega = 0) = \left( 1 - \frac{1}{N} \right) &&
\frac{N_0^2}{\rho_s k^2}~;\nonumber \\
x_1 \ll 1,&&~~~k \xi_J \ll 1,~~~ k \xi \gg 1 .
\label{goldstone}
\end{eqnarray}
The scale of the susceptibility has been set by the magnitude of $T=0$
staggered moment. The factor of $(1-1/N)$ arises from the fact that only
$N-1$ transverse modes contribute the Goldstone singularity, while the 
longitudinal mode is massive; after rotationally averaging this induces
a factor of $(N-1)/N$. We now demand that in the appropriate limit,
the scaling-form (\ref{finsize})
obey this Goldstone form. The key constraint is that (\ref{goldstone}) is
$T$ independent. It is easy to show that 
this can be satisfied by (\ref{finsize}) only if
\begin{eqnarray}
\Phi_{1s} ( \overline{k} , \overline{\omega} =0 , x_1 ) && = 
\frac{b}{\overline{k}^2 x_1^{\eta}}~; \nonumber \\
x_1 \ll 1, ~~~&& \overline{k} x_1 \ll 1,~~~\overline{k} e^{N/((N-2)x_1)}
\gg 1 ,
\label{phi1sscale}
\end{eqnarray}
for some constant $b$.
In the last restriction on $\overline{k}$ we have used (\ref{xit1}), valid
for the $O(N)$ sigma model and neglected pre-exponential factors of $x_1$.
The prefactor in (\ref{phi1sscale}) is of course arbitrary. We now make
the specific choice, $b = 1 - 1/N$ 
and thus specify the overall scale of $\Phi_{1s}$.
Comparing (\ref{finsize}), (\ref{phi1sphit}), (\ref{goldstone}), and 
(\ref{phi1sscale}), we can now fix the value of the prefactor $A$
\begin{equation}
A = \frac{N_0^2}{\rho_s} \left( \frac{N}{2 \pi \rho_s} \right)^{\eta} .
\end{equation}
Inserting this value of $A$ into (\ref{finsize}), we obtain (\ref{phi1s}) as
desired.

\subsection{Staggered susceptibility, $g > g_c$}
\label{phenomggtgc}

A similar analysis can be carried out in the spin-fluid phase
for the staggered susceptibility and its scaling function $\Phi_{2s}$ as 
in (\ref{phi2s}). 
We remind the reader that our theory is valid only for antiferromagnets with
a $3$-vector order parameter, in which case the quantum disordered phase is 
expected to have only integer spin excitations\cite{trieste,sun,spn}; 
in particular if spin-1/2 spinons
are present at intermediate scales, they are always confined at the longest distances.

At $T=0$ the equal-time order parameter correlation 
function will decay in space with a correlation length $\xi$. As the theory
has a Lorentz invariance, this implies that the gap towards spin-1
excitations, $\Delta$, is
\begin{equation}
\Delta = \frac{\hbar c}{\xi},
\end{equation}
where we have momentarily reinserted explicit factors of $\hbar$ and $c$.
An application of finite-size scaling, very similar to that for the ordered
side, now yields immediately the scaling function (we now return to
units in which $k_B = \hbar = c = 1$)
\begin{equation}
\chi_{s} ( k , \omega ) = \frac{\tilde{A}}{T^{2-\eta}} \Phi_{2s} \left( 
\overline{k}, \overline{\omega}, \frac{T}{\Delta} \right) .
\label{finsize2s}
\end{equation}
The only non-universal components on the right-hand-side are the amplitude
$\tilde{A}$ and the related overall scale of the function $\Phi_{2s}$. 
As before, we will fix this scale by matching with an experimental observable 
at $T=0$. Let us therefore think about the nature of the spectrum in the
spin-fluid phase at $T=0$. As all excitations have a gap, the spin 1 
quasiparticle should have an infinite lifetime for energies close enough to
the threshold $\Delta$ (this quasiparticle appears as the bound state of two spinons
in large $M$ theories of $SU(M)$ and $Sp(M)$ antiferromagnets\cite{sun,spn}). 
Further, the Lorentz invariance of the theory
implies that the dispersion spectrum, $\omega_k$ of this spin-1 quasiparticle
is given by
\begin{equation}
\omega_k = \sqrt{k^2 + \Delta^2}
\end{equation}
for small enough $k$. These facts combine to imply the following form
for $\chi_s$ at $T=0$ 
\begin{eqnarray}
\chi_s ( k, \omega ) = \frac{{\cal A}}{k^2 - (\omega + i \varepsilon)^2
+ \Delta^2}~; && \nonumber \\
T=0~~~, k \ll \Delta,~&&~~|\omega-\Delta| \ll \Delta ,
\label{chist0}
\end{eqnarray}
where $\varepsilon$ is a positive infinitesimal and ${\cal A}$ is the
spin-1 quasiparticle residue. This residue can be experimentally 
measured 
by examining the imaginary part of the local susceptibility, 
$\mbox{Im}\chi_L$. Using (\ref{childef}) and (\ref{chist0})
we find
\begin{equation}
\mbox{Im} \chi_L ( \omega ) = \frac{{\cal A}}{4} 
\theta( \omega - \Delta )
~;~~~T=0,~~~|\omega-\Delta| \ll \Delta .
\end{equation}
The discontinuity in the local dynamic susceptibility is therefore precisely
a quarter of the spin-1 quasiparticle amplitude.
We now demand that the scaling form (\ref{finsize2s}) satisfy (\ref{chist0})
as $T \rightarrow 0$. A little experimentation shows that this is
only possible if
\begin{eqnarray}
\Phi_{2s} ( \overline{k}, \overline{\omega} , x_2 ) = 
\frac{x_2^{2-\eta}}{(\overline{k} x_2 )^2 - (\overline{\omega} x_2 
+ i \varepsilon )^2
+ 1}~; \nonumber \\
x_2 \ll 1,~\overline{k} x_2 \ll 1,~|\overline{\omega} x_2 
-&& 1| \ll 1 .
\end{eqnarray}
Here we have arbitrarily set the overall scale
of $\Phi_{2s}$; this function is now completely universal.
Finally, the derivation
of (\ref{phi2s}) is completed by obtaining the amplitude $
\tilde{A}$ in (\ref{finsize2s}): 
\begin{equation}
\tilde{A} = \frac{{\cal A}}{\Delta^{\eta}} .
\end{equation}
The prefactor $\tilde{A}$ is expected to be non-singular as $g \searrow g_c$;
therefore the quasiparticle amplitude ${\cal A}$ must vanish as
\begin{equation}
{\cal A} \sim (g - g_c )^{\eta \nu} .
\end{equation}
This was noted earlier in (\ref{aen}).

\subsection{Uniform susceptibility}
The key ingredient in the determination of the scaling form for $\chi_u$
is the realization that $\chi_u$ is simply a stiffness related to twists in
boundary conditions on the system along the 
imaginary time ($\tau$) direction\cite{Fisher}.
A uniform magnetic field on the antiferromagnet causes a precession of all
the spins at the same rate. The relative angle between any two spins remains
unchanged. Therefore by transforming to a rotating reference frame almost
all vestiges of the magnetic field can be removed. However the partition
function in the laboratory frame had periodic boundary conditions along the
$\tau$ direction, implying that the system in the rotating frame has a twist
in its boundary condition. The susceptibility is the response to such a twist,
which is precisely the spin stiffness.

The Lorentz invariance of the theory now implies that the scaling 
properties of $\chi_u$ should be the same as those of $\rho_s$, which is
the stiffness for twists about the {\em spatial} boundary conditions. 
In other words, the scaling dimension of $\chi_u$ is exactly $D-2$,
and at $T=0$ the combination $\xi_J \chi_u$ approaches
a universal number as $g \nearrow g_c$.
The scaling laws (\ref{phi1u}), (\ref{phi2u}) are now a completely
straightforward consequence of the principles of finite-size scaling.
Unlike $\chi_s$, there is no need for any normalization condition to set
the overall scale of the scaling function. It is already fixed to a 
universal value by the hypothesis of two-scale factor 
universality\cite{stauffer,two-scale,matt}.

\subsection{Specific heat}

Consider the $D=2+1$ dimensional Lorentz invariant theory in a slab geometry
in the vicinity of $g=g_c$.
An early paper by Fisher and de Gennes\cite{degennes} argued by extending
(\ref{twoscalef}) to finite sizes, that the
free energy density ${\cal F}$ must have the following dependence on the size,
$L_{\tau}$ of the finite dimension
\begin{equation}
{\cal F}  = {\cal F}_0 + \frac{1}{L_{\tau}^D} \varphi \left( \frac{\xi_J}{L_\tau}
\right) ,
\label{degennes}
\end{equation}
where ${\cal F}_0$ is the bulk free energy density and
we have assumed that $g \leq g_c$.
The function $\varphi$ is universal at all $x$;
there are no non-universal metric
factors in either the argument or the scale of $\varphi$\cite{privman}. 
The scaling function (\ref{psi1}) for $C_V$ now follows immediately from
the thermodynamic relationship between ${\cal F}$ and $C_V$,
the relationship (\ref{twoscale}) between $\rho_s$ and $\xi_J$,
and the relationship (\ref{ltdef}) between $L_{\tau}$ and $T$. An entirely
analogous argument can be made for $g \geq g_c$.

We note that a related result has been discussed recently by
Castro 
Neto and Fradkin\cite{eduardo} in the 
context of 2+1 dimensional quantum systems;
they have also discussed an interesting connection to the 
Zamalodchikov's C-theorem\cite{zama}. 
We have chosen a numerical prefactor of the scaling function $\Psi_1$
in (\ref{psi1}) which is the specific heat of a 
single gapless bose degree of freedom with dispersion $\omega = c k$ in
2 dimensions. The number $\Psi_1 (T \rightarrow 0)$ is thus a measure of the 
effective number of such modes in the ground state.
For $g < g_c$, this number should equal
$N-1$, the number of spin-wave modes in the ordered state.
For $g=g_c$, the number $\Psi_1 ( \infty )$ is probably irrational and
will be calculated later in this paper to order $1/N$. 
Finally, in the quantum disordered phase, $g > g_c$, there are no gapless
modes and we should have $\Psi_2 ( 0 ) = 0 $.

A very similar connection between the effective 
number of gapless modes and $C_V$ was
established some time ago
for $1+1$ dimensional spin chains\cite{Blote}; in this case
$C_V \sim T$ with $C_V / T$ universal and related to the central charge of a
conformal field theory, which in effect measures 
the number of gapless quasiparticle modes.
 
\section{THE QUANTUM $O(N)$ NON-LINEAR SIGMA MODEL}
\label{NLsmodel}

In this section we will discuss the $O(N)$ quantum non-linear sigma model
field theory. This theory may be viewed as the simplest model which
displays a quantum phase transition in 2+1 dimensions. 
Moreover, a microscopic connection between weakly frustrated antiferromagnets
with short-range N\'{e}el order and the $O(3)$ sigma model can
also be established\cite{trieste}.

There are several subtle and difficult questions relating
to the consequences of Berry phase terms which are present in the
antiferromagnet but are absent in the sigma model\cite{sun}.
In this paper we will simply neglect the effects of the Berry phase
terms. 
In Appendix~\ref{appdiv} we will present some circumstantial
evidence, in computations for $SU(M)$ antiferromagnets,
which suggests that these Berry phases are irrelevant both in the
N\'{e}el phase and at the quantum critical point, but do significantly
modify the properties of the quantum disordered phase. In critical phenomena
terminology this implies that the Berry phases are `dangerously-irrelevant'.
As all of our scaling functions are properties of flows in the vicinity
of the quantum-critical point, we do not expect any modifications of 
our results by Berry phase effects in this scenario. 
Berry phases will however modify
the corrections to scaling.

In passing, we note that there is an alternative expansion which could
have been used to obtain the scaling functions of this paper: this is the
large $M$ 
expansion about antiferromagnets with 
$SU(M)$ or $Sp(M)$ symmetry~\cite{sun,spn,AA}.
However, 
the presence of a gapless gauge field in the perturbative $1/M$ corrections,
makes this expansion
somewhat more involved than the $O(N)$ expansion.
For general $M$, $N$, both approaches predict that the lowest-lying non-zero
spin excitation above the quantum-disordered ground state in an 
antiferromagnet with collinear order~\cite{spn} carries
spin $S=1$; the detailed structure of the spectrum at higher energies
is however different in the two theories.
The results of this paper show that the $O(N)$ expansion is numerically
much more accurate at $N=3$ than is the $Sp(M)$ or $SU(M)$ expansion
at $M=2$. This can be seen immediately by comparing the values of $\eta$
in the two theories: the large $M$ theory gives $\eta = 1 - {\cal O}(1/M)$
while the large $N$ theory has $\eta = 0 + {\cal O}(1/N)$ -
compare this with the known value\cite{holm} 
in the $D=3$ classical Heisenberg
model $\eta \approx 0.028$.
We shall see below that, at $N=3$,
the $1/N$ corrections in the $O(N)$ model to the universal scaling
functions
are almost always less than about 20\% of the leading $N=\infty$ term.
 
We will begin this section by a definition of the quantum $O(N)$
non-linear sigma model.
The first subsection will present its 
exact solution at $N=\infty$. 
This solution has also been discussed earlier by 
Rosenstein {\em et. al.\/}~\cite{Rosen}, although they did not emphasize the
universal scaling properties of the solution. 
For clarity, we will repeat some of the step in Ref.~\cite{Rosen} and will
then explicitly compute
all of the scaling functions introduced in
Section~\ref{intro} in the $N=\infty$ limit.
The $N=\infty$ computations are also similar to earlier
studies of finite-size scaling properties of the spherical 
model~\cite{spherical} - in our case the inverse temperature plays the
role of the finite-size along the time direction.
The second subsection will present complete formal expressions for the
staggered and uniform susceptibility which are correct to order
$1/N$; to the best of our knowledge,
these constitute the first computations of finite-size corrections
to two-loop order in any system.
Subsequent sections will manipulate these expressions
into the appropriate scaling forms
for the three regions of Fig.~\ref{phasediag}. 
The structure of the $1/N$ corrections 
is however rather involved
and the casual reader may be satisfied by studying only 
the $N=\infty$ solution of Section~\ref{NLsmodelNinfty}. 
Even this limited solution is quite 
rich and instructive; its main shortcoming is 
the absence of spin-wave damping and anomalous dimensions which appear
only at order $1/N$.

The $O(N)$
non-linear sigma model is defined by the functional integral
\widetext
\begin{equation}
Z = \int {\cal D} n_{\ell} \delta ( n_{\ell}^2 -1 ) 
\exp \left( - \frac{\rho_s^{0}}{2\hbar} \int d^2 {\bf r}
\int_0^{\hbar/k_B T} d {\tau} \left[ 
(\nabla_{{\bf r}} n_{\ell} )^2 + \frac{1}{c_0^2}(\partial_{\tau}
n_{\ell} )^2
\right ] \right),
\label{nlsz}
\end{equation}
\narrowtext
where the index $\ell$ 
runs from $1$ to $N$, $\rho_s^0$ is the bare spin stiffness,
and $c_0$ is the 
bare spin-wave velocity. Both $\rho_s^0$ and $c_0$ differ from
their renormalized values $\rho_s$ and $c$; however at $N=\infty$
we will find $c_0=c$, although 
the renormalization of $c_0$ will be quite crucial
in subsequent sections. Further analysis is simply expressed in terms
of the coupling constant,
\begin{equation}
g = \frac{N\hbar c_0}{\rho_s^0} ,
\end{equation}
which has the units of inverse-length. 
We will find that the quantum transition occurs at a $g$ of order
unity. This implies that we have to choose
$\rho_s^0 \sim N$ in the large $N$ limit. 
In the remainder of this section
we will use units such that $k_B = \hbar = c_0 = 1$.
The large $N$ analysis of $Z$ begins with the introduction of the
rescaled field,
\begin{equation}
\tilde{n}_{\ell} = \sqrt{N} n_{\ell} ,
\end{equation}
and the imposition of the constraint by a Lagrange multiplier
$\lambda$. This transforms $Z$ into
\widetext
\begin{equation}
Z = \int {\cal D} \tilde{n}_{\ell} {\cal D} \lambda 
\exp \left( - \frac{1}{2g} \int d^2 {\bf r}
\int_0^{\beta} d \tau \left[ 
(\nabla_{{\bf r}} \tilde{n}_{\ell} )^2 + (\partial_{\tau}
\tilde{n}_{\ell} )^2 + i \lambda (\tilde{n}_{\ell}^2 - N)
\right ] \right) .
\end{equation}
\narrowtext
This action is quadratic in the $n_{\ell}$, which can therefore be
integrated out. This induces an effective action for the $\lambda$
field which has the useful feature of having all its $N$ dependence
in a prefactor. Therefore, for large $N$ the $\lambda$ functional
integral can be evaluated by using its value at its saddle-point.
Terms higher-order in $1/N$ can be obtained by a systematic expansion
of the functional integral about this saddle-point.
We parametrize the saddle-point value of $\lambda$ by
\begin{equation}
i \langle \lambda \rangle = m^2 ,
\end{equation} 
where the `mass' $m$ is to be determined by solving the 
constraint equation $n_{\ell}^2 = 1$, order-by-order in $1/N$. 
This value of $m$ can then be used to obtain a $1/N$ expansion 
of the $n_{\ell}$-$n_{\ell}$ correlator
and hence of all the observables related to the staggered susceptibility.

Determination of the uniform susceptibility requires introduction of 
a slowly-varying magnetic field $\vec{B}$ into the non-linear sigma model.
In the $O(3)$ model such a field causes a precession of the 
local order parameter about the magnetic field axis. This precession
is realized in the $O(3)$ sigma model by the substitution\cite{Fisher}
\begin{equation}
\partial_{\tau} \vec{n} \rightarrow \partial_{\tau} \vec{n} 
- i \frac{g \mu_B}{\hbar} \vec{B} \times \vec{n}
\end{equation}
in functional integral $Z$; here $g \mu_B /\hbar$ is the Bohr magneton for
a single spin; in the remainder of this section we will measure the field in 
units of $g \mu_B /\hbar$ and hence omit explicit factors of $g \mu_B /\hbar$. 
For the general $O(N)$, the analog of the magnetic field is
a second-rank antisymmetric tensor $b_{\ell m}$ which causes a
precession of the spin-components lying in the plane defined by
the $\ell$, $m$ directions. For $N=3$, $b_{\ell m}$ is related to
$\vec{B}$ by
\begin{equation}
b_{\ell m} = \epsilon_{\ell m p} B_{p} .
\end{equation}
The full action in the presence of the $b$ field is defined by 
(\ref{nlsz}) and the substitution
\begin{equation}
\partial_{\tau} n_{\ell} \rightarrow \partial_{\tau} n_{\ell}
- i  b_{\ell m} n_m .
\end{equation}
The uniform susceptibility, $\chi_u$ is now obtained by evaluating $\ln Z$ 
in powers of $b$ and picking out the coefficient of the quadratic term:
\begin{equation}
\chi_u = \frac{1}{2T V} \frac{\partial^2}{\partial b_{\ell m}^2}
\log Z ,
\label{chiuz}
\end{equation}
where $V$ is the volume of the system.

\subsection{Solution at $N=\infty$}
\label{NLsmodelNinfty}

The $O(N)$ sigma model can be solved exactly at $N=\infty$. Closed
form expressions for all the scaling functions introduced in
Section~\ref{intro} can be easily obtained. We begin with the
staggered spin susceptibility which
is given by
\begin{equation}
\chi_s ( k , \omega ) = \frac{g\tilde{S}^2}{N} \frac{1}
{k^2 - (\omega + i \varepsilon)^2 + m_0^2 } .
\label{chisNinfty}
\end{equation}
We have introduced here the mass $m_0$ which is the value of $m$ at
$N=\infty$, and $\tilde{S} = Z_S S$ which is a rescaling factor between 
the actual susceptibility of a quantum spin-S antiferromagnet and the 
susceptibility of the unit $n$-field in the $O(3)$ sigma model.
The renormalization factor $Z_S$ accounts for the fluctuations at short
scales, which have to be integrated out in the derivation of the sigma model
from the original spin Hamiltonian.
Expressions for all the experimental observables dependent upon 
$\chi_s$ can now be easily obtained; we will refrain from giving explicit 
expressions. One quantity we will need is the correlation length $\xi$,
which we defined earlier from the long-distance decay $\sim e^{-r/\xi}$ of the
equal-time $n$-$n$ correlation function. Such a decay is present at all
finite $T$ for all values of $g$. At $N=\infty$ we find
\begin{equation}
\xi = \frac{1}{m_0} .
\label{xim}
\end{equation}

Now turn to the uniform spin susceptibility. There is no damping at
$N=\infty$ and hence the spin-diffusion constant is infinite. We
therefore consider only the value of the static susceptibility.
Evaluating (\ref{chiuz}) at $N=\infty$ we find
\begin{equation}
\chi_u^{{\rm st}} =  2 T \sum_{\omega_n} \int \frac{d^2 k}{4 \pi^2}
\frac{\epsilon_k^2 - \omega_n^2}{(\epsilon_k^2 + \omega_n^2 )^2} ,
\label{chiustwn}
\end{equation}
where $\epsilon_k = k^2 + m_0^2$, and $\omega_n$ is the Matsubara frequency
which takes the values $2\pi n T$ with $n$ integer. 
The frequency summation and the
subsequent momentum integration can be performed exactly. Moreover, the 
momentum integration is convergent in the ultraviolet, and final result
depends only on $m_0$ and $T$:
\begin{equation}
\chi_u^{{\rm st}} =  \frac{T}{\pi} \left[
\frac{m_0}{T} \frac{e^{m_0 / T}}{e^{m_0 /T} - 1} - \log( e^{m_0 / T} - 1)
\right].
\label{chiustm0}
\end{equation}
Note the absence of direct dependence on the coupling $g$.

Consider next the free energy density, ${\cal F}$, from which $C_V$
can be obtained by taking two temperature derivatives. At $N=\infty$,
${\cal F}$ can be directly obtained from the saddle-point value of
the effective action. The result is~\cite{Rosen,Dashen}
\begin{eqnarray}
{\cal F} &=& \frac{NT}{2} \sum_{\omega_n} \int \frac{d^2 k}{4 \pi^2}
\log ( k^2 + \omega_n^2 + m_0^2 ) - \frac{N m_0^2}{2 g} \nonumber \\
&=& N T\int \frac{d^2 k}{4 \pi^2} \log( 1 - e^{-\epsilon_k /T} ) \nonumber \\
&~&~~
+ \frac{N}{2} \int \frac{d^2 k d \omega}{8 \pi^3} \log ( k^2 +
\omega^2 + m_0^2 ) - \frac{N m_0^2}{2 g},
\label{frenNinfty}
\end{eqnarray}

It now remains to determine the dependence of $m_0$ on $g$ and $T$.
The constraint $n_a^2 = 1$ takes the following form at $M=\infty$:
\begin{equation}
T \sum_{} \int \frac{d^2 k}{4 \pi^2} \frac{ g}{ k^2 + \omega_n^2 +
m_0^2} = 1.
\end{equation}
It is easy to see that the momentum summation is
divergent in the ultraviolet,
and it is therefore necessary to introduce an cut-off.
We use a relativistic 
Pauli-Villars cut-off at the momentum scale $\Lambda$, assumed to be
much larger than the temperature. 
This transforms the constraint
equation into
\begin{equation}
T \sum_{} \int_0^{\infty} \frac{d^2  k}{4 \pi^2} \left( 
\frac{ g}{ k^2 + \omega_n^2 +
m_0^2} - \frac{g}{ k^2 + \omega_n^2 + \Lambda^2 } \right) = 1.
\end{equation}
The momentum integration is now convergent in the ultraviolet and can
be performed exactly. The subsequent frequency summation is also
tractable and yields 
\begin{equation}
\frac{gT}{2 \pi} \mbox{ln} \left(
\frac{\mbox{sinh}(\Lambda/2T)}{\mbox{sinh}(m_0 /2T)} \right) = 1.
\end{equation}
Finally, we can solve for the dependence of $m_0$ on $T, g,$ and
$\Lambda$ 
\begin{equation}
m_0 = \frac{1}{\xi} = 2T \mbox{arcsinh} \left( \exp\left(-\frac{2\pi}{gT}\right)
\mbox{sinh} \left(\frac{\Lambda}{2T} \right) \right).
\label{mval}
\end{equation}
In the limit $T \ll \Lambda$, this equation can be rewritten as~\cite{Rosen}
\begin{equation}
m_0 = 2T \mbox{arcsinh} \left[ \frac{1}{2}\exp\left(-\frac{2\pi}{T}
\left(\frac{1}{g} - \frac{1}{g_c} \right),
 \right) \right]
\label{mval2}
\end{equation}
where
\begin{equation}
g_c = \frac{4 \pi}{\Lambda}.
\end{equation}
By examining the $T \rightarrow 0$ limit of this equation, it is
immediately apparent that the behavior of $\xi$ is quite different
depending upon whether $g$ is smaller, larger, or close to the critical
value $g_c$.
For $g > g_c$, $\xi$ approaches a finite value as $T \rightarrow 0$,
while for $g \leq g_c$, $\xi$ diverges as $T \rightarrow 0$. The finite,
$T=0$ asymptote of $\xi$ for $g > g_c$ itself diverges as $g$ approaches
$g_c$. We find
\begin{equation}
\xi ( T=0, g > g_c ) \sim (g - g_c )^{-1} .
\end{equation}
This identifies the $N=\infty$ value of the exponent $\nu$ to be
\begin{equation}
\nu = 1 .
\end{equation}
For $g < g_c$ these is a Josephson scale\cite{josephson} 
$\xi_J$ which diverges with the same
exponent $\nu$ as 
$g$ approaches $g_c$. However, this scale is not present in the $N=\infty$
theory: this is because the exponent $\eta = 0$ at $N=\infty$,
making the critical and Goldstone
fluctuations indistinguishable.
At $g=g_c$, we have from (\ref{mval})
\begin{equation}
m_0 = \Theta T , 
\end{equation}
where the number
\begin{equation}
\Theta = 2 \log \left( \frac{\sqrt{5}+1}{2} \right) = 0.962424 
\end{equation}
will occur frequently in our analysis.
We see therefore that the correlation length scales with $1/T$ at the 
critical point.

We now examine the scaling limit of the above results. This limit
is obtained when the temperature $T$, and the deviation from criticality
$|g - g_c|/g_c^2$ are both much smaller than the upper cutoff $\Lambda$.
We will consider the cases of $g$ smaller or greater than $g_c$ separately.

\subsubsection{Scaling properties for $g \leq g_c$}
{}From the discussions in Sections~\ref{introneel} and~\ref{phenom} it is 
clear that the results become
simple after they have been expressed in terms 
of the $T=0$ ordered staggered moment
$N_0$ and the fully renormalized, $T=0$ spin-stiffness $\rho_s$. 
In Appendix~\ref{appent=0} 
we have performed a $1/N$ expansion of these $T=0$ variables.
At $N=\infty$ their exact values are
\begin{eqnarray}
\rho_s &=& N \left( \frac{1}{g} - \frac{1}{g_c} \right) , \nonumber \\
N^{2}_0  &=& \tilde{S}^2 \left( 1 - \frac{g}{g_c} \right) .
\label{rhosNinfty}
\end{eqnarray}
We now use these results with the expression (\ref{mval}) for $\xi$
to study the limit $\rho_s /N , T \ll \Lambda$. It is not difficult to show
then that $\xi$ takes the scaling form (\ref{scalexi})
where the scaling function $X_1$ is 
\begin{equation}
X_1 ( x_1 ) = 2 ~\mbox{arcsinh}~\left( \frac{e^{-1/x_1}}{2} \right).
\label{X1Ninfty}
\end{equation}
Thus, in the renormalized classical region $x_1 \ll 1$, we have
\begin{equation}
X_1 = e^{-1/x_1}~;~~~~~x_1 \ll 1,
\label{Xsmallx1}
\end{equation}
implying a correlation length which is exponentially large,
while in the quantum-critical region ($x_1 \gg 1$) we find
\begin{equation}
X_1 = 2 \log \left( \frac{\sqrt{5} + 1}{2} \right) - \frac{2}{\sqrt{5} x_1 }
~;~~~~~x_1 \gg 1,
\end{equation}
indicating that the correlation length, $\xi = 1/(T X_1)$ scales with $1/T$.

The scaling functions for all the other observables 
now follow in a straightforward
manner. The scaling function $\Phi_{1s}$ for the staggered susceptibility
in (\ref{phi1s}) and the exponent $\eta$ 
can be deduced from (\ref{chisNinfty}), (\ref{xim}), 
(\ref{rhosNinfty}), and (\ref{scalexi}) to be
\begin{eqnarray}
\eta &=& 0 ,\nonumber \\
\Phi_{1s} ( \overline{k}, \overline{\omega} , x_1 ) &=& \frac{1}{ \overline{k}^2
- (\overline{\omega} + i \varepsilon )^2 + X_1^2 ( x_1 )}.
\label{phi1sNinfty}
\end{eqnarray}
The scaling functions for the observables dependent upon $\chi_s$ now
follow immediately. For the structure factor $S(k)$ we have the scaling function
$\Xi_1$ in (\ref{sksig1}) which is
\begin{equation}
\Xi_1 ( \overline{k}, x_1 ) = \frac{1}{2 [\overline{k}^2 +
X_1^2 (x_1 )]^{1/2}} \coth \frac{[\overline{k}^2 + X_1^2 ( x_1 )]^{1/2}}{2}.
\label{XiNinfty}
\end{equation}
The local susceptibility $\mbox{Im} \chi_L ( \omega )$ obeys 
(\ref{chilscale}) with the scaling function $F_1$ given by
\begin{equation}
F_1 ( \overline{\omega} , x_1 ) = \frac{1}{4} \left[
\theta( \overline{\omega} - X_1 ( x_1 )) - \theta( -
\overline{\omega} - X_1 ( x_1 )) \right],
\end{equation}
where $\theta(x)$ is the unit step function. The presence of a gap in
$F_1$ at finite $T$ is an artifact of the $N=\infty$ theory and will
be cured upon including $1/N$ corrections.
A consequence of this gap is that there is no relaxation of nuclear
spins at $N=\infty$ and the scaling function for $1/T_1$ in
(\ref{nmrscale}) is
\begin{equation}
R_{1} ( x_1 ) = 0.
\end{equation}

The result for the uniform susceptibility, $\chi_u^{{\rm st}}$, 
(\ref{chiustm0}) can also be
collapsed into the scaling form (\ref{chistdef}). We use (\ref{chiustm0}),
the expression (\ref{mval}) for $m_0$, and (\ref{rhosNinfty}) to obtain
\begin{equation}
\Omega_{1} (x_1 ) = \frac{1}{\pi x_1} + \frac{\sqrt{4 + e^{-2/x_1}}}{
\pi e^{-1/x_1}} \mbox{arcsinh} \left( \frac{e^{-1/x_1}}{2} \right).
\label{Omega1func}
\end{equation}
In the renormalized classical limit $x_1 \ll 1$ this function has the 
limiting behavior
\begin{equation}
\Omega_{1} (x_1 ) = \frac{1}{\pi x_1} + \frac{1}{\pi} ~;~~~~~~~x_1 \ll 1,
\end{equation}
while in the quantum-critical region ($x_1 \gg 1$) it obeys
\begin{equation}
\Omega_{1} ( x_1 ) = \frac{\sqrt{5}}{\pi} \log \left( \frac{\sqrt{5} + 1}{2}
\right) \left[ 1 + \frac{4}{5 x_1} \right]~;~~~x_1 \gg 1.
\label{Omega1subleading}
\end{equation}

We turn now to the free-energy density, ${\cal F}$ and the specific
heat $C_V$. We will evaluate ${\cal F}(T) - {\cal F}(T=0)$ at a fixed
value of $g$. The calculations are performed most easily if we
use relativistic cutoff and write the value of 
$g_c$ in  the following form
\begin{equation}
\frac{1}{g_c} = \int \frac{d^3 P}{8 \pi^3} \frac{1}{P^2},
\label{gc3mom}
\end{equation}
where $P = ({\bf k}, \omega)$ is the relativistic 3-momentum and the
integral is suitably regulated in the ultraviolet. We now apply
(\ref{rhosNinfty}) and the result $m_0 (T=0) = 0$ to
(\ref{frenNinfty}) to obtain
\begin{eqnarray}
{\cal F} (T)&& - {\cal F}(0) = NT \int \frac{d^2 k}{4 \pi^2} \log ( 1 -
e^{-\epsilon_k /T} )\nonumber \\
 +&& \frac{N}{2} \int \frac{d^3 P}{8 \pi^3}
\left[ \log \left( \frac{P^2 + m_0^2}{P^2} \right) -
\frac{m_0^2}{P^2} \right] - \frac{m_0^2 \rho_s }{2}.
\end{eqnarray}
Using $m_0 = T X_1 ( x_1 )$, and simplifying the above integrals, we
get 
\begin{eqnarray}
{\cal F} (T) - {\cal F}(0) =&& N T^3 \left[ \int_{X_1 (x_1 )}^{\infty}
\frac{\epsilon d \epsilon}{2 \pi} \log(1 - e^{-\epsilon}) \right.
\nonumber \\
&&~~~~~~\left. - \frac{X_1^3 ( x_1 )}{12 \pi} - \frac{X_1^2 ( x_1 )}{4 \pi x_1}
\right] .
\end{eqnarray}
Finally we use the thermodynamic relation $C_V = - T \partial^2 {\cal
F}/\partial T^2$ to obtain the scaling function $\Psi_1 (x_1 )$ for
the specific heat as defined in (\ref{psi1})
\begin{eqnarray}
\Psi_1 ( x_1 ) = &&- \frac{N\pi}{3 \zeta (3)
x_1} \frac{d^2}{d x_1^2} \left[ x_1^3 
\left( \int_{X_1 (x_1 )}^{\infty}
\frac{\epsilon d \epsilon}{2 \pi} \log(1 - e^{-\epsilon})
\right.\right. \nonumber \\
&&~~~~~~~~~~~~\left.\left.
- \frac{X_1^3 ( x_1 )}{12 \pi} - \frac{X_1^2 ( x_1 )}{4 \pi x_1}
\right) \right].
\end{eqnarray} 
We recall that the function $X_1 ( x_1)$ is given in (\ref{X1Ninfty}).
It is clearly that the universal crossover function $\Psi_1 ( x_1 )$
is quite non-trivial even at $N=\infty$.
In the renormalized classical limit, $x_1 \rightarrow 0$ it has the
value 
\begin{equation}
\Psi_1 ( 0 ) = N ,
\end{equation}
which is the number of gapless spin wave modes (upto relative order
$1/N$). The normalization in (\ref{psi1}) was chosen to make this result
simple. 
In the quantum-critical limit, $x_1 \rightarrow \infty$,
the integrals cannot be analytically evaluated. However, one of us
has recently shown~\cite{polylog} that application of some unusual identities
of polylogarithmic functions can be used to prove that the
integrals reduce
a surprisingly simple result
\begin{equation}
\Psi_1 ( \infty ) = \frac{4N}{5}. 
\end{equation}
We have no physical understanding of why this number is rational.

The reduced scaling functions of the renormalized classical region,
describing the 
crossover from Goldstone to classical, thermal disorder (Fig.~\ref{komdep})
can also be easily obtained. 
{}From (\ref{RD1}), (\ref{fxi}), and (\ref{XiNinfty}) 
we find the scaling function
$f(y)$ for the structure factor $S(k)$ 
\begin{equation}
f(y) = \frac{1}{y^2 + 1} .
\end{equation}

\subsubsection{Scaling properties for $g \geq g_c$}
\label{Ninftyggtgc}
The first step is to obtain the value of the $T=0$ gap. From 
(\ref{mval2}) we have the exact $N=\infty$ result
\begin{equation}
\Delta = 4 \pi \left( \frac{1}{g_c} - \frac{1}{g} \right).
\label{deltaval}
\end{equation}
Second, we also need the spin-1 quasiparticle amplitude ${\cal A}$.
{}From (\ref{chisNinfty}) we obtain immediately at $N=\infty$
\begin{equation}
{\cal A} = \frac{g\tilde{S}^2}{N}.
\end{equation}

{}From (\ref{mval}) we can now deduce the scaling function $X_2 $ 
as defined in (\ref{scalexi}) for
the correlation length
\begin{equation}
X_2 ( x_2 ) = 2~ \mbox{arcsinh} \left( \frac{e^{1/(2x_2)}}{2} \right).
\label{X2Ninfty}
\end{equation}
(Recall that $x_2 = T/\Delta$.)
This function has the following asymptotic limits in the 
quantum disordered and quantum-critical regions respectively
\begin{equation}
X_2 ( x_2 ) = \left\{ \begin{array}{ll} {\displaystyle 
\frac{1}{x_2} + 2 e^{-1/x_2}} ; & 
~~~x_2 \ll 1 , \\
{\displaystyle 2 \log \left( \frac{\sqrt{5} + 1}{2} \right) +
 \frac{1}{\sqrt{5} x_2
}} ; & ~~~x_2 \gg 1 .\end{array}
\right. 
\end{equation}
These results imply a correlation length which is of order $\Delta^{-1}$
and $T^{-1}$ respectively.

The scaling function $\Omega_{2} (x_2 )$ for the uniform static susceptibility
in (\ref{chistdef}) can be obtained from (\ref{chiustm0}) combined with
(\ref{mval}) and (\ref{deltaval}). We find
\begin{equation}
\Omega_{2} (x_2 ) = -\frac{1}{2 \pi x_2} + \frac{\sqrt{4 + e^{1/x_2}}}{
\pi e^{1/(2x_2)}} \mbox{arcsinh} \left( \frac{e^{1/(2x_2)}}{2} \right).
\end{equation}
In the quantum-disordered region ($x_2 \ll 1$), this function is
exponentially small
\begin{equation}
\Omega_{2} ( x_2 ) = \frac{e^{-1/x_2}}{\pi x_2}~;~~~~~~~~x_2 \ll 1,
\end{equation}
while in the quantum-critical region it implies a $\chi_u^{{\rm st}}$
of order the temperature
\begin{equation}
\Omega_{2} ( x_2 ) = \frac{\sqrt{5}}{\pi} \log \left( \frac{\sqrt{5} + 1}{2}
\right) \left[ 1 - \frac{2}{5 x_2} \right]~;~~~x_2 \gg 1 .
\end{equation}

Next we consider the free-energy density, ${\cal F}$ and the specific
heat $C_V$. 
The following relation between the $T=0$ gap, $\Delta$, and the
coupling $g$ will be useful
\begin{equation}
\frac{1}{g} = \int \frac{d^3 P}{8 \pi^3} \frac{1}{P^2+\Delta^2},
\end{equation}
where $P = ({\bf k}, \omega)$ is the relativistic 3-momentum. Note
that we again use relativistic cutoff for the value of $g_c$.
Applying this result to 
(\ref{frenNinfty}), we obtain
\begin{eqnarray}
{\cal F} (T)&& - {\cal F}(0) = NT \int \frac{d^2 k}{4 \pi^2} \log ( 1 -
e^{-\epsilon_k /T} )\nonumber \\
 +&& \frac{N}{2} \int \frac{d^3 P}{8 \pi^3}
\left[ \log \left( \frac{P^2 + m_0^2}{P^2+\Delta^2} \right) -
\frac{m_0^2-\Delta^2}{P^2+\Delta^2} \right].
\end{eqnarray}
Using $m_0 = T X_2 ( x_2 )$, and simplifying the above integrals, we
get 
\begin{eqnarray}
{\cal F} (T) - {\cal F}(0) = && N T^3 \left[ \int_{X_2 (x_2 )}^{\infty}
\frac{\epsilon d \epsilon}{2 \pi} \log(1 - e^{-\epsilon}) \right. \nonumber \\
&&~~~~\left.
- \frac{X_2^3 ( x_2 )}{12 \pi} + \frac{3 x_2^2 X_2^2 ( x_2 )-1}{24 \pi x_2^3}
\right]. 
\end{eqnarray}
Finally we use the thermodynamic relation $C_V = - T \partial^2 {\cal
F}/\partial T^2$ to obtain the scaling function $\Psi_2 (x_2 )$ for
the specific heat as defined in (\ref{psi2})
\begin{eqnarray}
\Psi_2 ( x_2 ) =&& - \frac{N\pi}{3 \zeta (3)
x_2} \frac{d^2}{d x_2^2} \left[ x_2^3 
\left( \int_{X_2 (x_2 )}^{\infty}
\frac{\epsilon d \epsilon}{2 \pi} \log(1 - e^{-\epsilon}) \right.\right.
\nonumber\\
&&~~~~~~~\left.\left.
- \frac{X_2^3 ( x_2 )}{12 \pi} + \frac{3 x_2^2 X_2^2 ( x_2 )-1}{24 \pi x_2^3}
\right) \right].
\end{eqnarray} 
where the function $X_2 ( x_2)$ is given in (\ref{X2Ninfty}).
In the quantum disordered limit, $x_2 \rightarrow 0$, we have $\Psi_2
( 0 ) = 0$ corresponding to the absence of any gapless degrees of freedom.
Recall also that $\Psi_2 ( \infty ) = \Psi_1 ( \infty )$.

The scaling functions for the remaining observables are identical to
those obtained above for $g \leq g_c$ after the substitution
$X_1 ( x_1 ) \rightarrow X_2 ( x_2 )$. Thus {\em e.g.} the
scaling function $\Phi_{2s}$ for $\chi_s$ in (\ref{phi2s}) differs
from the function $\Phi_{1s}$ in (\ref{phi1sNinfty}) only by this
substitution.  

\subsection{$1/N$ corrections}
\label{NLsmodel1N}
We now present formal expressions for the modifications to the
uniform and staggered spin susceptibilities at order $1/N$. These
expressions will be used in the following sections to determine 
the scaling properties of the quantum-critical,
renormalized-classical and quantum-disordered regions. 

The formal structure of the $1/N$ expansion has been reviewed in the
book by Polyakov\cite{Polyakov} (see also ~\cite{Colem,Aref}). 
The leading corrections arise from considering the
fluctuations of the $\lambda$ field about its saddle point. The
contribution to the staggered susceptibility, or equivalently the
propagator of the $n_{\ell}$ field, is given by the Feynman graph in 
Fig.~\ref{figfeyn} This leads immediately to
\begin{equation}
\chi_s ( k , i \omega_n ) = \frac{g \tilde{S}^2}{N} \frac{1}
{k^2 + \omega_n^2 + m^2 + \Sigma
( k , i \omega_n ) },
\label{chissigma}
\end{equation}
where $\Sigma$ is the self-energy arising from the $\lambda$
fluctuations. It is convenient to absorb the value of $\Sigma ( 0,
0)$ into the mass $m^2$; all our expressions for $\Sigma$ will
therefore always include a subtraction which makes $\Sigma ( 0 , 0) =
0$. The leading contribution to $\Sigma$ is 
\begin{eqnarray}
\Sigma && ( k , i \omega_n ) = \nonumber \\
\frac{2}{N}
T && \sum_{\epsilon_n} \int \frac{d^2 q}{4\pi^2}
\frac{G_0 ({\bf k} + {\bf q}, i\omega_n + i\epsilon_n )-G_0 ( q,i\epsilon_n)}
{\Pi (  q, i\epsilon_n )},
\label{ressig}
\end{eqnarray}
where $1/\Pi$ is the propagator of the $\lambda$ field
\begin{eqnarray}
\Pi (  q, && i\epsilon_n ) = \nonumber \\
T \sum_{\Omega_n} && \int \frac{d^2 q_1}{4 \pi^2}
G_0 ( {\bf q} + {\bf q_1}, i\epsilon_n + i\Omega_n) 
G_0 ( q_1, i\Omega_n ),
\label{respi}
\end{eqnarray}
and $G_0$ is the propagator of the $n_{\ell}$ field at $N=\infty$ 
\begin{equation}
G_0 ( k, i\omega_n ) = \frac{1}{ k^2 + \omega_n^2 + m_0^2}.
\label{resg}
\end{equation}

It now remains to determine $m^2$. The value of $m^2$ is set by
solving the constraint equation $n_{\ell}^2 = 1 $, or
\begin{equation}
T \sum_{\omega_n} \int \frac{d^2 k }{4 \pi^2 } \chi_s ( k, i\omega_n )
= \frac{\tilde{S}^2}{N},
\label{chisconst}
\end{equation}
order by order in $1/N$. At $N=\infty$, the dependence of $m_0$ on
$g$ and $T$ is determined by solving the following equation exactly
\begin{equation}
g T \sum_{\omega_n} \int \frac{d^2 k }{4 \pi^2} G_0 ( k , i\omega_n )
= 1.
\label{constg0}
\end{equation}
This has of course already been done in the previous subsection. We
represent the $1/N$ corrections to $m^2$ by
$\delta m^2$, where $m^2 = m_0^2 + \delta m^2$. 
Upon examining the $1/N$ corrections to (\ref{chisconst}) we find
in a straightforward manner
\begin{equation}
\delta m^2 = - \frac{K_1 (T, m_0)}{K_2 (T, m_0)} ,
\label{resdm1}
\end{equation}
where the constants $K_1$, $K_2$ are
\begin{eqnarray}
K_1 (T , m_0) &=& T \sum_{\omega_n} \int \frac{d^2 k}{4
\pi^2 } \Sigma ( k , i\omega_n ) G_0^2 ( k , i\omega_n ) , \nonumber \\
K_2 (T , m_0) &=& T \sum_{\omega_n} \int \frac{d^2 k}{4 \pi^2} G_0^2 ( k ,
i\omega_n ) \nonumber \\
&~& = \frac{1}{8 \pi m_0} \coth \left( \frac{m_0 }{2 T} \right) .
\label{ress1s2}
\end{eqnarray}
Note that $K_1$, $K_2$ depend only upon $T$ and $m_0$, and
do not depend directly upon the value of the coupling $g$.
Comparing (\ref{ress1s2}) with (\ref{ressig}), 
(\ref{respi}) and (\ref{resg}), the
expression for $K_1$ can be manipulated into the following
\begin{eqnarray}
K_1 ( T, && m_0 ) =  - \frac{2}{N} T \sum_{\epsilon_n} \int \frac{d^2 q}{4 \pi^2}
\frac{1}{\Pi ( q , i\epsilon_n )} \nonumber \\
&& \left(
\frac{1}{4 m_0 } \frac{\partial \Pi ( q,i \epsilon_n )}{\partial m_0}
+ G_0 ( q , i\epsilon_n ) K_2 ( m_0 , T) \right).
\label{resdm2}
\end{eqnarray}
These results for $m^2$, $K_1$ and $K_2$ will be quite useful in 
subsequent sections.

Expressions for observables that depend upon $\chi_s$ can now be
obtained as before. In particular, the equal-time staggered
structure factor $S(k)$ is
given by
\begin{equation}
S(k) = T \sum_{\omega_n} \chi_s ( k , i\omega_n ).
\label{skchis}
\end{equation}
Using (\ref{chissigma}) we see that to order $1/N$ this can be rewritten as
\begin{equation}
\frac{S(k)}{S_0 (k)} = 
\left[ 1 + \frac{\sum_{\omega_n} ( \delta m^2 + \Sigma ( k , i\omega_n ))
G_0^2 ( k , i\omega_n )}{S_0 ( k ) } \right]^{-1} ,
\label{sksigg0}
\end{equation}
where $S_0 (k)$ is the structure factor at $N=\infty$:
\begin{equation}
S_0 (k) = \frac{g\tilde{S}^2}{N} T \sum_{\omega_n} G_0 ( k , i \omega_n).
\end{equation}   

The correlation length, $\xi$, is defined by the  
pole of $S(k)$ in the complex $k$ plane which is closest to the real
$k$ axis; this pole is at $k=i/\xi$. From (\ref{skchis}) it is
clear that, at any finite $T$, the poles of $S(k)$ are simply the
poles of $\chi_s ( k , i\omega_n )$ for all $\omega_n$. It is also
clear that the pole closest to the real-axis is that associated with
$\chi_s ( k , i\omega_n = 0)$. The location of this pole can be simply
determined from (\ref{chissigma}) in a $1/N$ expansion. We find
\begin{equation}
\xi^{-1} = m + \frac{1}{2 m_0} \Sigma ( k=im_0 , i\omega_n = 0).
\label{xim0}
\end{equation}
Finally, the residue of the pole in the structure factor, which sets
the overall scale for exponentially decaying correlations, can be
determined in a similar manner: 
\begin{displaymath}
S (k ) \approx \frac{g\tilde{S}^2}{N} 
T \left( 1 - \frac{\partial \Sigma ( k = i m_0 ,
i\omega_n = 0 )}{\partial k^2} \right) \frac{1}{k^2 + \xi^{-2}};
\end{displaymath}
\begin{equation}
~~~~~~~~~~~\mbox{$k$ close to $i \xi^{-1}$}.
\label{skresidue}
\end{equation}
The results for $\xi$ and $S(k)$ will be of great use to us
in the subsequent sections.

Now we turn to a consideration of the $1/N$ corrections to the static
uniform susceptibility, $\chi_u^{{\rm st}}$. This is a four-point function
of the $n_{\ell}$ field and the Feynman graphs which contribute
at order $1/N$ are shown in Fig.~\ref{figfeyn}. 
The evaluation of these graphs is,
in principle, quite straightforward, but rather tedious. After some fairly
lengthy frequency summations,
we obtained the following, surprisingly compact result:
\begin{equation}
\chi_u^{{\rm st}} =  \int \frac{d^2 k}{4 \pi^2} \left[
-2 n^{\prime}_k - \frac{n^{\prime\prime}_k}{\epsilon_k} ( m^2 - m_0^2)
\right] - K_3 ( T, m_0),
\label{chist1n1}
\end{equation}
where the energy $\epsilon_k = (k^2 + m_0^2)^{1/2}$ was introduced earlier,
$n_k \equiv n(\epsilon_k )$ is the Bose function
\begin{equation}
n ( \epsilon ) = \frac{1}{e^{\epsilon/T} - 1},
\end{equation}
and $n^{\prime}_k = dn(\epsilon_k)/d\epsilon_k$, etc. (note that the symbol
$n$ is used both for the Bose function and the non-linear sigma model field -
the appropriate meaning should however be clear from the context).
The very first term in (\ref{chist1n1}) is identical to the $N=\infty$
result of (\ref{chiustwn}), while the second term arises from 
effective mass renormalization of the the $N=\infty$ graphs. The remaining
$1/N$ corrections are contained in 
the function $K_3 (T,m_0)$ which is given by
\begin{eqnarray}
&& K_3 (T, m_0) = \frac{2}{N} T \sum_{\omega_n} \int \frac{d^2 q}{4 \pi^2}
\frac{1}{\Pi ( q, i \omega_n )} \int \frac{d^2 k}{4 \pi^2}
\frac{n_k^{\prime\prime}}{\epsilon_k} \nonumber \\
&&~~~~\left[
\frac{\epsilon_{k+q}^2 - \epsilon_k^2 + \omega_n^2}{
(\epsilon_{k+q}^2 + \epsilon_k^2 + \omega_n^2)^2 - 4 \epsilon_k^2 
\epsilon_{k+q}^2} - \frac{1}{\epsilon_q^2 + \omega_n^2} \right].
\label{resk3}
\end{eqnarray}
We will evaluate this expression in the later sections on 
the different critical regimes.

Lastly, the $1/N$ corrections to the free energy density, ${\cal F}$.
The $1/N$ corrections to (\ref{frenNinfty}) arise from the functional
determinant of the integration over the $\lambda$ field. The
propagator of the $\lambda$ field is $\Pi$ as defined in
(\ref{respi}) and the correction to the free energy in simply $(1/2)
\mbox{Tr} \log \Pi$. We have therefore
\begin{eqnarray}
{\cal F} = && \frac{NT}{2} \sum_{\omega_n} \int \frac{d^2 k}{4 \pi^2}
\left[ \log ( k^2 + \omega_n^2 + m_0^2 )
 \right.\nonumber\\
&&~~~~~~~~~~~~+\left. \frac{1}{N} \log ( \Pi ( k , i\omega_n )) \right] 
- \frac{N m_0^2}{2 g}.
\label{freen1N}
\end{eqnarray}
Note that we have not included the $1/N$ correction to the mass
$i\langle \lambda \rangle = m$. This is because by construction
$d {\cal F} / d \langle \lambda \rangle = 0$ at $N=\infty$. This
immediately implies that the correction to $\langle \lambda \rangle$
at order $1/N$ will modify ${\cal F}$ only at relative order $1/N^2$.
Thus may as well use $i\langle \lambda \rangle = m_0$ to compute
${\cal F}$ to order $1/N$.
 
\section{QUANTUM-CRITICAL REGION}
\label{secqc}

This section will present expressions for the scaling functions
in the quantum critical region to order $1/N$. We will restrict our discussion
here
to the critical point at $g=g_c$.
This point will be accessed by taking the $x_1 \rightarrow \infty$
limit from the ordered side. We will thus present explicit results
for the scaling functions $\Phi_{1s} ( \overline{k},
\overline{\omega}, x_1 = \infty)$ and 
$\Phi_{1u} ( \overline{k},
\overline{\omega}, x_1 = \infty)$. Those for $\Phi_{2s}$ and
$\Phi_{2u}$ can be obtained immediately from (\ref{zqdef}).

An important issue that arises at the outset of any calculation of
universal scaling functions is that of
proper choice of an ultraviolet cutoff. 
In the preceeding section, we chiefly used a Pauli-Villars cutoff to obtain the 
value of $g_c$. However, we will see that $g_c$ does not explicitly show up
in $1/N$ corrections and we are therefore free to choose the most 
convenient regularization scheme.
All of the computations in
this section were performed with two different cutoffs: 
\newline
({\em a\/}) Lattice-cutoff: The $n_{\ell}$ field was placed on a
square lattice of spacing $1/\Lambda$, but no restriction was placed
on the allowed values of the Matsubara frequencies $\omega_n$. 
The scaling functions were obtained in the limit $\Lambda \rightarrow
\infty$.
By
construction, in this cutoff scheme relativistic invariance is violated at
short scales and
is present only in the long-distance theory at $T=0$. 
Consequently, the $T=0$ spin-wave
velocity will be renormalized from its bare value, and care will have
to be taken to express the scaling functions in terms of the fully
renormalized spin-wave velocity.
\newline
({\em b\/}) Relativistic, hard cutoff: The momenta $k$,
and frequencies $\omega_n$ carried by the $n_{\ell}$ field were
restricted to satisfy
\begin{equation}
k^2 + \omega_n^2 < \Lambda^2.
\end{equation}
Unlike the previous cutoff, this scheme has full relativistic
invariance, 
and there will be no renormalization of the bare $T=0$ spin-wave
velocity. 
\newline
All of the remaining discussion in this section will use the second,
relativistic cutoff scheme; as a result we will not have to consider
explicitly the renormalization of the spin-wave velocity.
We emphasize however that all of our
numerical computations have been carried out with both schemes. While
many of the intermediate results of the two schemes were different,
the final results for the universal scaling functions were found to
be identical. This agreement provides strong support for
the complete universality of the scaling functions, and makes it
virtually certain that there are no numerical errors in our
computations in the quantum critical region.

The basic strategy for obtaining the scaling functions is straightforward.
We evaluate $\chi_s$ and $\chi_u$ to order $1/N$ as outlined in 
Section~\ref{NLsmodel1N}, and then 
invert Eqns (\ref{phi1s},\ref{phi1u}) to express
$\Phi_{1s,u}$ in terms of $\chi_{s,u}$. We will also need in this procedure
the $T=0$ value of the ratio $N_0^2 / \rho_s$.
The structure of the
$T=0$ theory for $g < g_c$ is discussed in Appendix~\ref{appent=0} 
where we found
\begin{equation}
\frac{N_0^2}{\rho_s} = \frac{g\tilde{S}^2}{N} \left[ 1 - \eta \log \left(
\frac{N \Lambda}{16 \rho_s} \right) \right],
\label{n02rs}
\end{equation}
where the number $\eta=8/(3\pi^2 N)$ 
will become the critical
exponent $\eta$ defined in Section~\ref{intro}. 

We will begin by noting some of the significant issues that arose in the
evaluation of the results of Sec~\ref{NLsmodel1N} in
 the quantum-critical region.
We will then proceed to a statement of the results for the
various scaling functions.

The first step was to develop a fast computer program for the rapid
evaluation of $\Pi ( q, i\epsilon_n )$, 
in (\ref{respi}) for $m_0 = \Theta T$. Simple power-counting shows
that $\Pi$ is convergent in the limit of the cutoff $\Lambda
\rightarrow \infty$. However, it is not clear a-priori that it is
valid to take the $\Lambda \rightarrow \infty$ limit at this
early stage. The point is that subsequent integrals will involve
values of $\Pi ( q, i \epsilon_n )$ with $q, \epsilon_n$ itself of
order $\Lambda$. However, we have shown by a detailed consideration
of the relevant integrals, that these potentially dangerous
contributions from $\Pi$ cancel out in the final results for all
universal quantities. Thus we will fearlessly evaluate $\Pi$ in the
limit of an infinite cutoff.

The evaluation of $\Pi$ began with analytic determination of the
integral over $q_1$ in (\ref{respi}). The summation over $\omega_n$
was then performed by a direct numerical evaluation. Terms upto some
large value of $\omega_n$ were explicitly evaluated, and the
remainder were summed by fitting to an inverse power series in
$1/\omega_n^2$ upto order $1/\omega_n^6$. A very similar procedure
was used to evaluate $\partial \Pi / \partial m_0$. Finally the
results were checked against the following computed asymptotic
expressions: 
\begin{eqnarray}
&&\Pi ( q, i \epsilon_n) = \frac{1}{8 (q^2 + \epsilon_n^2)^{1/2}}
~~~~~~~~~~~~~\nonumber \\
&&~~~~~~~~+ \frac{(2 \epsilon_n^2 - q^2 ) \Theta^3 T^3}{(q^2 +
\epsilon_n^2)^3} \frac{1 - 6 \gamma}{3 \pi} + {\cal O} \left(
\frac{T^5}{(q, \epsilon_n)^6} \right) , \nonumber \\
&&-\frac{1}{4m_0} \frac{\partial \Pi (q, i \epsilon_n )}{\partial m_0}
= \frac{\sqrt{5}}{8 \pi \Theta T} \frac{q^2 + \epsilon_n^2}{
(q^2 + \epsilon_n^2)^2 + 4 \Theta^2 T^2 \epsilon_n^2}
\nonumber \\
&&~~~~~~~~~~~~~~~+ {\cal O} \left(
\frac{T^3}{(q, \epsilon_n)^6} \right) ,
\label{qcapppi}
\end{eqnarray}
where
\begin{equation}
\gamma = \frac{1}{\Theta^3} \int_{\Theta}^{\infty} dx \frac{x^2}{e^x
- 1} = 2.32414354317 .
\end{equation}

Next, we evaluated the self-energy, $\Sigma$, and the constant $K_3$
defined in (\ref{resk3}). It is not difficult to show from
(\ref{qcapppi}) that both these quantities are logarithmically
divergent in the limit $\Lambda \rightarrow \infty$. Further the
coefficient of $\log ( \Lambda )$ can be easily determined
analytically. We numerically evaluated the integral over the momenta
and the summation over the frequency for a fixed $\Lambda$ and
subtracted the known $\log ( \Lambda )$ term. The remainder was found
to be independent of $\Lambda$ for large $\Lambda$, and was the
required finite part of the result. These computations yielded the following
catalog of useful results
\begin{eqnarray}
\frac{\partial \overline{\Sigma} ( \overline{k} = 0 
, i \overline{\omega} = 0)}{\partial \overline{k}^2} 
& = &\eta \log \left( \frac{\Lambda}{T} \right) - \frac{0.25266}{N} \nonumber\\
\overline{\Sigma} ( \overline{k} = 
i \Theta , i \overline{\omega} ) &=& - \Theta^2 \eta \log \left( 
\frac{\Lambda}{T} \right) + \frac{0.22616}{N} ,\nonumber \\
\frac{\partial \overline{\Sigma} ( \overline{k} = i \Theta 
, i \overline{\omega} = 0)}{\partial \overline{k}^2} 
& =& \eta \log \left( \frac{\Lambda}{T} \right) + \frac{0.69400}{N} ,
\nonumber \\
\overline{\Sigma} ( \overline{k} , \overline{\omega} \gg 1 ) 
&=& ( \overline{k}^2 + \overline{\omega}^2 ) \left[
\eta \log \left( \frac{\Lambda}{T} \right) \right. \nonumber \\
-\frac{\eta}{2}  \log \left ( \overline{k}^2 \right. &+& 
\left. \overline{\omega}^2
\right) + \left. \frac{8}{9 \pi^2 N} \right] ,\nonumber \\
\frac{K_3 ( T, m_0=\Theta T)}{T} &=&  - \frac{\eta \Theta^2}{2 \pi} \log \left(
\frac{\Lambda}{T} \right) + 0.17800.
\label{catalog}
\end{eqnarray}
where $\overline{\Sigma} ( \overline{k} , i \overline{\omega} )= 
\Sigma ( T \overline{k}, iT \overline{\omega} )/T^2$.

A little care is required in inferring the $1/N$ correction
to the mass $m$ at $g=g_c$. 
The point is that the critical coupling $g_c$ itself has
$1/N$ corrections, and in addition to the correction $\delta m^2 $ in
(\ref{resdm1}), there is an additional $T$-independent shift to $m_0$
itself. Accounting for the $1/N$ correction to the value of $g_c$, we
find the following result for $m^2$ at $g=g_c$
\begin{equation}
m^2 = m_0^2 - \frac{K_1 (T, m_0 ) - K_1 ( 0,0)}{K_2 (T,m_0)},
\label{mdefk1k2}
\end{equation}
where $K_1$, $K_2$ are to be evaluated using (\ref{resdm1}), 
(\ref{ress1s2}),  and (\ref{resdm2}) at
$m_0 = \Theta T$. The above result for $m^2$ differs from that in
Section~\ref{NLsmodel1N} by the $T$-independent correction $K_1 (0,0)$ which
in fact ensures that $\delta m^2 (T=0) =0$, as should be the case at
the gapless critical point. 

Now we need $K_1 (T, m_0 = \Theta T) $.
Simple power counting in (\ref{resdm2}) shows that $K_1$ is linearly
divergent for large $\Lambda$. Moreover, the linear $\Lambda$ term is
$K_1 (0,0)$, which from (\ref{mdefk1k2}), must be subtracted out.
However, upon using the explicit results in (\ref{qcapppi}) for $\Pi$
and $\partial \Pi/\partial m_0$ in (\ref{resdm2}), one finds that $
K_1$ is in fact only logarithmically divergent! The absence of any
terms of order $T/(q,\epsilon_n)^2$ (which are allowed by naive power
counting) in the asymptotic expansion of
$\Pi$ was crucial in obtaining this surprising result. The 
dangerous $T/(q,\epsilon_n)^2$ terms are, in fact, 
present at all values of
$m_0$ other than $\Theta T$. Even for this special value of $m_0$,
there is linear $\Lambda$ contribution to $K_1$ from $q, \epsilon_n$
of order $\Lambda$. However it was precisely these contributions that
were dropped when $\Pi$ was evaluated in the limit of infinite
cutoff. Since we are interested only in $K_1 ( T, m_0)  - K_1 (
0,0)$, we blithely neglect such contributions, and simply evaluate
$K_1$ as defined in (\ref{resdm2}) using the values of $\Pi$ and
$\partial \Pi/ \partial m_0$ obtained above. The integral is only
logarithmically divergent and can be evaluated in a manner similar to
$\Sigma$ and $K_3$.
At the end, we obtained from this the needed result for $m$:
\begin{equation}
\frac{m^2}{T^2} = \Theta^2  + \eta \Theta^2  \log \left(
\frac{\Lambda}{T} \right) + \frac{0.21346}{N}.
\label{catalogm}
\end{equation}

We will now describe the universal scaling results obtained by combining the
above with the results of Sec~\ref{NLsmodel1N}. The factors of $\log(\Lambda)$ were
found to cancel out of all universal quantities.

\subsection{Correlation length and structure factor}
\label{qccorr}

For the correlation length we obtained
\begin{equation}
\frac{1}{T \xi} \equiv X_1 ( \infty ) = \Theta \left( 1 + \frac{0.2373}{N} ,
\right)
\label{finalxi}
\end{equation}
The residue of the structure factor in the vicinity of the pole 
in the complex $k$ plane at $i/\xi$ is contained in the following
result for the scaling function $\Xi_1$ of the structure factor
(defined in (\ref{sksig1}))
\begin{eqnarray}
\Xi_1 ( \overline{k} , \infty) = && 
\left(1 + \frac{0.4415}{N} \right)
\frac{1}{\overline{k}^2 + X_1^2 ( \infty )}; \nonumber \\
&&~~~~~\mbox{for $\overline{k}$ near $i X_1 ( \infty )$}.
\end{eqnarray} 
Our numerical results for the full scaling function $\Xi_1 ( \overline{k},
\infty)$ for real $\overline{k}$ are presented in Fig.~\ref{figsfac}. 
Analytic forms can be obtained in the limit
of large and small $\overline{k}$
\begin{eqnarray}
\Xi_1^{-1} ( \overline{k}, \infty ) = &&
0.860818 + \frac{0.3697}{N}\nonumber\\
&&+ \overline{k}^2 \left( 
0.864674 - \frac{0.079}{N} \right)~;~\overline{k} \ll 1 ,
\label{xi1smallk}
\end{eqnarray}
\begin{equation}
\Xi_1 ( \overline{k}, \infty ) = \frac{\Gamma((1-\eta)/2)}{\Gamma(1-
\eta/2)} \frac{A_Q}{2 \sqrt{\pi} \overline{k}^{1-\eta}} ~;
~~~~~~~~\overline{k} \gg 1.
\label{xi1largek}
\end{equation}
This last result for the behavior of $\Xi_1 ( \overline{k} , \infty )$
for large $\overline{k}$ required knowledge of the asymptotic properties
of the scaling function $\Phi_{1s}$ for the dynamic staggered susceptibility
discussed in Section~\ref{qcstag} below;
the constant $A_{Q}$ 
is given in (\ref{valaq}).

For experimental comparisons, it is convenient to express $S(k)$
directly in terms of $k \xi$, where $\xi$ is the actual correlation
length.
{}From (\ref{flucdiss}), (\ref{xi1phi1s}), 
(\ref{xi1smallk}) and (\ref{xi1largek}),
we obtain
\begin{displaymath}
S(k) = \frac{N_0^2}{\rho_s} 
\left( \frac{N k_B T}{2 \pi \rho_s} \right)^{\eta}
\frac{\sqrt{5}}{2} \frac{\xi}{(1 + k^2 \xi^2)^{1/2}} 
\left( 1 - \frac{0.1925}{N} \right)
\end{displaymath}
\begin{equation}
~~~~\times
 \left\{ \begin{array}{cc}
1 - {\displaystyle  \left( 1 - \frac{0.100}{N} \right) 
\frac{\Theta}{\sqrt{5}} k^2} ;&
~~k \xi \ll 1 ,\\*[10pt]
{\displaystyle \frac{(k \xi)^{\eta}}{\sqrt{5}} 
\left( 1 + \frac{0.267}{N} \right)} ;&
~~k \xi \gg 1 .
\end{array} 
\right.
\label{static}
\end{equation}

\subsection{Uniform susceptibility}
For the scaling function 
$\Omega_{1}$, (see (\ref{chistdef})), of the static uniform susceptibility
we obtained
\begin{equation}
\Omega_{1} ( x_1 \rightarrow \infty ) = \frac{\sqrt{5}}{\pi} \log \left(
\frac{\sqrt{5} + 1}{2} \right) \left[\left(
1 - \frac{0.6189}{N}\right) + \frac{4}{5x_1} \right],
\label{Omega1}
\end{equation}
We have also performed Monte Carlo simulations of a lattice version
of the $O(3)$ sigma model which are described in Appendix~\ref{appmonte}.
This yielded $\Omega_1 ( \infty ) = 0.25 \pm 0.04$, in good agreement
with the above result.

\subsection{Staggered Susceptibility}
\label{qcstag}
We first describe some asymptotic limits of
the two-parameter scaling function
$\Phi_{1s} ( \overline{k}, \overline{\omega} , \infty)$.
For small $\overline{k}$ we have
\begin{eqnarray}
\mbox{Re} \Phi_{1s}^{-1} ( \overline{k}, 0, \infty ) = &&
\Theta^2 \left( 1 + \frac{0.4830}{N} \right) \nonumber\\
&&~~~+
\overline{k}^2 \left( 1 - \frac{0.0001}{N} \right)
~;~\overline{k} \ll 1.
\end{eqnarray}
Our numerical accuracy is not sufficient to rule out
zero $1/N$ correction to the coefficient of $\overline{k}^2$; the small value
obtained for this correction appears to be accidental.
At large arguments we found
\begin{eqnarray}
\Phi_{1s}^{-1} (\overline{k} , i\overline{\omega}, \infty) &&
=  (\overline{k}^2 + \overline{\omega}^2 ) \left[ 1 - \frac{\eta}{2}
\log ( \overline{k}^2 + \overline{\omega}^2 ) \right.\nonumber \\
&&\left. + \eta \left(
\log \left( \frac{8}{\pi} \right) + \frac{1}{3} \right) \right]
~;~\overline{k} , \overline{\omega} \gg 1.
\label{asympphi1s}
\end{eqnarray}
We expect that this logarithmic series can be exponentiated. Performing
the exponentiation, followed by an analytic continuation to 
real frequencies we get finally
\begin{equation}
\Phi_{1s} ( \overline{k} , \overline{\omega} , \infty ) = 
\frac{A_{Q}}{(\overline{k}^2 - \overline{\omega}^2 )^{1-\eta /2}}
~;~~~~~~~~~~\overline{k}, \overline{\omega} \gg 1,
\end{equation}
where the universal number $A_{Q}$ is given by
\begin{equation}
A_{Q} =  1 - \eta \left(
\log \left( \frac{8}{\pi} \right) + \frac{1}{3} \right) 
\label{valaq}
\end{equation}
to order $1/N$.

We turn finally to the determination of the scaling function $\Phi_{1s}  
(\overline{k}, \overline{\omega} , \infty)$ for arbitrary $\overline{k}$,
$\overline{\omega}$. As analytic evaluation is clearly impossible, we
will have to resort to numerical computations. Moreover, as it is not easy
to analytically continue numerical data, we will perform the numerical
computations directly at real frequencies. 
There are some interesting issues which arise from the interplay between the 
analytic continuation, finite-temperature effects, and ultraviolet divergences. 
These issues do not appear to have been discussed elsewhere before, 
and it appears
worthwhile to present some details.

Our strategy will be as follows: we will start with the key observation that
to order $1/N$ $\mbox{Im} \Sigma$ is free of ultraviolet divergences and that
\begin{equation}
\mbox{Im} \Phi^{-1}_{1s}( \overline{k}, \overline{\omega},\infty) = 
\mbox{Im}  \overline{\Sigma} ( \overline{k} , \overline{\omega} ) .
\label{Ext1}
\end{equation}
Therefore $\mbox{Im} \Phi^{-1}_{1s}$ can be obtained by a direct evaluation of
(\ref{Ext1}). All ultraviolet divergences and $\Lambda$ dependences are in fact
in the real part of self-energy. Next we note that $\Phi^{-1}_{1s}$ is analytic
as a  function of $\overline{\omega}$ in the upper-half frequency plane.
However, we will find that because $\mbox{Im} \Phi^{-1}_{1s} \sim
\overline{\omega}^2$ for large $\overline{\omega}$, its Kramers-Kronig
transform is not convergent and cannot be directly used to obtain the real
part. Instead, we will use the analytic information contained in the large
$\overline{\omega}$ behavior in (\ref{asympphi1s}) to perform an appropriate
subtraction, and take the Kramers-Kronig transform of the remainder.

First some details on the evaluation of $\mbox{Im} \overline{\Sigma}$.
{}From the results for $\Sigma$ in 
Section~\ref{NLsmodel1N}, we obtain after and 
analytically continuation to real frequencies
\begin{eqnarray}
\mbox{Im} \overline{\Sigma}( \overline{k}, && \overline{\omega})
 = \frac{1}{4\pi^2 N} 
\int d^2 \overline{q}   
\int_{0}^{\infty} d \overline{\Omega} \mbox{Im} \left( 
\frac{1}{\Pi ( \overline{q}, \overline{\Omega} )} \right)\nonumber \\
&& \times \frac{1}{\overline{\epsilon}_{\bf{\overline{q}} + \bf{\overline{k}}}}
 \left[
|n_{\bf{\overline{q}} + \bf{\overline{k}}} - n_{\overline{\Omega}}| 
\delta( \overline{\omega} - 
|\overline{\epsilon}_{\bf{\overline{q}} + \bf{\overline{k}}}-
\overline{\Omega}|)\right. \nonumber \\ 
&&~~\left. + (1 + n_{\bf{\overline{q}} + 
\bf{\overline{k}}} - n_{\overline{\Omega}})
 \delta( \overline{\omega} - 
\overline{\epsilon}_{\bf{\overline{q}} + \bf{\overline{k}}}-\overline{\Omega})
 \right] .
\label{Ext3}
\end{eqnarray}
Here  $ \overline{\epsilon}_{\bf{\overline{k}}} = ({\bf \overline{k}}^2 
+ \Theta^2 )^{1/2}$, $\overline{\omega} > 0$ ($\mbox{Im}  
\overline{\Sigma}$ is an odd function of $\overline{\omega}$), 
$n_{\overline{k}} = n(\overline{\epsilon}_{\overline{k}})$ is the Bose
function, and  
there is no cutoff in the $\overline{q}$ integration.
The propagator $1/\Pi ( \overline{q}, \overline{\Omega} )$
is simply $1/\Pi$ in rescaled variables. Thus from (\ref{respi}) we
get $\mbox{Im} \Pi$:
\begin{eqnarray}
&& \mbox{Im}\Pi ( \overline{q}, \overline{\Omega} ) 
 =  \frac{1}{16\pi} \int d^2 \overline{p}~~ 
\frac{1}{\overline{\epsilon}_{\bf{\overline{p}} + \bf{\overline{k}}}~ 
\overline{\epsilon}_{\bf{\overline{q}}}} \nonumber \\
&&~~~\left[
|n_{\bf{\overline{p}} + \bf{\overline{k}}} - n_{\bf{\overline{p}}}| 
 \delta( \overline{\Omega} - 
|\overline{\epsilon}_{\bf{\overline{p}} + \bf{\overline{k}}} - 
\overline{\epsilon}_{\bf{\overline{q}}} |) 
\right.\nonumber \\
&&~~~~+ \left. (1 + n_{\bf{\overline{p}} + \bf{\overline{k}}} +
 n_{\bf{\overline{p}}})
 \delta( \overline{\Omega} - 
\overline{\epsilon}_{\bf{\overline{p}} + \bf{\overline{k}}} - 
\overline{\epsilon}_{\bf{\overline{q}}} ) \right] ,
\label{Ext4}
\end{eqnarray}
where $\overline{\Omega} > 0$ ($\mbox{Im} \Pi$ 
is an odd function of $\overline{\Omega}$),
and there is no cutoff in the $\overline{p}$ integration.
The real part $\mbox{Re} \Pi$ can be obtained by a
Kramers-Kronig
transform of $\mbox{Im} \Pi$ (the frequency integral is free of 
ultraviolet divergences), which can then be used to determine 
$\mbox{Im} (1/\Pi)$.
Note all of the integrals above defining $\mbox{Im} \overline{\Sigma}$ are
pure numbers and amenable to a direct numerical evaluation, which we
undertook.
The presence of delta-functions in the integrand considerably speeded up
the numerical computations.

Next, we turn to
$\mbox{Re} \Phi_{1s}^{-1}$.
We deduce from 
(\ref{asympphi1s}) that for
 momenta $\overline{k}$ fixed, but $\overline{\omega} \rightarrow \infty$
we must have 
\begin{eqnarray}
\mbox{Re} \Phi_{1s}^{-1} 
&=& \frac{\eta}{2} \left( 
\overline{\omega}^2 \log \overline{\omega}^2 - 
A_Q^{-1} \overline{\omega}^2 -\mu_1 (\overline{k}) \log
\overline{\omega}^2  \right) \nonumber \\
&~&~~~~~~+ \mu_2 (\overline{k}) 
+ \ldots ,
\nonumber \\
\mbox{Im} \Phi_{1s}^{-1}
&=& - \frac{\pi\eta}{2} \mbox{sgn} \overline{\omega} ~\left( 
\overline{\omega}^2  - 
 \mu_1 (\overline{k}) + \ldots \right).
\end{eqnarray}
where the functions $\mu_{1,2} (\overline{k})$ are unknown. We fit the
numerically computed $\mbox{Im} \Phi_{1s}^{-1}$ to the above asymptotic form,
and thence obtained $\mu_{1} (\overline{k})$.
Then we defined the function
$P(\overline{k},\overline{\omega})$ by
\begin{equation}
P ( \overline{k},\overline{\omega}) \equiv \mbox{Im} 
\Phi_{1s}^{-1} ( \overline{k}, \overline{\omega} ) 
+ \frac{\pi\eta}{2}  \mbox{sgn} (\overline{\omega})  
 \left( \overline{\omega}^2 - \mu_1 (\overline{k}) \right).
\end{equation}
The terms on the right-hand side have been chosen so that
$P$ falls off sufficiently fast at large $\overline{\omega}$ for its
Kramers-Kronig transform to be well defined. Then, using the analyticity of
$\Phi_{1s}^{-1}$ in the upper-half plane, we can conclude
\begin{eqnarray}
\mbox{Re} \Phi_{1s}^{-1} ( \overline{k}, \overline{\omega} ) = && {\cal P}
\int_{-\infty}^{\infty} \frac{d\Omega}{\pi}
\frac{P(\overline{k}, \Omega)}{\Omega - \overline{\omega}} + 
\frac{\eta}{2} \left( \overline{\omega}^2 \log \overline{\omega}^2
\right. \nonumber \\
&&~~~~~\left. - 
A_Q^{-1} \overline{\omega}^2 -\mu_1 (\overline{k}) \log
\overline{\omega}^2\right) +  \mu_2 (\overline{k});
\end{eqnarray}
this determines $\mbox{Re} 
\Phi_{1s}^{-1} (\overline{k}, \overline{\omega})$ 
upto the additive 
frequency-independent function $\mu_2 ( \overline{k}) $. 
Finally, $\mu_2 (\overline{k}) $ was fixed
by evaluating $\mbox{Re} \Phi_{1s}^{-1} ( \overline{k}, \overline{\omega}=0)$ 
by an independent method:
we determined it directly from the expression 
(\ref{ressig}) - unlike the computations just
discussed, the frequency sums were evaluated by a direct summation along the
imaginary frequency axis
and a 
straightforward numerical evalutations of the relevant Feynman integrals.
 
This completes our discussion of the derivation $\Phi_{1s} ( \overline{k},
\overline{\omega}, \infty )$. The numerical computations are summarized in
Fig.~\ref{figphi1s}.

One important feature of Fig.~\ref{figphi1s}, 
which hasn't been discussed so far,
is the behavior of $\mbox{Im} \Phi_{1s}$ at small frequencies.
This can be determined directly from the expressions
(\ref{Ext3}) and (\ref{Ext4}) for $\mbox{Im} \overline{\Sigma}$.
We found
\begin{equation}
\mbox{Im} \Phi_{1s} (\overline{k}=0, \overline{\omega}, \infty) \sim \frac{1}{N}
\exp\left (-\frac{3 \Theta^2}{2 |\overline{\omega}|} \right),
\label{exp_sing}
\end{equation}
while 
\begin{equation}
\mbox{Im} \Phi_{1s} ( |\overline{\omega}| < \overline{k}, \infty) \sim \frac{1}{N} 
\overline{\omega} \exp\left(-\frac{3 \Theta^2 }{2|\overline{k}|}\right).
\label{exp_sing2}
\end{equation}
These peculiar singularities are probably artifacts of 
 large $N$
expansion, because they arise from the strong
contraints imposed by the delta functions in (\ref{Ext3},\ref{Ext4})
and the difficulty in satisfying them at small momenta; in other words
energy-momentum conservation drastically reduces the phase space for
emission/absorption of spin-waves with the spectrum $\overline{\omega}  = 
( \overline{k}^2 + \Theta^2 )^{1/2}$.  Actually, for self-consistent
calculations, we have to include the damping of intermediate excitations,
which lifts the  restrictions imposed by
the delta functions. This should probably give 
 $\mbox{Im} \Phi \sim \overline{\omega}$ for
small $\overline{\omega}$, as in naive expectations.
Note that in an exactly solvable $1+1$ dimensional model of a 
quantum phase transition, where analogous scaling functions can be computed,
no such singularities appear\cite{cardy}.

\subsection{Local Susceptibility and NMR Relaxation}
Having obtained scaling results for the staggered susceptibility,
we can now easily obtain properties of the local susceptibility $\chi_L$.
The scaling function, $F_1$, for $\mbox{Im} \chi_L$ was defined in
(\ref{chilscale}). We determined $F_1$ by performing the
integration in (\ref{f1def}) numerically. Our result for $N=3$
was shown in Fig.~\ref{figchil}.
{}From our numerical computation we find for small frequencies
\begin{equation}
F_1(\overline{\omega}) = \mbox{sgn}
(\overline{\omega})\frac{0.06}{N} |\overline{\omega}|^{1-\eta} ~;~~~~~
\mbox{$|\overline{\omega}| \ll 1$}.
\label{F1smallom}
\end{equation}
The power of $\overline{\omega}$ at small $\overline{\omega}$ is fixed by the requirement that $\mbox{Im} (
\chi_L ( \omega )) \sim \omega$ for small $\omega$. 
For large $\overline{\omega}$, we use the large $\overline{k}$, 
$\overline{\omega}$ result for $\Phi_{1s}$ in (\ref{asympphi1s}), and
(\ref{f1def}) to deduce that 
\begin{eqnarray}
F_1 (\overline{\omega}) &=& \frac{1}{|\overline{\omega}|^{1-\eta}} 
\int \frac{d^2 \overline{k}}{4\pi^2}
\frac{A_{Q}}{(\overline{k}^2 - \overline{\omega}^2)^{1-\eta/2}}  \nonumber \\
&=& \frac{A_{Q}}{4\pi} \sin \left(\frac{\pi\eta}{2}\right) 
\frac{\mbox{sgn}(\overline{\omega})}{|\overline{\omega}|^{\eta}} 
\int_0^{|\overline{\omega}|} 
\frac{\overline{k} 
d\overline{k}}{(\overline{\omega}^2 - \overline{k}^2)^{1-\eta/2}}
 \nonumber \\
&=& \mbox{sgn}(\overline{\omega}) 
\frac{A_{Q}}{4} \frac{\sin ( \pi \eta /2 )}{\pi \eta /2}~;~~~~~~~~~~
|\overline{\omega}| \gg 1.
\label{F1largeom}
\end{eqnarray}
Thus $F_1 (\overline{\omega})$ tends to a universal 
constant for large $\overline{\omega}$.

{}From (\ref{r1qdef}) and (\ref{F1smallom}) 
we find that the scaling function $R_1 ( x_1 )$
for the NMR relaxation in (\ref{nmrscale}) satisfies
\begin{equation}
R_1 ( x_1 = \infty ) = \frac{0.06}{N}.
\end{equation}

\subsection{Specific Heat}

We consider evaluation of the expression (\ref{freen1N}) for the free energy
density ${\cal F}$ at $g=g_c$. We will of course only be interested in 
${\cal F}(T) - {\cal F}(0)$ which is finite as $\Lambda \rightarrow \infty$.

We will need the value of $\Pi ( k, i\omega )$ at $T=0$. From the result
(\ref{qcapppi})
\begin{equation}
\Pi ( k , i \omega ) |_{T=0} 
= \frac{1}{8 ( q^2 + \omega^2 )^{1/2}}~;~~~~~~~~~~~q,\omega \ll 
\Lambda .
\end{equation}
Using this result, and the representation (\ref{gc3mom}) for $g_c$ we get from
(\ref{freen1N})
\widetext   
\begin{eqnarray}
{\cal F} (T) - {\cal F} (0) = && 
\frac{NT}{2} \sum_{\omega_n} \int \frac{d^2 k}{4 \pi^2}
\left[ \log ( k^2 + \omega_n^2 + m_0^2 )
+ \frac{1}{N} \log ( \Pi ( k , i\omega_n )) \right]
\nonumber \\
&& - \frac{N}{2} \int \frac{d \omega d^2 k}{8 \pi^3} \left[
\log ( k^2 + \omega^2 )
- \frac{1}{2N} \log (8 ( k^2 + \omega^2 ))
 + \frac{m_0^2}{k^2 + \omega^2}\right].
\end{eqnarray}
Repeated applications of Poisson summation formula~\cite{Dashen}
 simplifies this result to
\begin{eqnarray}
{\cal F} (T) - {\cal F}(0) = && NT \int \frac{d^2 k} { 4 \pi^2} \left[
\log \left( 1 - e^{-\sqrt{k^2 + m_0^2}/T} \right) - \frac{1}{2N} 
\log\left( 1 - e^{-|k|/T}
\right) \right] \nonumber \\
&&~~~~~~~ + \frac{T}{2} \sum_{\omega_n} \frac{d^2 k }{4 \pi^2} \log \left(
8 (\omega_n^2 + k^2 )^{1/2} \Pi ( k , i \omega_n ) \right) \nonumber \\
&&~~~~~~~ + \frac{N}{2} \int \frac{d^3 P}{8 \pi^3} 
\left[ \log \left( \frac{P^2 + m_0^2}{P^2} \right)
- \frac{m_0^2}{P^2} \right].
\label{ftqc}
\end{eqnarray}
\narrowtext
All the integrals and summations
in this last expression can 
be shown to be finite in the limit $\Lambda \rightarrow
\infty$ after using the asymptotic expansion of $\Pi$ in (\ref{qcapppi}).
The absence of any terms of order $T/(q,\epsilon_n)^2$ in (\ref{qcapppi})
is again quite crucial; the structure of the $T^3$ term in
(\ref{qcapppi}) is also such that all potentially dangerous $\log (
\Lambda )$ terms cancel.
The frequency summation and momentum integration in (\ref{ftqc}) were
performed numerically and led to 
a result proportional to $T^3$. 
We then evaluated the specific heat and
obtained
\begin{equation}
\Psi_1 ( \infty ) = \frac{4N}{5} - 0.3344.
\label{psi1qc}
\end{equation}

\section{RENORMALIZED CLASSICAL REGION}
\label{secrc}

We now proceed to the calculation of the scaling functions 
in the region where  the ground state is ordered ($g<g_{c}$)
 and the temperature satisfies $Nk_{B}T \ll 2\pi\rho_{s}$, 
i.e.  $x_{1}\ll 1$. Under these conditions, the Josephson correlation length 
$\xi_{J} \sim \rho^{-1}_{s}$ is much smaller than $\hbar c/k_{B}T$.
At the shortest scales, the antiferromagnet possesses
$D=2+1$ critical spin fluctuations which continue to be described by
the scaling function $\Phi_{1s}$ in (\ref{phi1sqc}).
The crossover to  
the Goldstone region (see Fig.~\ref{komdep}) occurs at
scales $k \xi_{J} \sim 1 ~ (\omega \xi_{J} /c \sim 1)$.
In this regime, the dynamics is governed by rotationally-averaged spin-wave 
fluctuations about a Neel-ordered ground state. 
We will focus in this section on the second crossover (Fig.~\ref{komdep}),
which occurs at the correlation length $\xi$ ($\xi \gg \xi_J , \hbar
c /(k_B T)$) to a fully disordered antiferromagnet.
At scales
larger than $\xi$, all correlations decay exponentially in space and the
low-energy dynamics is purely relaxational. 
For a qualitative description,
one can neglect quantum effects in the vicinity of this crossover,
and study a simplified, purely classical
version of the problem - the
 classical lattice rotor model\cite{CHN}.
We will see, however, that for a complete quantitative 
description, quantum effects cannot 
be neglected at {\em any} $k$. 

A detailed study of the staggered spin correlations in the renormalized 
classical region was performed by Chakravarty {\em et. al.\/}\cite{CHN} in the 
framework of the perturbative renormalization group 
approach  for a classical rotor model.
In this approach, one starts the 
description at relatively short spatial and time 
scales where there is a perfect short-range Neel order, and  one can
 distinguish between the longitudinal and transverse susceptibilities.
 At these short scales,
 the thermal coupling constant $g_{T} = k_{B}T/ \rho_{s}$,
 which measures the strength of thermal fluctuations, is small 
(notice that in two dimensions, $g_{T}$  is a dimensionless quantity).
One then performs RG calculations to see how $g_T$ grows at
larger scales. The scale where the running coupling constant becomes
of the order of unity is identified with the inverse correlation length 
$\xi$. At larger scales, perturbative approach is unapplicable but it is 
assumed\cite{Pol,CHN}
that $\xi$ is the only large scale in the problem and
the behavior at $k<\xi^{-1}$ is not very different from that at 
$k = \xi^{-1}$. This assumption has been explicitly verified
by the Bethe-ansatz solution of the $O(3)$
sigma model in $2$ dimensions\cite{Wiegmann}.

The $1/N$ expansion, which we use here, attacks the same problem but from a
different perspective. 
 We start with the Green's function which satisfies the 
mean-field equation for the saddle point.  The structure of the
saddle-point equation
in Section~\ref{NLsmodelNinfty} shows that the symmetry remains
unbroken at all finite $T$.
The mean-field solution 
 has a gap for 
quasiparticle excitations which we identified, at $N=\infty$, with
 the inverse correlation length. 
In other words, the correlation length is finite
from the very beginning. An obvious consequence is that the 
spin susceptibilities are isotropic functions
in the spin space: $\chi_{ij} (q,\omega) \propto \delta_{ij}$. This is indeed
what we expect from the true scaling functions in 
2D antiferromagnet at finite $T$ (Eqn \ref{phi1s}). On the other
hand, the temperature dependences of observables are not necessarily
 correctly reproduced by the infinite $N$ theory. We  will
 show below that the perturbative $1/N$
expansion for $x_{1} \ll 1$ is actually an expansion 
in $1/(N-2) \log x_{1}$. 
We assume that the logarithmic terms  can be exponentiated;
the $1/N$ expansion thus yields corrections in the form of
extra powers of temperature: 
$x^{1/(N-2)}_{1}$. For $N=3$, the power $1/(N-2) = 1$, and there
will be substantial changes in  the temperature dependences of the
observables, and in particular, of the correlation length.

Most of our results for the staggered dynamic susceptibility
agree with the results of the RG treatment for the classical lattice rotor 
model (some minor discrepancies with Chakravarty {\em et.
al.\/}\cite{CHN} are found
however). We are also able to go beyond previous results and obtain
explicit expressions for various pre-factors and scaling functions up
to the two-loop level.
For the 
uniform susceptibility, 
we have calculated 
the temperature dependence of 
$\chi_{u}^{{\rm st}}$ in a quantum
antiferromagnet and found a
 linear $T$ dependence at low $T$, with a universal slope. We emphasize that 
the temperature dependence of $\chi_{u}^{{\rm st}}$
is a purely quantum effect. It was absent in previous studies of
the
classical lattice rotor 
model~\cite{Andreev,CHN}
 which had (for $N=3$) $\chi^{{\rm st}~\alpha \beta}_{u} = 
\chi_{\perp} (\delta_{\alpha \beta} -
\langle n_{\alpha} n_{\beta}\rangle) 
\equiv \frac{2}{3} \chi_{\perp} \delta_{\alpha \beta}$ 
($\chi_{\perp}$ is the transverse susceptibility at $T=0$). 

We now proceed to a more detailed discussion of $1/N$ expansion.
As in the quantum-critical region, we will 
use the result (\ref{n02rs}) for the value of $N^{2}_{0}/\rho_{s}$ at $T=0$,
to provide a counterterm for the $\log \Lambda$
contributions to the universal function $\Phi_{1s}$. However,
unlike the quantum-critical region, temperature no longer
sets the overall scale for fluctuations, and we
 find it useful to introduce  a function $\tilde{\Phi}_{1s}
(k, \omega)$ related to $\Phi_{1s}$ by 
\begin{equation} 
\tilde{\Phi}_{1s} (k, \omega)  = \left (\frac{\hbar c}{k_{B}T} \right )^2 
\left (\frac{N k_{B}T}{2\pi \rho_{s}}\right )^{\eta} 
\Phi_{1s}(\overline{k}, \overline{\omega}).
\label{RC1}
\end{equation}
In terms of this function
\begin{equation}
\chi_{s} (k, \omega) = \frac{N^{2}_{0}}{\rho_s} ~\tilde{\Phi}_{1s} (k, \omega).
\label{RC2}
\end{equation}
Next, 
at $x_{1} \ll 1$, the typical frequencies $\omega \sim c\xi^{-1}$ are much 
smaller than $k_{B}T/\hbar$ and equal-time 
structure factor $S(q)$ is simply related to 
$\tilde{\Phi}_{1s} (k,\omega =0)$: 
\begin{eqnarray}
S(k) & = & \frac{T}{\pi} \int^{\infty}_{-\infty} \frac{\mbox{Im} \chi_s
(k,\omega)}{\omega} d\omega 
\nonumber \\
& = & k_{B}T  \chi_{s}(k,\omega=0) =
\frac{k_{B}T N^{2}_{0}}{\rho_{s}} \tilde{\Phi}_{1s} (k,\omega =0).
\label{RC3}
\end{eqnarray}
The result for $\tilde{\Phi}_{1s}$  to order $1/N$  follows from
(\ref{chisNinfty}) and (\ref{n02rs}):
\begin{eqnarray}
\tilde{\Phi}^{-1}_{1s} (k, \omega) =  && \left( 1 - \eta \log \left(
\frac{N \Lambda}{16 \rho_s} \right) \right) \nonumber \\
&&~~~~~~~~ \left( k^2 + \omega^2 + m^2 +
\Sigma (k, \omega) \right),
\label{RC4}
\end{eqnarray}                    
where self-energy and mass renormalization terms ($m^2 = m^{2}_{0} + \delta
m^2$) are to be calculated as in Sec \ref{NLsmodel1N}. 

We will now describe, in outline, the computations to order $1/N$ for $\xi$, $m$,
and the structure factor. We will then proceed, in the subsequent subsections, to present
more precise and detailed results for these and other static and dynamic observables.

At $N=\infty$, we have from (\ref{Xsmallx1})
\begin{equation}
m_{0} = \frac{k_{B}T}{\hbar c} \exp \left( -\frac{2\pi\rho_{s}^{N=\infty}}{N 
k_{B}T} \right ) ,
\label{RC5}
\end{equation}
where $\rho_s^{N=\infty}$ is given by (\ref{rhosNinfty}).      
We emphasize  that $m_{0}$ is expressed here in 
 terms of the  $T=0$ spin-stiffness
constant $\rho_{s}$ computed at $N=\infty$. There are
however $1/N$ corrections at $T=0$ as well, and the fully renormalized
$\rho_s$ indeed differs from (\ref{rhosNinfty}). 
When reexpressed in terms of 
the total $\rho_{s}$ at $T=0$, $m_{0}$ by itself acquires a 
correction of the order of $1/N$; this correction will be important
later as a counterterm for 
the leading  ultraviolet divergence of $m=1/\xi$. 

At $N=\infty$, the value of the correlation length, $\xi$, is given
simply by $\xi=1/m_0$.
Comparing Eqn (\ref{Xsmallx1}) with the results of previous 
perturbative approaches, we see 
 that neither the numerical factor in the exponent nor 
the temperature dependence of the prefactor in $\xi$ 
agree with the results of the 
two-loop RG analysis of Chakravarty {\em et. al.\/}\cite{CHN}.
As we already discussed above, this is
not surprising  because 
 their analysis was done for the particular case of $N=3$.
 The two-loop $\beta$-function for arbitrary $N$ was first 
calculated by Brezin and 
Zinn-Justin \cite{Brezin-Zinn} in a perturbative expansion about the ordered
state, and their result for the 
correlation length is
\begin{equation}
\xi \sim \frac{\hbar c}{k_{B}T} \left (\frac{k_{B}T (N-2)}{2\pi\rho_{s}}
\right )^{1/(N-2)} \exp \left( \frac{2\pi\rho_{s}}{(N-2) k_{B}T}
\right) .
\label{RC6}
\end{equation}
At $N=\infty$, this expression coincides with $m_{0}$, as it should. Further, 
if we formally expand the r.h.s. of (\ref{RC6}) in $1/N$, 
we find terms of the form $(1/(N-2))
\log (k_{B}T/(\hbar c m_{0}))$ and $(1/(N-2)) \log \left
 (\log (k_{B}T /\hbar c m_{0}) \right )$.  We therefore anticipate
that the same terms 
should appear in the $1/N$ expansion. 
This is indeed what we found in our calculations,
as we now demonstrate.

We first observe that all $\log ({k_{B}T/\hbar c m_{0}})$  terms
in the $1/N$ expansion
come from integration over spatial scales which are much larger than the
scale set by the temperature.  Accordingly, performing the calculations 
 with the logarithmic accuracy, 
 we can restrict  the evaluation of $\Sigma$ and $\delta m^2$ to a classical 
lattice rotor model; this implies that we restrict the summation 
over $\omega$ to the contribution at $\omega =0$ only and 
set  $k_{B}T/\hbar c$ as the upper cutoff in the 
momentum integration. This substantially simplifies 
the calculations and we easily obtain from (\ref{respi})
\begin{equation}
 \Pi({q},0) = \frac{k_{B}T}{\pi q
 \sqrt{q^2 + 4 m^{2}_{0}}} \log{\frac{q + \sqrt{q^2 + 4 m^{2}_{0}}}{2m_{0}}}.
\label{RC7}
\end{equation}
Substituting this expression into  (\ref{ressig}), performing the 
integration, and using (\ref{RC5}) for $m_0$, we obtain
\begin{equation}
\Sigma_{k,0} = \frac{k^2}{N} \log {\lambda_{k}} + \ldots;~
 \lambda_{k} = \frac{
\log[k_{B}T/(\hbar c m_0)]}
{\log[\sqrt{k^2 + m{^2}_0 }/m_0]} ,
\label{RC8}
\end{equation}
where dots stand for the terms of higher order in $1/N$, 
nonlogarithmic classical contributions, and for quantum contributions.
Note that as written, (\ref{RC8}) is valid 
 only for $k \gg m_0$; for $k = {\cal O}(m_0)$, 
 we have with the logarithmic accuracy 
 $\lambda_{k} = \lambda_m = \log[k_{B}T/(\hbar c m_0)]$.
We now substitute (\ref{RC8}) into
(\ref{ress1s2}) and, using the smallness of $\hbar c m_0 /k_{B}T$, 
obtain after some algebra
\begin{equation}
\delta m^2 = \frac{m^{2}_{0}}{N} \left [-4  
\log{\frac{k_{B}T}{\hbar c m_{0}}} + 3 \log \lambda_m \right ]  + \ldots .
\label{RC9}
\end{equation}
 
Next, we have to show that ({\em i\/}) the actual expansion holds in $1/(N-2)$
rather than in $1/N$ and ({\em ii\/}) 
the logarithmic series are  geometrical
and therefore can be exponentiated. 
In principle, to prove any of these 
statements, one has to examine the structure of the perturbative expansion
 up to infinite order in $1/N$. This is a rather difficult problem to
 analyze and we will be content with demonstrating that both expectations are 
 consistent with the perturbative  results up to order $1/N^2$.
 Specifically, we computed $1/N^2$ logarithmic corrections to 
$\Sigma_{k,0}$. The computational procedure is 
tedious but straightforward. We followed
the general procedure described by Polyakov\cite{Polyakov}: we identified 
 various regions of virtual momenta which contribute to 
 $\Sigma_{k,0}$ with logarithmic accuracy
 and, evaluating the integrals, obtained:
\begin{eqnarray}
k^2 + \Sigma_{k,0} & = & k^2 \left [1 + \frac{1}{N} \left (1 + 
\frac{2}{N} + \ldots\right )  
\log \lambda_{k} \right. \nonumber \\
&~&~~~~~~~~~~~~~~~~~~~~~~\left. + \frac{1}{2N^2} \log^{2} \lambda_{k} 
+ \ldots \right]
\nonumber \\
&\rightarrow&  
k^2 (\lambda_{k})^{1/(N-2)} ,
\label{RC10}
\end{eqnarray}
precisely as we anticipated.
 We didn't perform $1/N^2$ calculations for $\delta m^2$, but it is very
 likely that the expansion for the mass is similar to that for 
 $\Sigma_{k,0}$. We assume that this is the case, and 
  assemble the $1/N$ 
 corrections to $m^2$ into  logarithmic and double logarithmic series.
 This yields
\begin{eqnarray}
&& m^2 = \left (\frac{k_{B}T}{\hbar c}\right )^2 
 \left (\frac{\hbar c m_{0}}{k_{B}T}\right )^2 
 \left (1 - \frac{2}{N-2} \log \left (\frac{k_{B}T}{\hbar c m_{0}}
\right )^2   
\right.\nonumber \\ 
&&~~~~~~~~~~~~~+ \ldots\Biggr) 
 \left (1 + \frac{3}{N-2} \log \lambda_m  + \ldots \right )
\nonumber \\
&& = 
 \left (\frac{k_{B}T}{\hbar c}\right )^2 \left (\frac{\hbar c m_{0}}{k_{B}T}
 \right )^{2N/(N-2)} 
 \left (\lambda_m \right )^{3/(N-2)}.
\label{RC11}
\end{eqnarray}
In writing the last line we also assumed that the value of
$m_0$ in double
logarithmic terms can be replaced by the total mass $m$. The verification
of this assumption requires a computation of finite contributions to order
$1/N^2$ which we didn't perform.
We now assemble the contributions 
in (\ref{RC4}), (\ref{RC10}) and (\ref{RC11})
  and obtain 
\begin{equation} 
\tilde{\Phi}_{1s} (k,\omega =0) 
\propto \frac{\lambda_{k}^{-1/(N-2)}}{k^2 + \xi^{-2}} ,
\label{RC12}
\end{equation}
where with logarithmic accuracy
\begin{eqnarray}
\xi \propto && \left (\frac{\hbar c}{k_{B}T}\right ) \left (\frac{k_{B}T 
(N-2)}{2\pi \rho_{s}}\right )^{1/(N-2)} \nonumber \\ 
&&~~~~~~~~~~~\exp \left(
-\frac{2\pi \rho_{s}}{(N-2)k_{B}T}\right).
\label{RC13}
\end{eqnarray}
This agrees with the two-loop RG result (\ref{RC6}).
{}From (\ref{RC12}) and (\ref{RC3}), the equal-time structure factor is
\begin{eqnarray}
S(k) \propto && \xi^{2} \left (\frac{k_{B}T (N-2)}{2\pi 
\rho_{s}}\right )^{(N-1)/(N-2)} \nonumber \\
&&~~~~~~~~~~
\,\,  \frac{(1 + \frac{1}{2} \log (1 + (k\xi )^2 ))^{1/(N-2)}}{1 + 
(k\xi )^2 }.
\label{RC15}
\end{eqnarray}
For $N=3$, this  coincides with the result of Chakravarty 
{\em et. al.\/}\cite{CHN}. 

An advantage of the $1/N$ expansion is that we can go further than
(\ref{RC12}-\ref{RC15}) and calculate not only logarithmic
contributions in $1/N$, but also the regular ones. 
For this type of calculation, the classical approximation is not sufficient 
and one has to consider the full quantum problem. The computations are lengthy
but straightforward. We will skip the details here: interested readers can obtain 
a fuller description of the intermediate steps directly from the authors.
We will list here only 
a catalog of the results similar to Eqn
(\ref{catalog})
\widetext

\begin{equation}
m^2  =  Z m^{2}_{0} \left[ 1 - \frac{4}{N} \log 
\frac{k_{B}T}{\hbar c m_{0}} ~+~ \frac{4}{N} 
\log \log \frac{k_{B}T}{\hbar c m_0} ~+~ \frac{2}{N} (3 \log 2 -1 + C +
0.3841)
\right] ,  
\label{RRR}
\end{equation}
\begin{eqnarray*}
m^{2}_{0} + \Sigma (k = i m_0, 0) & = & 
Z^{-1} m^{2}_{0} \left[1 ~-~  \frac{2}{N} 
\log \log \frac{k_{B}T}{\hbar c m_0}~-~ \frac{2}{N} 0.3841 \right] , \\
1 + \frac{\partial \Sigma (k \rightarrow 0, 0)}{\partial k^2} & = & Z  \left[ 1 ~+~
  \frac{2}{N} \log \log \frac{k_{B}T}{\hbar c m_0}~+~ \frac{2}{N}
0.3518\right] , \\
k^2 + \Sigma( k \gg \xi^{-1}, 0) & = & Z k^2 \left[ 1 ~-~ \frac{1}{N} \log 
\left(1 + \frac{1}{2} \log (1 + (k\xi )^2 )\right) ~+~  
 \frac{2}{N} \log \log \frac{k_{B}T}{\hbar c m_0} 
 ~+~ \frac{1.9561}{N} \right] , \\
1 - \frac{\partial \Sigma (k = im_0, 0)}{\partial k^2} & = & Z^{-1} \left[1 ~-~
 \frac{2}{N} \log \log \frac{k_{B}T}{\hbar c m_0}~+~ \frac{0.2385}{N} \right] ,
\end{eqnarray*}
\begin{equation}
Z  =  \left(1 + \eta \log \frac{N \Lambda}{16 \rho_s}\right) \left[ 1 ~-~
 \frac{1}{N} \log \log \frac{k_{B}T}{\hbar c m_0}~-~ \frac{0.9561}{N} \right] .
\label{RC16}
\end{equation}
\narrowtext
Here $C$ is the Euler constant $C \approx 0.5772$.
As before, we use these results to evaluate universal functions for various
observables.

\subsection{Correlation length and equal-time structure factor}
\label{xiren_cl}
We start with the scaling function for the correlation length. From
(\ref{RRR}, \ref{RC16}), we find
\begin{eqnarray}
\xi = && \xi_{0} \,
  \left (\frac{\hbar c}{k_{B}T}\right )
\left (\frac{k_{B}T 
(N-2)}{2\pi \rho_{s}}\right )^{1/(N-2)}\,\,\nonumber\\ 
&&~~~~~~~~~~~~~\times\exp
\left(-\frac{2\pi \rho_{s}}{(N-2) k_{B}T}\right),
\label{RC17}
\end{eqnarray} 
where $\xi_{0}$ has a rather simple form
\begin{equation}
\xi_{0} = 1 - \frac{(3\log{2} - 1 + C)}{N}.
\label{RC18}
\end{equation}
The same result was recently obtained in a different way 
by Hasenfratz and
Niedermayer~\cite{hasen1}. They  deduced $\xi$ by comparing the free energy
of the Heisenberg antiferromagnet in moderate magnetic fields with the
Bethe-ansatz solution for the $O(N)$ 
sigma model~\cite{Wiegmann,Pol_Wiegmann}.
Moreover, they argued on the basis of the numerical 
analysis, that the $1/N$ result for $\xi_0$ is in fact the first term
in the $1/N$ expansion for 
\begin{equation}
\xi_0 =  \left(\frac{e}{8}\right)^{1/(N-2)}~~ \Gamma(1 + 1/(N-2)),
\label{RC19}
\end{equation}
 where $\Gamma(...)$ is
the Gamma-function. The numerical evidence for (\ref{RC19})
  is rather convincing and
we will use (\ref{RC19}) for the experimental 
comparisons in Sec~\ref{comp_exp}.

Further, the equal-time structure factor is given by (\ref{RC3}) and using
(\ref{RC16}) we obtain
\begin{equation} 
S(k) = S(0) ~f(k\xi),
\label{RC20}
\end{equation}
where 
\begin{eqnarray}
S(0) = && 2\pi N^{2}_{0} \left (\frac{k_{B}T}{2\pi \rho_{s}}\right )
\mbox{} \,\left (\frac{(N-2) k_{B}T}{2\pi \rho_{s}}\right )^{1/(N-2)}
\nonumber\\
&&~~~~~~~~~~~~~~~~ \times
\xi^2 \left (1 + \frac{0.188}{N}\right ), 
\label{RC21}
\end{eqnarray}
 and  $f(k\xi)$ is a universal scaling function
introduced first by   Chakravarty {\em et. al.\/}\cite{CHN}, for which
we obtain 
\widetext
\begin{equation}
f(k\xi)= \frac{1}{1+ (k\xi)^2} \times \, \left \{ \begin{array} {ll}
	 {\displaystyle 1 + \frac{0.065}{N} (k\xi)^2 ~;}  & k\xi \ll 1, \\
	 {\displaystyle \frac{N -1}{N}  \left (1 - \frac{0.188}{N} \right ) 
	 \left (1 + \frac{1}{2} \log (1 + (k\xi )^2)\right )^{1/(N-2)}~;}
	 & \xi^{-1}
	 \ll k \ll \xi^{-1}_J .
	 \end{array}
      \right.
\label{RC22}
\end{equation}
\narrowtext
At $k\xi \gg 1$, i.e., in the `Goldstone' region,  (\ref{RC20})
 reduces to 
\begin{equation}
S(k) \approx \frac{ k_{B}T N^{2}_{0}}{\rho_{s} k^2} \mbox{}\, \left (
\frac{N-1}{N}\right ).
\label{RC23}
\end{equation}
For $N=3$ this agrees with the result obtained by   
Chakravarty {\em et. al.\/}\cite{CHN} (their definition
for $S(k)$ differs from ours by a factor of $N$).
Clearly, Eqn (\ref{RC23}) is nothing but the rotationally averaged result
for the Neel-ordered antiferromagnet. In the ordered state, there are $N-1$
gapless transverse spin waves which contribute to the low-energy part of
$S(k)$ and one  longitudinal 
component of fluctuations which has a finite gap
and does not contribute at low energies. Rotational averaging then
gives a factor $(N-1)/N$, as in (\ref{RC23}). 

Though our results are very similar to the scaling theory~\cite{CHN},
we observe   that the two limits
in (\ref{RC22}) cannot be assembled 
into the single one-parameter interpolation form for
$f(k\xi)$ suggested by Tyc {\em et. al.\/}\cite{Tyc}. This
is not surprising however because the one-parameter scaling function was
introduced as
 a convenient, but approximate  way to interpolate between 
 $k\xi \gg 1$ and $k\xi \ll 1$, where the 
behavior of $S(k)$ is known from the hydrodynamic considerations.

Further, we emphasize that  $\rho_{s}$ and $N_{0}$ in (\ref{RC21}) are  the 
fully renormalized
spin-stiffness and sublattice magnetization  at $T=0$.  Only in this case,
all ultraviolet $\log \Lambda$ divergences in $1/N$ corrections to 
 $\xi$ and $S(k)$ are canceled out.  
Finally, 
 in obtaining the  universal functions (\ref{RC17}, \ref{RC22}), 
we actually didn't use
the condition $\rho_s \ll J$. It thus appears that at least to
  first order in $1/N$,
the universal behavior holds for arbitrary $\rho_s$, i.e., 
in the whole renormalized classical region. 
This is a remarkable property of the quantum sigma model, and 
the the arguments that the universality at all $g$ may hold at arbitrary $N$ 
 were elegantly displayed in the analyses of 
Hasenfratz {\em et. al.\/}\cite{hasen1,hasen2}. 
In any event, this implies that our results for the
renormalized classical region, which were obtained in a theory valid
near $g=g_c$, are in fact valid at small $T$ for all $g < g_c$.

\subsection{Uniform susceptibility}

We turn next to the calculation of the static uniform susceptibility.
The expression for $\chi_u$ valid at arbitrary $x_1$ was given in 
(\ref{chist1n1}) and (\ref{resk3}). In the renormalized classical region,
it is convenient to
 introduce $\tilde{\Phi}_{1u} \equiv k_{B}T~\Phi_{1u} (0,0,x_{1})$ such that
\begin{equation}                                                
\chi_{u}(0,0) = (\frac{g \mu_{B}}{\hbar c})^2 \, \tilde{\Phi}_{1u} .  
\end{equation}
We then use $n_{m_0} \approx k_{B}T/\hbar c m_0$, and  obtain  for 
$\tilde{\Phi}_{1u}$
\begin{equation}  
\tilde{\Phi}_{1u} = \tilde{\Phi}^{\infty}_{1u} -
 K_3 ~- \frac{k_{B}T}{m^{2}_0}~(m^2 - m^{2}_0) ,
\end{equation}
where   $\tilde{\Phi}^{\infty}_{1u}$, given in (\ref{chiustm0}), is
the contribution at $N=\infty$ and the remaining terms are $1/N$ corrections. 
For $\tilde{\Phi}^{\infty}_{1u}$  we have 
\begin{equation}
\tilde{\Phi}^{\infty}_{1u} = -2 (\hbar c)^2 ~ \int 
\frac{d^2 k}{4\pi^2} n^{\prime}_{k} =
\frac{k_{B}T}{\pi}~ (\log {\frac{k_{B}T}{\hbar c m_0}} + 1) .
\label{RC24}
\end{equation}
Using the expression for $m_0$ at small $x_1$, we can rewrite the $N= \infty$
result for $\chi_{u}^{{\rm st}}$ as 
\begin{equation}
\chi_{u}^{{\rm st}} =
(\frac{g \mu_{B}}{\hbar} )^2 
(\frac{2}{N} \chi_{\perp} + \frac{1}{\pi c^2} k_{B}T ),
\label{RC25}
\end{equation}
where $\chi_{\perp} \equiv \rho_s c^2$ is the transverse susceptibility.
We see that in the limit of vanishing temperature,   $\chi_{u}^{{\rm st}}$ 
is precisely the rotationally
averaged uniform susceptibility of the ordered antiferromagnet. This is
likely to be the exact result, and we expect that all $1/N$ corrections
at $T=0$ can be assembled into the renormalization of $\chi_{\perp}$. On the 
other hand, the temperature dependence of $\chi_{u}^{{\rm st}}$ 
is a purely
 quantum effect (it is entirely due to the second term in (\ref{RC24})) and 
 $1/N$ corrections to $d\chi_{u}^{{\rm st}}/dT$ 
are indeed possible. We also observe that
the slope of $\chi_{u}^{{\rm st}}$ 
versus $T$ at $N=\infty $ is twice as large as in 
the mean-field Schwinger boson 
approach\cite{AA,Drusha}. This is not surprising however, because 
the mean-field Schwinger boson theory 
is the $N=\infty$ limit for a  $\sigma$- model
of the $N$-component {\it{complex}} unit field 
defined on the $CP^{N-1}$ manifold. 
This model is isomorphic to $O(3)$ sigma-model
only at a particular value of 
$N=2$ and there is no reason why the $N \rightarrow \infty$ limits of
the two theories should be the same. One of main technical points of
this paper is that the $1/N$ corrections in the $O(N)$ theory are
numerically quite small, making it a superior approach to make
precise numerical predictions.

The  corrections to $\tilde{\Phi}_{1u}$ 
were computed in the same way 
 as for the correlation length and equal-time 
structure factor. We skip the details
of calculations and discuss only the results.
As before, we found that all divergent
contributions in the ultraviolet are canceled out when the the result 
is expressed in terms of the fully renormalized transverse susceptibility at
$T=0$. We have explicitly checked that there are no other corrections at $T
\rightarrow 0$ besides the renormalization of $\chi_{\perp}$. 
Moreover, we did not find any logarithmic corrections 
to the temperature
dependent part of $\chi_{u}^{{\rm st}}$ 
which might have change the power of the leading $T$-dependent
correction. This
result is  likely to be valid to
arbitrary order in $1/N$ although the proof is lacking.
 At the same time, we
did find the finite correction to the $\tilde{\Phi}_{1u}$  and our result
for $\chi_{u}^{{\rm st}}$ valid to order $1/N$ is
\begin{equation}
\chi_{u}^{{\rm st}}=
\left (\frac{g \mu_{B}}{\hbar} \right )^2 
 \,\left (\frac{2}{N} \chi_{\perp} + \left (\frac{N-2}{N}\right )
 \frac{k_{B}T}{\pi c^2} \right ). 
\label{RC26}
\end{equation}
The renormalization factor, 
$(N-2)/N$, of the linear $T$ term is likely to be correct to all
orders in $1/N$.
The argument is that in the $XY$ case ($N=2$), the 
spin-wave
analysis for $\chi_{u}^{{\rm st}}$ 
is free from divergences\cite{Kag_Chub}  and 
predicts a cubic, rather than linear dependence on $T$~
 $\chi_{u}^{{\rm st}} = \chi_{u}^{{\rm st}}(0) + {\cal O}(T^3)$. 
We also note that 
for  $N=3$, the renormalization factor
is 1/3 and the slope of $\chi_{u}^{{\rm st}}$ 
is therefore significantly reduced 
from the $N=\infty$ result and now differs substantially from
the slope of $\chi_{u}^{{\rm st}}$ in
the quantum critical regime.   We will use this later 
 in Sec. \ref{comp_exp} for the 
interpretation of the experimental data.

\subsection {Dynamic staggered structure factor} 

The calculations of the dynamic susceptibility proceed along the same lines
as in Sec.\ref{secqc}. We use 
integral representation of the polarization operator
continued to the real axis, and compute the retarded self-energy. 
The calculations are rather lengthy, so we skip the details and focus only
on the results: details of the intermediate steps 
can be obtained directly from the 
authors. At $k\xi \gg 1$, we obtained 
\begin{equation}
\chi_{s}(k,\omega) = \frac{N-1}{N} \,\,
\frac{(\hbar c_k)^2 N^{2}_{0}}{\rho_{s}}\,\,
\mbox{}\frac{\rho^{k}_s}{\rho_s}~~\frac{1}{\epsilon^{2}_{k} - (\omega +
i\gamma_{k,\omega})^2} ,
\label{RC43}
\end{equation}
where $\epsilon_{k} = (\hbar c_k)^2 (k^2 + \xi^{-2})$,  
$c^{2}_{k} = \rho^{k}_{s}/ \chi^{k}_{\perp}$. We have introduced here the $k-$dependent spin
stiffness, $\rho_s^k$,
and spin susceptibility, $\chi_{\perp}^k$, which are values 
of these observables at the 
momentum scale $k$. In the present $1/N$ expansion, they are given by
\begin{equation}
\rho^{k}_{s} = \frac{(N-2) k_{B}T}{2 \pi} \left [\varrho + \frac{1}{2} 
\log{(1 + (k\xi)^2)}\right ] ,
\label{RC39}
\end{equation}
and
\begin{equation}
\chi^{k}_{\perp} = \chi_{\perp} \left [\frac{2}{N} + \frac{N-2}{N} \,
\left (\frac{\rho^{k}_{s}}{\rho_{s}}\right )^{N/(N-2)} \right ] .
\label{RC40}
\end{equation}
The dimensionless, numerical variable 
$\varrho$ was found, to first order in $1/N$, 
to be
$\varrho = 1$ and {\it{independent}} of 
the ratio of $\omega /\hbar c k$ as long 
as this ratio is less than 1.
Eqn.(\ref{RC39}) (with $\varrho = 1$) 
coincides with the one-loop result for
the running spin-stiffness inferred from the RG equation for the
static coupling
constant~\cite{CHN}. It
was suggested by Tyc {\em et. al.\/}\cite{Tyc} 
that the two-loop corrections may 
lead to $\varrho \neq 1$. In our approach, the same is likely to happen
from higher order terms in the $1/N$ expansion; dependence 
of $\varrho$ on the ratio $\omega/\hbar c k$
is also possible.  Furthermore, we have checked that 
Eqn (\ref{RC40})  coincides with the 
perturbative result for the transverse susceptibility measured on
the momentum scale $k$ (and frequency scale $ck$). The first
term in (\ref{RC40}) is the exact result at $T=0$, which we already
obtained in the previous subsection.
 The second term (which, we emphasize, is also a 
classical contribution) actually accounts for the difference between 
transverse and 
longitudinal susceptibility measured at finite momentum {\it{and}}
frequency. In this situation,
 we probe the system at finite spatial 
and time scales where the system appears N\'{e}el ordered.
The temperature 
dependence of this term for $N=3$
 is the same as in the scaling approach~\cite{CHN}
and, in fact, can be deduced directly  from the diagrammatic
expression of $\chi_{u}^{{\rm st}}$ in Sec. \ref{NLsmodel1N}
 if $k$ and $\omega$ are both small but 
finite. 

The $1/N$ result for the damping rate $\gamma^{1/N}_{k,\omega}$ 
is given by 
\begin{eqnarray}
\gamma_{k,\omega} = && \frac{\pi}{2} \frac{\hbar c_k k}{N-2} 
\eta_{k,\omega} \,\,
\left(\frac{(N-2)k_{B}T}{2 \pi \rho^{k}_{s}}\right)^{2} \nonumber\\ 
&&~~~~~~~~~\left ( 
\log{\frac{k_B T}{\hbar c k}} \right ),
\label{RCxx}
\end{eqnarray}
where $\eta_{k,\omega}$ (=1 for on-shell excitations) is a smooth function
of the ratio $\omega/ c_k k$
This result for the damping rate agrees with the lowest-order perturbative
calculations by Tyc and Halperin~\cite{TycHalp}.
They, however, have shown that the logarithmic dependence on the quasiparticle
momentum $k$ in (\ref{RCxx}) is actually an artifact of the Born approximation.
Significant corrections to the self-energy term arise from the damping of
intermediate excitations. Neglecting this damping 
is a legitimate
approximation only if the damping rate is 
smaller than the energy of the incoming
quasiparticle. 
Let us define the momentum scale for intermediate states, 
$q_m$, such that $\gamma_{q_m} = \epsilon_k$. 
Clearly then, the lowest-order
calculations are valid for $q <q_m$, but damping of intermediate states
must be included for $q > q_m$.
A simple estimate based on Eqn. (\ref{RCxx}) yields   $q_m \sim k (\rho^{k}_s
/T)^2$.  A careful consideration~\cite{TycHalp} then demonstrated that
$q_m$ has to be taken as the upper cutoff in the momentum integral leading
to (\ref{RCxx}), and $k-$dependent logarithm in (\ref{RCxx}) has to be
substituted by the self-consistent expression 
\begin{equation}
\log{\frac{k_B T}{\hbar c k}} \rightarrow \log{q_m /k} \rightarrow
\log\left [\frac{2}{\pi}~\left(\frac{2\pi \rho^{k}_{s}}{(N-2) k_{B}T}\right)^2
\right]
\end{equation}
The self-consistent result is then 
\begin{eqnarray}
\gamma_{k,\omega} = && \frac{\pi}{2} \frac{\hbar c_k k}{N-2} 
\bar{\eta}_{k,\omega} \,\,
\left(\frac{(N-2)k_{B}T}{2 \pi \rho^{k}_{s}}\right)^{2} \nonumber\\ 
&&~~~~~~~~~\left [2 
\log{\frac{2\pi \rho^{k}_{s}}{(N-2) k_{B}T}} + \Gamma \right ] ,
\label{RC42}
\end{eqnarray}
where $\bar{\eta}_{k,\omega}$ (=1 for on-shell excitations) is a smooth function
of the ratio $\omega/ c_k k$, which generally differs from $\eta$, and
$\Gamma$ is a pure number of order unity.

We return to consideration of the dynamic susceptibility $\chi_s (k, \omega )$
but now at momenta other than $k \xi \gg 1$. At $k \xi \sim 1$ the naive $1/N$
expansion is not terribly useful because the scale $q_m \sim {\cal O} (\xi^{-1})$,
and the damping of intermediate excitations cannot be neglected at any momentum.
We have therefore little to add here to the results of Ref~\cite{CHN,TycHalp}
who used the dynamic scaling hypothesis~\cite{Hohen} to study this region.
The content of this hypothesis is that 
there is only one large spatial 
scale in the problem, namely the correlation length, and
the damping of excitations becomes comparable to the real part of the 
spectrum at $k\xi =
{\cal O}(1)$. 
Indeed, if we use our result
(\ref{RC42}) 
for $\gamma_{k,\omega}$, which is strictly speaking valid only for 
$k \xi \gg 1$, and extend it down to $k \xi {\cal O}(1)$, this is precisely
what happens. The ratio  
$\gamma_{k,\omega}/\epsilon_{k}$, which behaves as
$(T/\rho^{k}_{s})^2 \log \rho^{k}_{s}/T$  at large $k\xi$, 
becomes of the order of unity at
the scale of $k\xi = {\cal O}(1)$.
 This happens because  $\rho^{k}_{s}$ in (\ref{RC39}) 
is renormalized substantially downward at small $k$ and
eventually becomes $\sim k_{B}T$~\cite{extra}. 

Notice also that  
$\chi^{k}_{\perp}$ tends to a finite value as $k$ approaches zero.
 As a result, $c_{k}$ decreases with $k$, and for $k\xi = {\cal O}(1)$, 
we have $c_{k} \sim c \,
\sqrt{k_{B}T/2\pi \rho_{s}}$ {\it {independent}} of $N$. 
This last result illustrates an important feature of the $1/N$ expansion 
in the
renormalized-classical region. The $N$-independent result 
for $c_k$ is inconsistent with
our earlier $N=\infty$ analysis in 
Sec \ref{NLsmodelNinfty} which is completely
 Lorentz-invariant and has a $T$-independent value of $c$.
 This apparent contradiction
is a consequence of the non-commutativity of the 
$T\rightarrow 0$ and $N\rightarrow \infty$
limits. The key point is that the  $T=0$
static uniform susceptibility in the $O(N)$ sigma model 
scales  as $O(1/N)$  (Eqn (\ref{RC26})) and
 therefore vanishes at $N=\infty$. In view of this,
one has to  proceed to the next order in the expansion over $1/N$ (as we did)
 to check that there is indeed 
a breakdown of Lorentz-invariance at small but finite
$T$.  

\subsection{NMR relaxation}

The assumption about the functional form of $\chi_s(k,\omega)$ at small $k$
is a key ingredient which makes possible the calculation of the 
antiferromagnetic contribution to the spin-lattice relaxation
rate. Using the definition of $1/T_1$ in  (\ref{1T1}) and
performing the integration in (\ref{RC43}), we obtain
\begin{eqnarray}
\frac{1}{T_{1s}} = && \lambda \,\,\frac{A^{2}_{\pi} N^{2}_{0}}{(N-2) \hbar}
 \,\, \left(\frac{2}{N}\right)^{1/2} \,\,\left(\frac{\xi}{\hbar c}\right) \,\,
\nonumber \\ 
&&~~~~~~~~~~~~~~~~\left(\frac{(N-2) k_{B}T}{2 \pi
\rho_{s}}\right)^{N/(2N-4)} ,
\label{RC45}
\end{eqnarray}
where $\lambda$ is a numerical factor whose calculation requires us to know
the precise form of $\chi_s ({k,\omega})$ at 
$k\xi = O(1)$. The functional form of
$1/T_{1s}$ in (\ref{RC45}) is identical to that obtained by Chakravarty and
Orbach~\cite{Chak-Orbach}
 on the basis of the scaling approach of Chakravarty {\em et al}.
 They also estimated the value of numerical
factor  to be $\lambda N^{2}_0 \approx 0.61$ (for $N=3$) by
fitting the scaling forms of Chakravarty {\em et al\/}\cite{CHN}
 and Tyc {\em et al\/}\cite{Tyc} to the numerical 
simulations on a classical lattice rotor model.

\subsection{Specific heat}

A simple inspection of the free energy (\ref{freen1N}) 
in the renormalized classical
regime shows that the dominant contribution to ${\cal F}(T) - {\cal F}(0)$
comes from the region of magnon frequencies comparable with the temperature.
In this region, $m_0$ in (\ref{freen1N}) can be neglected compared to $k^2 + 
\omega^2$, and we obtain 
\begin{equation}
\Pi (k, i\omega) = \frac{2 \rho_s}{N} ~ \frac{1}{k^2 + \omega^2}.
\end{equation}
The free energy (\ref{freen1N})
is then easily seen to be 
precisely the same as that of a gas of $ N-1$ gapless Bose degrees of freedom. 
By definition,  (\ref{psi1}), $\Psi_1 (T \rightarrow 0)$ is
a measure of the number of such modes in the ground state. We have therefore
\begin{equation}
\Psi_1 (x_1 \ll 1) = N-1 .
\end{equation}
It can be shown using the above results that $x_1$ dependent
corrections to $\Psi_1$ are suppressed by a factor $e^{-1/x_1}$.

\section{QUANTUM DISORDERED REGION}
\label{secqd}

The section will present  computations of $1/N$ corrections to the
staggered susceptibility. We will not compute $1/N$ corrections to
the uniform susceptibility and the specific heat: both these
quantities are suppressed by factors of $e^{-\Delta/(k_B T)}$ at low
temperatures, and are well described by the $N=\infty$ theory in
Section.~\ref{Ninftyggtgc} - $1/N$ corrections will lead to innocuous
numerical prefactors.

As in the previous sections, the computations will be carried out
with a relativistic cutoff scheme which restricts
$\omega_n^2 + q^2 < \Lambda^2$. We will begin with a description of
results at $T=0$, followed by a discussion of the finite $T$
corrections. 

\subsection{$T=0$}
An immediate simplification here is that all correlators are
completely relativistic. In fact, as the ultraviolet cutoff is also
relativistic, this is also true at frequencies and momenta of the
order of $\Lambda$. All Green's functions are therefore dependent
only upon a relativistic 3-momentum $Q$, related to the usual
momentum, $q$,  and real frequency, $\omega$, by
\begin{equation}
Q^2 = q^2 - (\omega + i \varepsilon)^2 .
\end{equation}
Here $\varepsilon$ is a positive infinitesimal.

An important feature of $\chi_s ( Q)$ has already been discussed in 
Section~\ref{phenomggtgc}: there is a perfect spin-1 quasiparticle
pole at $\omega = \sqrt{q^2 + \Delta^2}$. From (\ref{chissigma}) we
deduce that the residue, ${\cal A}$ at this pole is (to order $1/N$)
\begin{equation}
{\cal A} = \frac{g \tilde{S}^2}{N} \left ( 1 - \frac{\partial
\Sigma ( Q^2 = - \Delta^2) }{\partial Q^2} \right).
\end{equation}

We turn therefore to an evaluation of $\Sigma$. 
This will be obtained from the results of Section~\ref{NLsmodel1N}
evaluated at $T=0$ with $m_0 = \Delta$.
{}From (\ref{respi})
it is easy to obtain the exact result for $\Pi$
\begin{equation}
\Pi ( Q) = \frac{1}{4 \pi Q} \mbox{arctan} \left( \frac{Q}{2\Delta}
\right).
\label{QD1}
\end{equation}
Next, (\ref{ressig}) can be reduced to a one-dimensional integral
for $\Sigma (Q)$
\begin{eqnarray}
\Sigma (Q) = && \frac{1}{2 \pi^2} \int_0^{\Lambda} \frac{P^2 dP}{\Pi
(P)} \left[ \frac{1}{2PQ} \log \left( \frac{ (P+Q)^2 +
\Delta^2}{(P-Q)^2 + \Delta^2} \right) \right. \nonumber \\
&&~~~~\left. - \frac{2}{P^2 + \Delta^2} \right].
\end{eqnarray}
{}From this result, numerical and analytic manipulations show
\begin{eqnarray}
\frac{\partial \Sigma (Q^2 = - \Delta^2)}{\partial Q^2} 
&=& \eta \log \left( \frac{\Lambda}{\Delta} \right) 
- \frac{0.4817408231574}{N} \nonumber \\
\Sigma ( Q^2 \gg \Delta^2 ) &=& Q^2 \left[ \eta \log \left(
\frac{\Lambda}{Q} \right) + \frac{8}{3 \pi^2 N}\right].
\end{eqnarray}
The first result can be combined with (\ref{n02rs}) to obtain the
result (\ref{zqval}) for $Z_Q$, while the second is closely related
to the fourth equation in (\ref{catalog}).

\subsection{$T > 0$}
We present here results for the local susceptibility $\mbox{Im}
\chi_L (\omega )$ at very small $\omega$. Clearly, because of the
presence of gap, $\Delta$, we have 
$\mbox{Im} \chi_L ( \omega < \Delta )|_{T=0} = 0$. Thus the entire
contribution below comes from thermally excited quasiparticles.
{}From the definition (\ref{childef}) of $\chi_L$ and (\ref{chissigma})
we have to order $1/N$
\begin{equation}
\mbox{Im} \chi_L ( \omega \rightarrow 0) = 
\frac{g\tilde{S}^2}{N} \int \frac{d^2 k}{4 \pi^2} \frac{\mbox{Im}
\Sigma ( k, \omega )}{(k^2 + \Delta^2)^2}
\label{QD2}
\end{equation}
Using the integral representation of the polarization operator,
 it is not difficult to obtain that
in the quantum disordered region ($T \ll \Delta$)
\begin{equation}
\mbox{Im} \Sigma ( k, \omega \rightarrow 0 ) = \frac{2\omega}{NT} 
\int \frac{d^2 q}{4 \pi^2}
\frac{e^{-\epsilon_{k+q}/T}}{\epsilon_{k+q}} 
\mbox{Im} \Pi^{-1} ( q, \epsilon_{k+q} ),
\label{QD3}
\end{equation}
where, as before $\epsilon_k = \sqrt{k^2 + m_0^2}$. Note that
$\epsilon_k$ and $\Pi$ now have to be evaluated
 at $m_0 = \Delta$.
 
It now remains to evaluate $\mbox{Im} \Pi^{-1}$. We will begin by
considering $\mbox{Im} \Pi$. From the generalization of (\ref{Ext4})
to the quantum disordered region we find the following important
contributions to $\mbox{Im} \Pi$ for $T \ll \Delta$
\begin{eqnarray}
\mbox{Im} && \Pi ( q, \epsilon_{k+q} ) = \mbox{Im} \Pi ( q,
\epsilon_{k+q} ) |_{T=0} + \nonumber \\
&&+ \frac{1}{8\pi} \int d^2 p \frac{n_p - n_{p+q}}
{\epsilon_p \epsilon_{p+q}} \delta ( \epsilon_{p+q} - \epsilon_p -
\epsilon_{k+q} ),
\label{QD4}
\end{eqnarray}
where $\Pi |_{T=0}$ was obtained earlier in   
(\ref{QD1}). In principle, the $T$ dependent corrections to $\mbox{Re} 
\Pi$ can be obtained by a Hilbert transform of (\ref{QD4}); however,
we found that these contributed subdominant 
corrections to $\mbox{Im} \chi_L / \omega$ in the limit $T \ll
\Delta$. 

We can now compute $\mbox{Im} \chi_L$ by combining 
(\ref{QD1}), (\ref{QD2}), (\ref{QD3}) and (\ref{QD4}). We omit the
details and give the final result
\begin{eqnarray}
\mbox{Im} \chi_L ( \omega \rightarrow 0 && ) = \omega 
\frac{{\cal A} T}{4N\Delta^2 } e^{-2\Delta/T} \left[
\frac{1}{\mbox{arctan}^2 (1/\sqrt{2})} \right. \nonumber \\
+&&\left. \frac{4}{\log^2 (\Delta/T)} + {\cal O} \left(
\frac{1}{\log^3 ( \Delta/T )} \right) \right].
\end{eqnarray}
Note that the first term in the brackets is due to the
temperature dependent part of $\mbox{Im} \Pi$, while the logarithmical
terms come from the real and imaginary parts of $\Pi |_{T=0}$. 
This result can be combined with the definitions in
Section~\ref{secdefchil} to obtain the small argument
limit of the scaling function $F_2$.

\section{COMPARISON WITH EXPERIMENTS}
\label{comp_exp}

In this Section, we compare our theoretical results with the available 
experimental data for undoped, and weakly doped $La_{2-x}Sr_{x}CuO_{4}$ and
the numerical results
for 2D $S=1/2$ Heisenberg antiferromagnets on a square lattice.
But first, let us briefly summarize our findings. 

We presented above the 
general forms for uniform and staggered susceptibilities in
a two-dimensional quantum antiferromagnet which has $\rho_{s} \ll J$. The
explicit crossover functions were calculated at $N=\infty$ and the 
 $1/N$ corrections were examined in the limiting cases of $x_{1} \gg 1$ and
$x_{1} \ll 1$, where $x_{1} = N k_{B}T/(2 \pi \rho_{s})$ is a parameter
which governs the crossover between renormalized classical and quantum critical
regions (for $x_{1} \ll 1$, the system is in the renormalized classical
region, while for $x_1 \gg 1$ it is in the quantum critical region).
 We found that for large values of $x_{1}$, the perturbative expansion
is regular in $1/N$. Moreover, the corrections were numerically rather small; 
so we expect that for the physical case of $N=3$, 
the results obtained in the first order in 
 $1/N$ are already quite close to the exact values of observables. 
 On the other
hand, at small $x_{1}$, the $1/N$ expansion is logarithmically singular - it
holds in $ \log x_{1} /N $ and eventually changes 
the leading singularity in some of the scaling functions at $x_{1} \rightarrow
0$; the final low-T behavior is the same as  in the renormalized-classical
scaling theory of Chakravarty {\em et.al.\/}\cite{CHN}.    
 The crossover between small and large-$x_{1}$ behavior
should occur at $x_{1}$ around unity, though 
not necessarily  at the same $x_{1}$  
for various observables.  For $\rho_{s} \ll J$, $x_{1} \sim 1$ is 
within the validity  of the long-wavelength description, and we expect that
there should be 
a temperature range where our formulas for the quantum critical region 
describe the experimental data better than the renormalized classical theory. 
Strictly speaking, the renormalized classical theory
should be valid only  for $x_{1} \ll 1$ 
when  $\log x_{1}$ terms dominate regular perturbative corrections.

We now proceed to describe the data. 
Let us first discuss undoped antiferromagnet.
We know from elementary spin-wave analysis that zero-point fluctuations are
not divergent in two dimensions and therefore the $T=0$ renormalization of
spin-stiffness, spin-wave velocity and sublattice magnetization 
come primarily
from the lattice scales, where we can rely on the results of spin-wave
calculations. At present, two-loop spin-wave expressions are
available~\cite{Igarashi}.  
For $S=1/2$, they yield the values of $\rho_{s}, \, \chi_{\perp}$ 
and $N_{0}$, which are practically undistinguishable from the 
results obtained in numerical simulations~\cite{Singh}:
\begin{equation}
\rho_{s} = 0.181 J; \,\, \chi_{\perp} = \left(\frac{g \mu_{B}}{\hbar}
\right)^2 \, 
\frac{0.514}{8 J a^2} ; \,\, N_{0} = 0.307,
\label{E1}
\end{equation}
where $a$ is the lattice spacing.
Hence $2\pi \rho_{s} \approx 1.13J$ and we therefore 
 expect that quantum critical
expressions should work for $k_{B}T >
0.35J\div 0.4J$.
 The expected crossover temperature  is indeed
 not very small,  but it is still 
significantly lower than $J$; in other words we may reasonably expect 
out long-wavelength description to continue to
be valid for $x_{1} \geq 1$.  Notice that our result for $\rho_{s}$ 
differs from $\rho_{s} \approx 0.15$ used by 
 Chakravarty {\em et.al.\/}\cite{CHN}.  The reason is that
 Chakravarty {\em et.al.\/} extracted
the value of $\rho_{s}$ from one-loop spin-wave results extended to $S=1/2$,
while our  estimate of $\rho_{s}$ is based 
on a two-loop spin-wave calculations. 

We now discuss what happens at nonzero doping. First, a mean-field 
analysis based on Hubbard model predicts that antiferromagnetism is 
destroyed at arbitrary
small concentration of holes\cite{MFHub}.
However, more sophisticated considerations show\cite{borislast,Frenk}
that there is in fact no discontinuity in the immediate vicinity of
half-filling, and one needs a finite, though small, concentration of holes to
destroy antiferromagnetic ordering.  There are several possible scenarios
of the doping-induced loss of N\'{e}el order: 
\newline
({\em i\/}) There is a 
$T=0$ transition from the N\'{e}el state to a incommensurate
magnetically ordered state~\cite{boris,borislast,Frenk,Col_Gan_Andrey}; or 
\newline
({\em ii\/}) The N\'{e}el is state
is destroyed by quantum fluctuations and the system enters a quantum
disordered spin-fluid with commensurate correlations. Only at a larger
doping do incommensurate correlations appear.
\newline
A recent self-consistent two-loop
calculation\cite{jinwuunpub} on the Shraiman-Siggia\cite{boris} model  
displays both sequences of transitions depending on the strength of the
coupling between fermions and $n-$field.
\newline
The two scenarios differ primarily in their predictions for the
behavior in the non N\'{e}el-ordered state; in both approaches, 
the spin-stiffness decreases
with doping and vanishes at the critical point, while the spin-wave
velocity remains finite at the transition. Thus, in any event,
the main effect of small doping is simply to decrease the bare spin-stiffness
of the antiferromagnet. Moreover, the self-consistent analysis mentioned 
above~\cite{jinwu,jinwuunpub}
shows that the holes do not modify the critical properties of a 
N\'{e}el to quantum spin-fluid transition.
Thus, the universal scaling functions computed in this paper can be
used unchanged to describe this transition in doped antiferromagnets.

A further, important, complication that has to be considered in realistic
doped antiferromagnets is the effect of randomness. All cuprate
antiferromagnets have randomly placed dopant ions which will
perturb the properties of the two-dimensional antiferromagnet.
Randomness has been argued to be a  relevant
perturbation near the pure  phase transition 
of the $O(3)$ sigma model~\cite{jinwu}.
It thus necessarily 
changes the universality class of the fixed point and also pushes
the transition to smaller doping concentrations. 
In Appendix~\ref{apprandom} we have presented a phenomenological 
discussion of the expected scaling properties of the random fixed point.
 
However, we may
conjecture (though we have no strong theoretical arguments for this) that 
the effects of randomness are important only at low temperatures
in the immediate vicinity
of the quantum transition.  At higher
$T$ the dominant effect of doping is solely the change 
in $\rho_{s}$ and the properties of the antiferromagnet should be controlled
by the pure fixed-point. 
We will soon see that neglecting randomness at moderate temperatures
is consistent with the
available experimental data in weakly doped  $La_{2-x}Sr_{x}CuO_{4}$.
Of course, this simple approach will eventually fail at large
doping. Crudely, we may take the largest possible $x$ as one where
$\rho_s$ would vanish in the absence of randomness.  We will estimate from 
the data that
this $x$ to be somewhat larger than 0.04.

An important consequence of the decrease of $\rho_s$ with doping
is that 
$x_{1}$ becomes larger at the same $T$. Hence the crossover between
renormalized-classical and quantum-critical behavior occurs at lower $T$.
Consequently, there should be a wider temperature range where our predictions
for quantum-critical region should describe experiments better than the
renormalized-classical theory. 

We now consider separately the experimental data for various observables. 

\subsection{Uniform susceptibility}

We start with the uniform susceptibility, $\chi_u^{{\rm st}}$.
This quantity  has no
logarithmic corrections in the renormalized classical region, and is
therefore an ideal candidate to test the predictions of the $1/N$ expansion.
We consider first the numerical results for $\chi_{u}^{{\rm st}}$ 
in the square-lattice
$S=1/2$ antiferromagnet. There have been high-temperature series
expansions~\cite{Singh-Gelfand},
quantum Monte-Carlo~\cite{Makivic} and finite-cluster 
calculations~\cite{Sokol}
 of $\chi_{u}^{{\rm st}}$. Their
results all show that $\chi_{u}^{{\rm st}}(T)$ 
obeys a Curie-Weiss law at high $T$,
reaches a maximum at $k_{B}T \sim J$ and then falls down 
to a finite value at $T=0$
which is close~\cite{Singh_chi}
 to the rotationally averaged 1/S result $(\hbar a/g \mu_{B})^2
\chi_{u}^{{\rm st}}(T=0) = ({2}/{3}) \, ({0.514}/{8 J}) \approx 0.04/J$
($a$ is the lattice spacing).
 The data at low $T$ are not
accurate enough to make a reliable theoretical fit 
but at higher $T \, (0.35J <
T < 0.55J)$, both series expansions and 
Monte-Carlo calculations report a linear
temperature dependence of $\chi_{u}^{{\rm st}}$. 
The best fit to the Monte-Carlo data
gives  $(\hbar a/g \mu_{B})^2
 J \chi_{u}^{{\rm st}} = 
0.037 x_{1} (1 + 0.775/x_{1})$ (Fig.~\ref{figunisus}). 
We compared this
behavior with the theoretical prediction 
 for the quantum critical region,  which over the range
of $x_{1}$ values used in the figure, is well 
approximated by the large-$x_{1}$
formula Eqn (\ref{chistdef}, \ref{Omega1subleading},\ref{Omega1})
: $(\hbar a/g \mu_{B})^2
 J \chi_{u}^{{\rm st}} = 
0.037 x_{1} (1 + \alpha/x_{1})$ where $\alpha = 0.8 + {\cal O}(1/N)$.
(We do not know the $1/N$ corrections to $\alpha$; we only know
that the $1/N$ corrections modify
the $1/x_1$ term to $1/x_1^{1/\nu}$.)
The agreement between the slopes of the two results is remarkable.
On the contrary, the slope for the renormalized classical region, 
Eqn (\ref{RC26}), is 
$0.014 x_{1}$ in clear disagreement with the numerical data at $T > 0.35J$.  

We consider next measurements of $\chi_{u}^{{\rm st}}(T)$ in weakly doped 
$La_{2-x}Sr_{x}CuO_{4}$. The interpretation of the experimental data requires
caution~\cite{Johnson,Andy} because one has to subtract Van-Vleck, 
core and dia- and paramagnetic
contributions from valence fermions 
from the measured $\chi_{u}^{{\rm st}}(T)$. Besides, as we said
above, the effects of randomness are clearly important at low $T$.
The subtraction of extra contributions neglecting the effects of
randomness was first done by Johnson~\cite{Johnson}
who actually estimated the strength of the Van-Vleck contribution by
{\it{assuming}} that at zero doping, the susceptibility should
be  the same as in the Monte-Carlo studies of 2D antiferromagnet.
Nevertheless, his results for $\chi_{u}^{{\rm st}}$ 
clearly show~\cite{Johnson,Andy} that at  small doping 
concentrations, the susceptibility is linear in $T$ with the
slope of $(\hbar a/g \mu_{B})^2
 J \chi_{u}^{{\rm st}}$ vs. $x_{1}$ 
about 0.043 which is close to our result (0.037). 
Unfortunately, we cannot  definitely conclude from the experimental data 
whether the linear  behavior the with universal slope 
stretches up to lower $T$ as the doping increases, which has to be the case if
$\rho_{s}$ decreases with the doping. We will observe this effect, however, 
in the data for the spin-lattice relaxation rate.  Note also that at 
 higher doping concentrations $x_{1} \sim 0.1$, the experimentally measured 
$\chi_{u}^{{\rm st}}(T)$ vs $T$ flattens~\cite{chiexpdop}. However, 
at such $x_{1}$, the system is already in the metallic phase, 
where our approach
clearly has to be modified.

\subsection{NMR relaxation rate}

The simple Mila-Rice-Shastry model~\cite{Mila-Rice} for hyperfine coupling in 
$La_{2}CuO_{4}$ predicts that the hyperfine coupling constant for $^{63}Cu$ 
is finite at the antiferromagnetic ordering momentum $(\pi,\pi)$, where the 
dynamic susceptibility is peaked.  The region around $(\pi,\pi)$ thus gives 
the dominant contribution to $^{63}Cu$ relaxation and we  may use the 
long-wavelength theory for experimental comparisons. At the same time, the 
value of the hyperfine coupling constant in the sigma model approach 
cannot be inferred directly from the Knight shift measurements as in microscopic theories~\cite{MMP}.  Instead, we have to integrate out all 
intermediate scales in the microscopic model for the hyperfine interaction, 
to obtain the  coupling constant between the nuclear spin and the unit vector 
field in the sigma model. This renormalization is not singular, however, and 
the  fully renormalized coupling, which we label as $A_{\pi}$, should not 
be very different from the microscopic one. 

At very low $T$ ($x_{1}$ small), the system is in the renormalized 
classical region. For $N=3$, our theoretical result Eqn. ({\ref{RC45}}) is the 
same as in the hydrodynamic theory of  Chakravarty {\em et.al.\/} \cite{CHN} 
\begin{equation}
\frac{1}{T_{1}} = \sqrt{\frac{2}{3}}~ \lambda ~\frac{A^{2}_{\pi} 
N^{2}_0 }{\hbar}~ \frac{\xi}{\hbar c} \, \left(\frac{k_{B}T}{2 \pi \rho_{s}}
\right)^{3/2}.
\label{E2}
\end{equation}
Here $\lambda$ is the numerical factor which is difficult to calculate 
analytically. Chakravarty and Orbach~\cite{Chak-Orbach} estimated it to be 
$\lambda N^{2}_0 \approx 0.61$ by fitting the scaling forms of  
Chakravarty {\em et.al.\/}\cite{CHN} and Tyc {\em et.al.\/}\cite{Tyc} to the 
numerical simulations on a classical lattice rotor model. 

On the other hand, at higher temperatures ($x_{1} \geq 1$) we expect the 
quantum critical theory to work and  $1/T_{1}$ should behave as in 
Eqn (\ref{nmrscale}, \ref{r1qdef}) 
\begin{equation} 
\frac{1}{T_{1}} = R_1 (\infty) \, \frac{2 A^{2}_{\pi} N^{2}_0 }{\hbar \rho_{s}} 
\, \left(\frac{3 k_{B}T}{2\pi \rho_{s}}\right)^{\eta}. 
\label{E3}
\end{equation}
Comparing these two forms for $1/T_{1}$, we observe that the predicted 
spin-lattice relaxation rate  rapidly decreases with the temperature at 
low $T$, passes through a minimum at $x_{1} \sim 1$ and then slightly 
increases with $T$. In practice, $\eta$ is very small for clean systems
($\eta \approx 0.028$ ~~\cite{holm}) and therefore  $1/T_{1}$ should be 
nearly independent on temperature in the quantum-critical region.  

We now turn to the data. The spin-lattice relaxation rate for nuclei 
coupled to the antiferromagnetic order parameter of 2D $S=1/2$ antiferromagnet 
was numerically studied in high-$T$ series expansions~\cite{Singh-Gelfand} 
and finite cluster calculations~\cite{Sokol}. In both cases, $1/T_{1}$ 
rapidly decreases with increasing $T$ at high temperatures ($T \geq J/2$) and
becomes weakly temperature dependent around $T \sim J/2$. 
In finite-cluster calculations, the subsequent growth of $1/T_{1}$ at 
lower temperatures has also been observed. 
Clearly, this behavior is consistent with our theoretical observations.  

Experimental measurements of $1/T_{1}$ in undoped and weakly doped 
$La_{2-x}Sr_{x}CuO_{4}$ have recently been performed by 
Imai {\em et.al.\/}\cite{Slichter} in the temperature range $20 - 900K$. 
For undoped system, the behavior above $T_{N} = 308K$, but below $700K$ 
is well fitted by (\ref{E2}) with $A_{\pi} = (1.33 \pm 10)~ 10^{-2} K$ 
and $J = 1590 \pm 140 K$ (our estimate for $A_{\pi}$ is 
slightly different from theirs\cite{Slichter} because we use the exact 
prefactor for the correlation length). 
However, at about $650K$, $1/T_{1}$ flattens and remains practically 
independent of temperature up to $900K$ - the largest temperature reported 
in by Imai {\em et. al. \/}\cite{Slichter}.  This is indeed what we expect 
from $1/T_{1}$ in the quantum-critical region.  
Furthermore, the experimental $T$ range over which $1/T_{1}$ is nearly 
$T$ independent increases with doping: at $x=0.04$ it stretches 
nearly up to $500$ K. We interpret this result as an evidence that 
$\rho_{s}$ indeed decreases with doping, thus pushing the system into 
larger $x_{1}$ for the same $T$.

For a quantitative comparison with the data, we need to know the value of 
$R_1 (\infty)$. Direct $1/N$ calculations give 
$R_1 (\infty) = 0.06/N$, which is too small to account for the 
experimental result for $1/T_1$. However, we already observed in 
Sec. \ref{qcstag} that $\mbox{Im} \Phi_{1s}$ calculated to leading order 
in $1/N$ has peculiar exponential singularities at small frequencies 
(Eqn. (\ref{exp_sing}, \ref{exp_sing2})) which are probably artifacts of the 
large $N$ expansion. These singularities substantially reduce the 
slope of $F_1 (\overline{\omega})$ at the smallest $\overline{\omega}$ 
(see Fig. \ref{figchil}). On the other hand, no such low-frequency 
suppression of $F_1 (\overline{\omega})$ was found in numerical 
studies~\cite{Sokol} and in the experiments on weakly doped $La$ 
compounds~\cite{keimer2}. In view of this, it appears reasonable to 
estimate the value of $R_1 (\infty)$ from our result for the scaling 
function $F_1$ at slightly larger frequencies.  Inspection of the numerical 
result for $F_1$ (Fig. \ref{figchil}) shows that $F_1 (\overline{\omega})$ is 
linear in $\overline{\omega}$ for $\overline{\omega}$ between $0.5$ and $1$, 
and the slope yields $R_1 (\infty) \approx 0.22$. 
Substituting this result into (\ref{E3}) and using the values of $A_{\pi}$ and
$J$ from the low-$T$ (renormalized-classical) fit, we 
obtain $1/T_{1} = (3.2 \pm 0.5)\times 10^{3} sec^{-1}$; this is in a 
good agreement with the experimental result 
$1/T_{1} \approx 2.7\times 10^{3} sec^{-1}$.
 
\subsection {Correlation length}

 Detailed measurements of $\xi (T)$ in pure $La_{2}CuO_{4}$ have 
been performed~\cite{Yamada,birg,keimer2} at low T, where the system is in the renormalized classical 
region. Here Eqn.~(\ref{RC17}) is valid and using (\ref{E1}) we obtain 
\begin{equation}  
\xi(T) \approx 0.50 a \exp \left\{\frac{1.13J}{k_{B}T}\right\} .
\label{E4}
\end{equation}
 Chakravarty {\em et.al.\/}\cite{CHN} 
 used a different numerical prefactor in (\ref{RC17})
 and a different value for the
spin-stiffness constant. The combination of the two yielded nearly the same 
value of prefactor as in (\ref{E4}), but the numerical factor in the 
exponent was slightly different ($0.94$ instead of $1.13$). This discrepancy
is not crucial however, and both Eqn.~(\ref{E4}) and 
the analogous expression
of Chakravarty {\em et. al.\/}~\cite{CHN}
fit the experimental data between $350K$ and $560K$ rather
well. The value of $J$ has been estimated by high-energy neutron scattering
measurements of the spin-wave velocity~\cite{gabe} to 
be $J \sim 1560K$. Hence
 $x_{1} \sim T/590K$, and all of the experimentally accessible temperature 
range  is within the renormalized classical region. Nevertheless, we
estimated the value of $\xi$ at the 
highest experimentally accessible temperature
$ 560K$ by using the $N=\infty$ expression for the crossover function $X_{1}$
in (\ref{X1Ninfty}); we
obtained $\xi^{-1} = 0.023 \dot{A}^{-1}$, which is not far from the
experimental value~\cite{keimer2} of $0.03 \dot{A}^{-1}$.  

At finite doping, we expect that 
the crossover between the two regimes will 
occur at smaller temperatures;
quantum-critical behavior should therefore be observable
at temperatures above and around $500K$.
Deep in the quantum-critical region, we expect that $\xi$ behaves as
\begin{equation}
\xi^{-1} = 1.039 ~\frac{k_{B}T}{\hbar c a} ~ \left (1 -
 \frac{\gamma}{x_{1}}\right ), 
\label{E5}
\end{equation}
where $\gamma \approx 1$. 
The $1/N$ corrections have been included in the slope but are not known for
$\gamma$: they also change the subleading term to $x_1^{1/\nu}$.
We fitted the data of Keimer {\em et. al.\/}~\cite{keimer2} 
at $x=0.04$ by this 
formula and  found satisfactory
 agreement with the data over the temperature range between $300K$ and
$550K$. The value of $\rho_{s}$ extracted from the fit: $ 2\pi \rho_{s} \sim 
150K \div 300K$,
 is still positive, but of course is much smaller than at zero doping.
Note that Keimer {\em et. al.\/}\cite{keimer2}, 
used a phenomenological form for $\xi^{-1}(T)$
which combined the renormalized-classical result at zero-
doping and temperature-independent correction due to finite doping; this form
agreed well with the experimental data for doping concentrations 
$x = 0 \div 0.04$, but the theoretical arguments behind it are unclear to us. 

We also compared our results for the 
quantum-critical region with the numerical
data for $\xi$ in a pure 2D $S=1/2$ Heisenberg 
antiferromagnet~\cite{Makivic,Manous-MC,Man}.
Numerical
simulations were performed up to temperatures of about 
 $4J$; if quantum-critical behavior for $\xi$ is present in the Heisenberg
antiferromagnet in some temperature range, it should have been detected. 
It turns out, however, that up to $k_{B}T \sim J$, 
 numerical data are well fitted by the renormalized
classical theory, although the best fit~\cite{Makivic}
 gives the value for the prefactor 
 which is nearly half of that in (\ref{E4}). On the other hand, 
 for $0.4J < k_{B}T < 0.6J$ (where $1/T_1$ levels off to a constant value),
 our pure quantum-critical result for  $\xi$
 is close to the numerical one~\cite{Makivic,Manous-MC}.
The interpretation of the numerical data at higher $T$ requires cauton as 
these data at $k_B T > 0.6J$ are equally well
fitted by the quantum-critical result
$\xi^{-1} \propto x_{1}(1 - \gamma x_{1})$ where $\gamma$ is close
to one~\cite{Sasha}.
 However, the prefactor  in the fit is nearly twice as that in
(\ref{E5}). We argue in a separate publication~\cite{t2paper} 
that these discrepancies in the fit to the quantum-critical 
theory are chiefly due to nonuniversal corrections $\sim T/J$ which cannot
be neglected above $0.6J$.

Separate Monte-Carlo calculations
 of the correlation length precisely at $\rho_{s} =0$
have been performed by Manousakis and Salvador~\cite{Man}. They
simulated the quantum $O(3)$ sigma model directly, rather than the
spin-1/2 antiferromagnet. Their results
 yielded the value of the universal 
function $X_{1}(\infty) = 1.25$ which 
is not far from the result of Chakravarty {\em et. al.\/}~\cite{CHN},
or our result in (\ref{E5}): 
$X_{1}(\infty) = 1.04 $.

\subsection{Equal-time structure factor}

There are,  to our knowledge, only few data  available on $S(k)$.
The behavior of $S(k)$ vs $k\xi$ in a 2D antiferromagnet was studied by a 
Quantum Monte-Carlo by Makivic and Jarrell~\cite{Jarrell}.
 They fitted their data at $k_{B}T= 0.35J
\div 0.42J$ by the renormalized-classical scaling formula of Chakravarty 
{\em et.al.\/}\cite{CHN}. We present another interpretation for the
data. The key point is that at 
 $k_{B}T = 0.42J$, 
the correlation length is about$5.7 a$
and  the magnon energy is therefore $\hbar c k/k_{B}T \sim
0.7 (k\xi)$. We see that it becomes larger 
than $k_{B}T$ already
at $k\xi \approx 1.5$, and the use of 
the classical description at larger $k$ is hardly justified. We rather
have to use the full quantum expression for $S(k)$ to fit the data. Clearly,
 at  $k_{B}T \sim 0.4J$, the Josephson correlation length is not very 
different from $\xi$, that is
at $\hbar c k > k_{B}T$ the system should be in the quantum-critical
regime. However, we found above that the $1/N$ corrections to $S(k)$ in
this regime are very small, even for $N=3$, and for experimental comparisons
we may well use the 
scaling function for $S(k)$ computed at $N=\infty$. For $k_B T = 0.42 J$, we
have the following theoretical prediction, valid in both the quantum-critical
and renormalized-classical regions, from (\ref{sksig1}) and (\ref{XiNinfty})
\begin{equation}
S(k) = \bar{S} \, \,  \frac{\mbox{coth}[\alpha \sqrt{1 + (k\xi)^{2}}]}{\sqrt{1 + 
(k\xi)^{2}}}, 
\label{E6}
\end{equation}
where $\alpha = \hbar c/2 k_B T \xi \sim 0.35$ and $\bar{S} =
(N^{2}_{0}/\rho_s) (\hbar c \xi /2 a^2) \approx 2.48$. 
We fitted the Monte-Carlo data by (\ref{E6}), using the overall factor $\bar{S}$
as the only adjustable parameter; we found very good agreement with the
simulations not only for $\hbar c k > k_B T$, but for all $k$
(Fig.~\ref{figsk}).
However, the value of $\bar{S}$ in the fit is $\bar{S} = 3.61$, which is
somewhat larger than our $N=\infty$ result of $2.48$. As yet, we have no
explanation for the discrepancy, and therefore cannot judge from the data
whether at intermediate $k$ the system is in the quantum-critical or in the
Goldstone regime ($S(k)$ in the quantum-critical regime is very close to the
$N=\infty$ result, while in the Golstone regime, $S(k)$ chiefly differs
from its value at $N=\infty$ by the factor $(N-2)/N = 2/3$). At the same time,
the good agreement we found in the momentum dependence of $S(k)$ is
clearly consistent with  our conjecture
 that at $k_{B}T \sim 0.4J$, the antiferromagnet is very near the  
crossover  between renormalized-classical
and quantum-critical regimes.

Experimental data for energy-integrated $S(k)$ are available~\cite{keimer2} for 
pure $La_{2}CuO_{4}$. Clearly, the experimental
temperature dependence for the correlation length was inferred from these
data. The
experiments\cite{keimer2} were performed in the temperature range of $T < 560K$, 
where $x_{1} < 1$. The momentum
dependence of  $S(k)$ was 
reported to be well described by a simple Lorentzian $S(k) = S(0)/(1 +
(k\xi)^{2})$, and the temperature dependence of $S(0)$ agreed with the
renormalized-classical result $S(0) \propto T^{2} \xi^{2}$. The 
absolute value of $S(0)$ was not determined in the experiments, so we 
cannot compare the experimental result  at highest accessible temperature with
the theoretical expression in the quantum-critical region, as we did for the
correlation length. The data on $S(0)$ at finite doping have not been reported,
 to the best of our knowledge.                                           

\subsection{Local susceptibility}

Finally, we consider the experimental data on the momentum-integrated
dynamical susceptibility $\mbox{Im} \chi_{L}(\omega) = \int d^2 k 
\mbox{Im} \chi_s (k,\omega)$. Extensive experimental measurements were 
done~\cite{birg,keimer2}
 for $La_{1.96}Sr_{0.04}CuO_{4}$ in a frequency range $\omega = 2 \div 45
\mbox{mev}$ and for temperatures $10 < k_{B}T < 500K$. 
They showed that the experimental data
at all values of $\omega$ and $k_{B}T$, obeyed the following
functional form to reasonable accuracy
\begin{equation}
\mbox{Im} \chi_L ( \omega ) = I ( |\omega| ) F ( \omega / T ).
\end{equation}
This form is in agreement with the theoretical
scaling form (\ref{chilscale}, \ref{f1def}) for clean systems, or
(\ref{appchil}, \ref{appchilsg}) from Appendix A for random systems; further, 
we arrive at the
theoretical prediction that $I \sim |\omega|^{\mu}$. The exponent
$\mu = \eta > 0 $ in clean systems, while we expect $\mu < 0$
in random systems\cite{trieste,jinwu}.
Experimentally, it was found that $I$ was approximately $\omega$
independent at the larger frequencies; this is consistent with our
results for the clean antiferromagnet in which $\mu = \eta $ is very
small. At smaller frequencies, $I$ showed significant frequency
dependence which could be well fit by
an exponent
$\mu = -0.41 \pm 0.05$. Such a behavior is clear evidence of the
importance of randomness at low frequencies and low temperatures~\cite{jinwu}.
Note also that the scaling plot to test the $\omega / T$ dependence
of the function $F$ shows rather good collapse of the data
for a wide range of values of $\omega / T$~\cite{birg,keimer2}, thus
giving strong evidence of the presence of quantum-criticality.
The universal scaling functions at
the pure and random fixed points probably 
have a rather similar shape, and thus the presence of disorder
does not effect the scaling
plot very much.
It is only the exponent $\mu$ which is particularly sensitive
to disorder.

We emphasize, then, that this experimental data clearly shows
the presence of quantum-criticality and suggests that
the effects of randomness are important only at low energies; at high
enough $\omega$ or $k_{B}T$ one can suscessfully fit the data at $x=0.04$
by the quantum-critical theory
for a clean antiferromagnet. 

\section{CONCLUSIONS}
\label{concl}

We conclude the paper by recalling some highlights of our results.

We have presented the general theory
 of clean, two-dimensional, quantum Heisenberg
antiferromagnets which are close to the zero-temperature quantum transition
between ground states with and without long-range N\'{e}el order.
While some of our discussion was more general, the bulk of our theory
was restricted to antiferromagnets in which the N\'{e}el order
is described by a $3$-vector order parameter.
For N\'{e}el-ordered states, `nearly-critical' means that the
ground state spin-stiffness, $\rho_s$, satisfies $\rho_s \ll J$, where
$J$ is the
nearest-neighbor exchange constant, while `nearly-critical' quantum-disordered
ground states have a energy-gap, $\Delta$, towards excitations with 
spin-1, which satisfies $\Delta \ll J$.
The allowed temperatures, $T$, are also smaller
than $J$, but {\em no} 
restrictions are placed on the values of $k_B T /\rho_s$ or 
$k_B T/\Delta$.

Our results followed from some very general properties of the $T=0$
quantum fixed point separating the magnetically-ordered and
quantum-disordered phases. These properties are expected to be valid in both
undoped and doped antiferromagnets, though not in the presence of randomness.
They are ({\em i\/}) the spin-wave velocity, $c$, is non-singular at the
fixed-point (i.e., the dynamical critical
exponent $z =1$); for the case of a vector
order parameter this implies that the critical field-theory has the
Lorentz invariance of 2+1 dimensions; ({\em ii\/})
on the ordered side of the transition, there is 
a Josephson correlation length,
$\xi_J$, related to the $T=0$ spin stiffness  $\rho_s$ 
by the hypothesis of `two-scale factor'
universality which implies that
$\rho_s = \hbar c \Upsilon/\xi_J$, where the
 number $\Upsilon$ is dimensionless and universal; ({\em iii\/})
turning on a finite temperature places the critical field theory in a 
slab geometry which is infinite in the two spatial directions, but of 
finite length, $
L_{\tau} = \hbar c /(k_B T)$, 
in the imaginary time ($\tau$) direction. The consequences of a finite $T$
can therefore be deduced by the principles of finite-size scaling.

Under these circumstances, we showed that the 
wavevector/frequency-dependent uniform and staggered spin 
susceptibilities, and the specific heat, are completely universal functions
of just three thermodynamic parameters. On the ordered side, these
three parameters are $\rho_s$, the $T=0$ spin-wave velocity
$c$, and the ground state staggered moment $N_0$;
previous works have noted the universal dependence of the susceptibilities
on these three parameters 
only in the more restricted regime of $k_B T \ll \rho_s$.
On the disordered
side the three thermodynamic parameters 
are $\Delta$, $c$, and the spin-1 quasiparticle residue ${\cal A}$.

We have calculated the  universal scaling functions  
by a $1/N$ expansion on the $O(N)$ quantum non-linear sigma model,
and by Monte Carlo
simulations. For $\rho_s$ finite, these scaling functions demonstrate
the crossover behavior between the renormalized classical regime, when
thermal fluctuations are dominant, to the
quantum-critical regime, where the dynamics is govern by the renormalization
group flows near the $T=0$ quantum fixed point. 
For small $k_B T/\rho_s$, the $T$-dependence of our results was similar
to those already obtained by Chakravarty {\em et al}~\cite{CHN}. For large
$k_B T/\rho_s$, most of our results were new. We found that the
crossover between the renormalized-classical and quantum-critical
 regimes occurs at $x_1 \sim 1$, where $x_1 = N k_B T/ 2\pi
\rho_s$.
In a square lattice, $S=1/2$ Heisenberg antiferromagnet, $2\pi \rho_s \approx
1.13J$, and quantum-critical behavior therefore should be seen at $ k_B T \geq
0.4 J$. We compared our  quantum-critical results with
  a number of numerical simulations and 
experiments on undoped and lightly-doped
$La_{2-\delta} Sr_{\delta}
Cu O_4$ in the intermediate temperature range, and found good agreement
with the data, particularly for uniform susceptibility, NMR relaxation rate
and equal-time structure factor.
It appears, therefore, that the use of a `small' $\rho_s$ point-of-view
is quite reasonable even for a pure square lattice, $S=1/2$ Heisenberg
antiferromagnet - while ordered at $T=0$, this system is evidently 
close to the point where long-range order vanishes. 
A small $\rho_s$ approach also appears to be appropriate for
lightly-doped antiferromagnets.

\acknowledgements
This research was
supported by NSF Grants No. DMR-8857228 and DMR-9224290,
and by a fellowship from the A.P. Sloan Foundation.
We are pleased to thank R. Birgeneau, S. Chakravarty,
D. Fisher, E. Fradkin, P. Hasenfratz, 
T. Imai, B. Keimer, M. Makivic, 
E. Manousakis, A. Millis, H. Monien, A. Ramirez, 
N. Read,  A. Sokol and R. Shankar 
for useful discussions and communications.

\appendix

\section{Random antiferromagnets and anisotropic scaling}
\label{apprandom}
This appendix will present a brief discussion of the extension of the
results of this paper to the case of random antiferromagnets.
We will restrict our attention to systems in which the randomness
preserves the Heisenberg spin symmetry {\em i.e.\/} antiferromagnets
with random exchange constants $J_{ij}$.

Random impurities induce perturbations on the clean system which are
uncorrelated in the spatial directions but fully correlated along the
imaginary time direction. This has the immediate consequence of
breaking the long-distance Lorentz invariance of the pure system. 
It therefore becomes necessary to allow for anisotropic scaling in
space-time at the quantum fixed point. It is conventional to
introduce a dynamic scaling exponent $z$ such that characteristic
frequencies, $\omega$, scale with the characteristic wavevector $k$
as 
\begin{equation}
\omega \sim k^z
\end{equation}
in the critical region. 

It is important to distinguish to two distinct classes of magnetically
ordered phases that can occur in random antiferromagnets: these 
are ({\em i\/}) N\'{e}el ordered and ({\em ii\/}) spin-glass ground
states. In the first of these the ordering moment has a definite
orientation on each of the lattice sites. In the spin-glass, there is
a long-lived moment on each site, but its orientation is random. 
We will consider the properties of the transition of
these two magnetically-ordered
states to a spin-fluid in turn:

\subsection{N\'{e}el order}

We present here some phenomenological scaling ansatzes for the
quantum phase transition from a N\'{e}el ordered state to a
spin-fluid in a random antiferromagnet. The scaling arguments are
similar to those employed for the superfluid to bose-glass
transition by Fisher {\em et. al.\/}\cite{weichmann}, although they
did not discuss the issue of universal amplitudes. We will
restrict our attention to the magnetically-ordered side $g < g_c$.
The magnetically disordered side is expected to be
gapless\cite{bhatt_lee} and its
properties will not be discussed here. We also note that little
explicit reference to randomness will be made here, its main role
being the introduction of a $z \neq 1$. The results should therefore
also be applicable to other quantum phase transitions in {\em clean}
systems which have $z \neq 1$ and are below their upper critical
dimensions. We also assume below that the antiferromagnet is below its upper
critical dimension and hyperscaling is valid.

As in clean systems, we expect that $T=0$ ordered state is
characterized by a Josephson length scale, $\xi_J$, separating the
Goldstone and critical regions. As $g$ approaches $g_c$, this scale
must diverge as
\begin{equation}
\xi_J \sim (g_c - g)^{-\nu} .
\label{appxij}
\end{equation}
At scales larger than $\xi_J$, the system should exhibit conventional
Goldstone fluctuations with a well defined spin-wave velocity, $c$.
However, because $z \neq 1$, the spin-wave velocity should exhibit
non-trivial critical behavior as $g$ approaches $g_c$:
\begin{equation}
c \sim (g_c - g)^{\nu ( z -1 )} .
\label{appc}
\end{equation}
The ground state is also characterized by an average ordered moment $N_0$
which vanishes as
\begin{equation}
N_0 \sim (g_c - g)^{\beta} ,
\end{equation}
and a spin-stiffness $\rho_s$ which satisfies\cite{weichmann}
\begin{equation}
\rho_s \sim (g_c - g)^{\nu ( d + z - 2)} ,
\label{apprhos}
\end{equation}
where $d$ is the spatial dimensionality. The exponent identity
\begin{equation}
2 \beta = ( d + z - 2 + \eta ) \nu
\label{appbeta}
\end{equation}
generalizes (\ref{betaeta}) and will be useful to us below.

In the main part of the paper we showed that the three properties
$N_0$, $\rho_s$, and $c$, of the $T=0$ state, completely determine
the entire finite temperature form of the susceptibilities and the
specific heat in clean antiferromagnets. We argue below that,
remarkably, this continues to be true even in random antiferromagnets
which have $z\neq 1$. The value of $c$ helps determined the
appropriate scaling between space and time even though $c$ itself has
non-trivial critical behavior.

The following discussion will specialize explicitly to the case of
$d=2$, although the generalization to arbitrary $d$ is quite
straightforward. We will also assume that $z < 2$, otherwise, the universality
in the spectrum 
near the critical point will be broken by the higher order analytic terms in an
expansion in $k$. As before, we will use units in which $\hbar = k_B = 1$.
Of course, we are no longer free to set $c=1$!

The application of finite-size scaling to quantum systems requires
that one determine two length scales characterizing the effects of
({\em i\/}) the
deviations from criticality, and ({\em ii\/}) 
the temperature, and write down
universal functions of their ratio. The length scale characterizing
deviations from criticality is clearly $\xi_J$. In clean systems,
the length scale $\xi_T$ characterizing the effect of a finite $T$
was given by $\xi_T = c/T$. 
Scaling suggests that in systems with $z \neq 1$ we should
have $\xi_T \sim T^{-1/z}$. 
The scaling functions will now depend on the ratio $\xi_J / \xi_T$.
Using (\ref{appxij}) we see that this ratio is measured by the
value of
$(g_c - g)^{\nu z} /
T$. However from (\ref{apprhos}) this is the same (in $d=2$) as
$\rho_s / T$. Moreover, since this ratio also has engineering
dimension 0, there are {\em no} non-universal metric factors that can
appear. Thus as in clean systems, the scaling functions will have a
universal dependence upon $\rho_s / T$. For similar reasons, they can
also depend only on $\omega / T$. 

It remains to consider the wavevector dependence of the scaling
functions. By scaling, this must appear in the combination $q \xi_T
\sim q / T^{1/z}$. Let us now try and determine the metric factor in
front of this combination. There are two basic rules: ({\em i\/})
the metric factor should involve combinations of observables whose
scaling dimension is 0; and ({\em ii\/}) the entire combination which
appears in the argument of the scaling function should have
engineering dimension 0. Once these rules are satisfied, one is
guaranteed, by the principles of scaling, that no non-universal
pre-factors remain. A little experimentation shows that the following
satisfies these criteria:
\begin{equation}
\xi_T = \frac{c}{T} \left( \frac{T}{\rho_s} \right)^{1-1/z}.
\end{equation}

We now have enough information to use the same steps as were used in 
Section~\ref{phenom} and obtain universal scaling functions of the
observables. We will omit most of the intermediate steps here and go
directly to the results.

Consider first the susceptibility measuring fluctuations of the order
parameter, $\chi_s$. We find
\begin{eqnarray}
\chi_s ( k , \omega ) = && \frac{N_0^2}{\rho_s} \left( \frac{c}{T}
\right)^{2} \left( \frac{T}{\rho_s} \right)^{2 + (\eta-2)/z} \nonumber \\
&&~~~~
\Phi_{1s} \left( \frac{c k}{T} \left( \frac{T}{\rho_s}
\right)^{1-1/z} , \frac{\omega}{T} , \frac{T}{\rho_s} \right) ,
\end{eqnarray}
where $\Phi_{1s}$ is a completely universal function and there are
no non-universal metric factors. 
We have chosen the prefactors to satisfy the 
convention that scaling functions should 
remain finite as $g$ approaches $g_c$.
One can verify from (\ref{appc}),
(\ref{apprhos}) and (\ref{appbeta}) that the pre-factor of the
scaling function, and the coefficient of $q/T^{1/z}$ are 
non-singular as $g$ approaches $g_c$. 
The scaling results for all the observables dependent upon $\chi_s$
can now be obtained in a manner similar to that used for clean
antiferromagnets. We will display explicit expressions for only two
of them: the correlation length $\xi$ satisfies
\begin{equation}
\xi^{-1} = \frac{c}{T} \left( \frac{T}{\rho_s} \right)^{1-1/z}
X_1 \left( \frac{T}{\rho_s} \right),
\end{equation}
while the local susceptibility $\mbox{Im} \chi_L$ is given by
\begin{equation}
\mbox{Im} \chi_L ( \omega ) = \frac{N_0^2}{\rho_s } 
\left( \frac{T}{\rho_s} \right)^{\eta/z} F_1 \left( \frac{\omega}{T},
\frac{T}{\rho_s} \right) .
\label{appchil}
\end{equation}
Again $X_1$ and $F_1$ are completely universal functions chosen such
that $X_1 ( \infty )$ and $F_1 ( \overline{\omega} , \infty )$ are finite. 

The properties of the uniform susceptibility and the specific heat
follow from an understanding of the hyperscaling properties of the
free energy density, ${\cal F}$. 
A simple generalization of the arguments of 
Privman and Fisher\cite{privman} to anisotropic systems yields (in $d=2$)
\begin{equation}
{\cal F} = {\cal F}_0 + T \xi_T^{-2} \varphi \left( \frac{T}{\rho_s}
\right),
\end{equation}
where ${\cal F}_0$ is the ground state energy, and $\varphi$ is a
universal function. As in Section~\ref{phenom} the following results
for static uniform susceptibility $\chi_u^{{\rm st}}$ and the 
specific heat $C_V$ now follow
\begin{eqnarray}
\chi_u^{{\rm st}}  &=& (g \mu_B)^2 \frac{T}{c^2} \left( \frac{T}{\rho_s}
\right)^{-2 + 2/z} \Omega_1 \left( \frac{T}{\rho_s} \right) ,
\nonumber \\
C_V &=& \frac{T^2}{c^2} \left( \frac{T}{\rho_s} \right)^{-2+2/z}
\Psi_1 \left( \frac{T}{\rho_s} \right) ,
\end{eqnarray}
where $\Omega_1$, $\Psi_1$ are universal functions, with $\Omega_1 (
\infty )$ , $\Psi_1 ( \infty )$ finite. Note, in particular, that
the Wilson ratio, $W$, (Eqn. (\ref{Wilson})) continues to remain a fully
universal function of $T/\rho_s$, and has a universal value at $g=g_c$. 
The universality of the Wilson ratio, even in the presence of
anomalous powers in $\chi_u^{{\rm st}}$ and $C_V$, was also noted recently
(using very different arguments)
in the boundary critical theory of overscreened Kondo fixed 
points\cite{ludwig}.

\subsection{Spin-glass order}
We now consider the properties of a quantum transition from 
a spin-glass ground state to a spin fluid. Clearly many of the properties 
discussed in the previous section are special to N\'{e}el ordered states
and do not generalize. However measurements which are spatially local
do have 
similar critical properties. As we do not wish to discuss
the nature of the spin-glass state itself, we 
will restrict ourselves here to behavior in the quantum-critical region
where $T \gg (g_c - g)^{z \nu}$.

A very useful measure of the local spin correlations is provided
by the local spin susceptibility, $\chi_L ( \omega )$ defined in
(\ref{childef}).
Along the imaginary frequency axis, this local susceptibility is
given at the Matsubara frequencies $\omega_n$ by
\begin{eqnarray}
\chi_L ( i \omega_n ) 
 &=& \int_0^{1/T} d
\tau e^{i \omega_n \tau} C(\tau), \nonumber \\ 
C(\tau) &=&
\overline{ \langle {\bf S}_i (0) \cdot {\bf
S}_i ( \tau ) \rangle } ,
\label{ctau}
\end{eqnarray} 
where the bar represents an average over all the sites $i$.
The function $C(\tau )$ can be used to distinguish the spin-glass and
spin-fluid states. In the spin-fluid state $C( \tau )$ will decay to 
zero for large $\tau$, will in the $T=0$
spin-glass phase~\cite{young_binder} we must have 
\begin{equation}
\lim_{\tau\rightarrow\infty} C(\tau ) = q_{EA} > 0 ,
\label{appsg1}
\end{equation}
with $q_{EA}$ the Edwards-Anderson order-parameter. 
It is conventional to define the order-parameter exponent $\beta$ by
the behavior of $q_{EA}$ as $g$ approaches $g_c$ :
\begin{equation} 
q_{EA} \sim (g_c - g)^{\beta}.
\label{appsg2}
\end{equation}
The value of $q_{EA}$ thus fixes the behavior of $C( \tau )$ at
infinite time. We can obtain the behavior of $C( \tau )$ for finite
times $\tau$ by a simple application of the dynamic scaling
hypothesis. We expect at $T=0$ that
\begin{equation}
C ( \tau ) = (g_c - g)^{\beta} 
h_1 ( \tau |g - g_c |^{z\nu} )~;~~~~~~~~T=0 ,
\label{appsg3}
\end{equation}
where $h_1$ is a scaling function and $z$ is the dynamic scaling
exponent. Clearly we must have $h_1 ( x \rightarrow \infty )$ be a
finite non-singular constant to satisfy (\ref{appsg1}) and
(\ref{appsg2}). For $ \tau \ll (g_c - g)^{-z \nu}$, standard critical
phenomena lore requires that $C( \tau )$ become independent of $g_c -
g$. This only possible if $h_1 (x) \sim x^{-\beta/z\nu}$ for small $x$.
Putting this together with (\ref{appsg3}) we can determine the
behavior of $C(\tau)$ at $T=0$ and $g=g_c$
\begin{equation}
 C( \tau ) \sim \frac{1}{\tau^{\beta/(z\nu)}}~;~~
T=0,~~\tau \ll (g_c - g)^{-z\nu}.
\end{equation}
We can now use finite-size scaling to determine the behavior of
$C(\tau)$ at finite $T$
\begin{equation}
 C( \tau ) = \frac{1}{\tau^{\beta/(z\nu)}} h_2 (T \tau ) ~;~~
T^{-1},\tau \ll (g_c - g)^{-z\nu} , 
\end{equation}
where $h_2$ is yet another scaling function.
Finally we use (\ref{ctau}), 
to determine $\chi_L ( i \omega_n )$. In general we may find that
the fourier transform is dominated by non-universal contributions
at short times. However, upon analytically continuing to real
frequencies, we expect all these non-universal contributions to
affect only the real part of $\chi_L$, while the imaginary part is
dominated by the universal long-time behavior.
We therefore 
obtain the following generalization of 
(\ref{appchil}) to systems with spin-glass order
\begin{equation}
\mbox{Im} \chi_L ( \omega ) = |\omega |^{\mu}
F_1 \left( \frac{\omega}{T} \right)
~;~~~
T,\omega \gg (g_c - g)^{z\nu} ,
\label{appchilsg} 
\end{equation}
where
\begin{equation}
\mu = -1 + \frac{\beta}{z\nu} ,
\end{equation}
and $F_1$ is a universal scaling function with a single non-universal
overall scale. For small argument, we have $F_1 ( \overline{\omega} )
\sim \mbox{sgn} ( \omega ) |\omega |^{1-\mu}$, while 
$F_1 ( \overline{\omega} \rightarrow \infty )$ is finite and non-singular.

In addition to the local susceptibility, the thermodynamic properties
of the quantum phase transition which can be deduced from the 
hyperscaling properties of the free energy, are very similar in the
spin-glass and the N\'{e}el ordered systems. In particular, we expect
the following temperature dependence of $\chi_u^{{\rm st}}$ and $C_V$ in the
quantum-critical region ($T \gg (g_c - g)^{z\nu}$ 
of the spin-glass to spin-fluid transition:
\begin{eqnarray}
\chi_u^{{\rm st}} &\sim& T^{-1 + d/z} ,\nonumber\\
C_V &\sim& T^{d/z} .
\end{eqnarray}
It follows then that the Wilson ratio $W$ (Eqn. (\ref{Wilson}))
is a universal number at $g=g_c$.

\section{Berry phases and dangerously irrelevant couplings}
\label{appdiv}

We have assumed in this paper that the $O(3)$ sigma model is sufficient
to determine the quantum-critical scaling functions of quantum
antiferromagnets. A key step in this assumption 
is that the Berry phases present in the
antiferromagnet can be neglected. This assumption is based on the
following circumstantial evidence. After some rather involved
calculations, which have been discussed at length
elsewhere\cite{sun,murthy}, it was shown that the primary effect of
the Berry phases was to induce spin-Peierls ordering in the
quantum-disordered phase of non-even-integer-spin antiferromagnets.
This was shown in the context of large $M$ calculations for $SU(M)$
antiferromagnets. Further, a key feature of this calculation was the
appearance of two well-separated length scales which characterized
the fully-gapped spin-fluid phase. The first of these scales was
the two-spin correlation length $\xi$ which determined the
exponential decay of the equal-time spin-spin correlation function.
The second was $\xi_{SP}$ the length at which Berry phases first
became effective in inducing spin-Peierls ordering. In the large $M$
limit these two were found to be related by
\begin{equation}
\xi_{SP} \sim \xi^{4M\varrho_1} ,
\end{equation}
where $\varrho_1 = 0.062296 + {\cal O}(1/M)$. It is clear that
for sufficiently large $M$ (which is the only region in which we know
how to perform these calculations) we have $\xi_{SP} \gg \xi$.

It was then pointed out to us by Daniel Fisher\cite{daniel} that the appearance of
two length-scales at a second-order phase transition, one of which is
a power of the other, is a characteristic property of systems with
dangerously irrelevant coupings. A dangerously irrelevant coupling
is defined as one which is irrelevant at the critical fixed point
separating the two phases, but is relevant at the 
fixed point which
controls the nature of the phase one is studying. Crudely speaking,
the coupling decays to a very small
value at the first length scale while the system is controlled by
the critical fixed point, but grows again to a value of order unity
at the second, larger, length scale. 

For the antiferromagnet, our assumptions then, are the following.
We assume that the dangerous-irrelevancy of the Berry phase
effects, found in the $SU(M)$ models, survives in the
$O(3)$ sigma model.
The coupling to the Berry phase terms in the action
decays to a negligibly small value after renormalizing out to a scale
of order $\mbox{max} ( c/T, c/\Delta )$. At scales larger than this,
the coupling grows again, as the renormalization group flows are now in
the vicinity of the strong-coupling fixed point controlling the
quantum-disordered phase. However, this coupling
significantly modifies only those correlation functions which are
directly sensitive to the presence of spin-Peierls ordering. For all
other spin-correlations (which includes all we have considered in
this paper) the Berry phases can be neglected in
determining the leading quantum-critical behavior.

To clarify this issue, we now present a pedagogical
discussion of a simple statistical mechanical model with a
dangerously irrelevant coupling. Unlike the quantum antiferromagnet,
the overall structure of the renormalization group flows is
well-understood in this model.

We consider the finite temperature properties of a classical,
$XY$ model on a cubic lattice. 
At each site we introduce a four-fold anisotropy field
$h_4$, which we will find is dangerously irrelevant.
The model is described by the partition function $Z$ 
\begin{displaymath}
Z = \int {\cal D} \theta e^{S} ,
\end{displaymath}
\begin{equation}
S = \frac{1}{T} \sum_{<ij>} \cos ( \theta_i - \theta_j )
+ \sum_i h_4 \cos (4 \theta_i ) ,
\label{XYmodel}
\end{equation}
where the sites $i,j$ lie on a 3-dimensional cubic lattice.
This model will have a phase transition at some $T=T_c$ from a high
temperature paramagnetic phase to a low temperature ordered phase.
It is well known that the four-fold anisotropy $h_4$ is irrelevant
near $T=T_c$ and the phase transition is therefore in the
universality class of the pure three dimensional $XY$ model.
However, it is also clear that the field $h_4$ surely cannot be
neglected in the ordered phase. It breaks the $O(2)$ symmetry of the 
$XY$ model, and must therefore destroy the Goldstone modes. Further,
the common mean orientation at each site must be one of
$\theta = 0 , \pi/2 , \pi , 3\pi/2$ and cannot be arbitrary as in the
$XY$ model. The apparently conflicting properties of the critical
point and the ordered phase are reconciled by the concept of a
dangerously irrelevant coupling.

Let us examine the structure of the renormalization group flows of
this model for $T$ close to $T_c$: we measure the deviation from
criticality by the reduced temperature variable
\begin{equation}
t= \frac{T_c - T}{T}.
\end{equation}
A schematic of the renormalization group flows projected onto the 
$T, h_4$ plane are shown in Fig.~\ref{figberry}. 
The critical fixed point is at
$t=0$ and $h_4  = 0$. The initial growth of $t$ away from this fixed
point is given by
\begin{equation}
t(\ell ) = t e^{\ell/\nu},
\label{divt}
\end{equation}
where $0 < t \ll 1$, 
$e^{\ell}$ is the length rescaling factor, and $\nu$ is the
usual thermal critical exponent. The field $h_4$ is irrelevant at
this critical point and will therefore decay exponentially
\begin{equation}
h_4 ( \ell ) = h_4 e^{-\omega \ell / \nu} ,
\end{equation}
where $h_4$ is of order unity, 
$\omega$ is the crossover exponent associated with $h_4$ at the
critical fixed point. The system will emerge from the critical region
at the Josephson length scale $\xi_J$ where 
$t ( \ell = \ell_1^{\ast} ) \approx 1$.
{}From (\ref{divt}) we see that
\begin{equation}
\xi_J = e^{\ell_1^{\ast}} = t^{-\nu}.
\end{equation}
For $\ell > \ell_1^{\ast}$, the flow of $h_4$ will now be controlled
by the $T=0$ fixed point. Let us assume that $h_4$ is relevant at the
$T=0$ fixed point with eigenvalue $\phi > 0$ (the value of $\phi$
will be determined later). Then we have
\begin{eqnarray}
h_4 ( \ell > \ell_1^{\ast} ) &\approx& h_4 ( \ell_1^{\ast} ) e^{\phi
(\ell - \ell_1^{\ast})} \nonumber \\
&=& h_4 e^{- (\omega / \nu + \phi) \ell_1^{\ast} + \phi
\ell}.
\end{eqnarray}
Thus $h_4$ will return to a value of order unity when the argument of
the exponent is zero. This defines a second length 
scale $\xi_4 = e^{\ell_2^{\ast}}$
where
\begin{equation}
\ell_2^{\ast} = \left( 1 + \frac{\omega}{\phi \nu} \right)
\ell_1^{\ast} . 
\end{equation}
The effects of the $h_4$ field thus become important at length scales
of order $\xi_4$: this is the scale at which the Goldstone modes are
destroyed, and the condensate gets locked at one of $\theta = 0 , \pi
/2 , \pi , 3 \pi /2$. The scale $\xi_4$ is related to $\xi_J$ by
\begin{equation}
\xi_4 = \xi_J^{1 + \omega/(\phi \nu)} .
\label{divxi4}
\end{equation}

Additional insight can be gained by considering the scaling form for
the transverse susceptibility near the transition. Assume the
condensate points at $\theta = 0$. Then the transverse susceptibility
satisfies the scaling form
\begin{equation}
\left\langle | \theta (k) |^2 \right\rangle = \frac{N_0^2}{\rho_s}
\xi_J^2 \varphi ( k \xi_J , c_1 h_4 \xi_J^{-\omega/\nu} ),
\end{equation}
where $k$ is the wavevector, $\varphi$ is a universal function, 
and $c_1$ is the only non-universal metric factor. For most values of
$k \xi_J$ the term proportional to $h_4$ can be treated as a small
perturbation which makes a subdominant correction to the leading
critical behavior. Only at extremely small values of $k \xi_J$ does
the $h_4$ term become important. Matching to the expected form of the
incipient Goldstone modes, we should have
\begin{equation}
\varphi ( \overline{k} , \overline{h} ) = \frac{1}{\overline{k}^2 
+ \overline{h}}~;~~~~~~~~~~~\overline{k} \ll 1 .
\end{equation}
Thus the $\overline{h}$ term is significant for all $\overline{k}
< \overline{h}^{1/2}$ or for
\begin{equation}
k^{-1} > \xi_4 \sim  \xi_J^{1 + \omega/(2 \nu)} .
\end{equation}
Comparing with (\ref{divxi4}) we see that $\phi = 2$. 
The key point, of course, is that $\xi_4 \gg \xi_J$. In particular,
the crossover from critical to Goldstone fluctuations occurs 
at a scale of order $\xi_J$ and is described by the scaling
function of the pure $XY$ model $\varphi ( \overline{k} , 0)$. Only
at much larger scales does it become necessary to include the effects
of the $h_4$ field.

\section{Monte Carlo evaluation of quantum-critical uniform susceptibility}
\label{appmonte}

It was shown in Section~\ref{intro} that the high
temperature behavior of the uniform, static spin susceptibility is
given by (see (\ref{chistdef})
\begin{equation}
\chi_u^{{\rm st}} (T) = \left( \frac{g \mu_B }{\hbar c} \right)^2 k_B T ~
\Omega_{1} ( \infty ) .
\end{equation}
The universal number $\Omega_1 ( \infty )$ has been evaluated so far
in a $1/N$ expansion with the result (\ref{Omega1}). In this
appendix we will describe a determination of $\Omega_1 ( \infty )$ 
at $N=3$ by
Monte Carlo simulations.

The quantum $O(3)$ non-linear sigma model is expected to be in the
same universality class as the classical, Heisenberg ferromagnet on a
cubic lattice. Our simulations were therefore performed at the
critical point of this latter model. We considered the ensemble
defined by the following partition function
\begin{displaymath}
Z = \int \prod_i d {\bf S}_i e^{-{\cal H}} ,
\end{displaymath}
\begin{equation}
{\cal H} = -K \sum_{<ij>} {\bf S}_i \cdot {\bf S}_j ,
\label{appmontecarlo}
\end{equation}
where $i,j$ extend over the sites of a cubic lattice, and ${\bf S}_i
= (S_{x,i} , S_{y,i} , S_{z,i} )$ is a 3-component vector of unit
length. We used a lattice with $L \times L \times L_{\tau}$ sites,
with periodic boundary conditions in all three directions. The Wolff
single-cluster algorithm\cite{wolffalgo} was used to sample the states. This
simulation was carried out at the critical value $K= K_c$ at
which this model has a second-order phase transition. The value of
$K_c$ is known very accurately from recent high precision Monte Carlo
simulations\cite{holm}:
\begin{equation}
K_c = 0.6930 .
\end{equation}

It has been argued that $\chi_u^{{\rm st}}$ is related to the
stiffness, $\rho_{\tau}$, of this system to twists along the $\tau$
direction. Upon examining the response of ${\cal H}$ to a field that
generates rotations in the $x-y$ plane, we obtain the following
expression for $\rho_{\tau}$
\begin{eqnarray}
\rho_{\tau} = && \frac{1}{L^2 L_{\tau}} \left\langle
\sum_{i} K \left(S_{x,i} S_{x,i+\hat{\tau}} + S_{y,i}
S_{y, i + \hat{\tau}} \right) \right.\nonumber \\
&& \left. - \left[ \sum_i K \left(
S_{x,i} S_{y,i+\hat{\tau}} - S_{y,i}
S_{x, i + \hat{\tau}} \right) \right]^2 \right \rangle ,
\label{apprhodef}
\end{eqnarray}
where the expectation value is to be evaluated in the ensemble
defined by $Z$. Finally, the universal number $\Omega_1 ( \infty )$
is defined by
\begin{equation}
\Omega_1 ( \infty ) = \lim_{L_{\tau} \rightarrow \infty} \left[
\left( \lim_{L \rightarrow \infty} \left. 
L_{\tau} \rho_{\tau} \right|_{K=K_c} \right)
\right] .
\end{equation}
It is crucial that the $L \rightarrow \infty$ limit be taken first,
to model a quantum system which is infinite in the
spatial directions. The subsequent $L_{\tau} \rightarrow \infty$
places the quantum system at zero temperature.

The results of our simulations are contained in Table~\ref{tabmontecarlo}.
Three independent simulations of 70000, 70000, and 210000, flips per
spins were performed in systems upto $L=30$ and $L_{\tau} = 10$.
A polynomial extrapolation to $L=\infty$ at $L_{\tau}$ fixed yielded
the results shown in the last column of Table~\ref{tabmontecarlo}. Finally, a second
polynomial extrapolation to $L_{\tau} = \infty$ was performed to
yield the following estimate for $\Omega_1 ( \infty )$
\begin{equation}
\Omega_1 ( \infty ) = 0.25 \pm 0.04 .
\end{equation}

\section{Computations in the N\'{e}el state at $T=0$}
\label{appent=0}

In this Appendix, we derive the expressions for the 
$T=0$ sublattice magnetization 
$N_{0}$, and the spin-stiffness $\rho_{s}$ 
 to order $1/N$. These results will be used in the derivations
of the universal scaling functions for the uniform and staggered susceptibilities
in both the quantum-critical and renormalized classical regions.
As in Sec \ref{NLsmodel}, our starting point is
 the functional integral for the $O(N)$ sigma model.
At  zero temperature, the spin rotation symmetry is broken and the
perturbative $1/N$ expansion has to be modified to account
for the nonzero expectation value of the order parameter. This $1/N$ 
expansion has been
developed by Brezin and Zinn-Justin~\cite{Brezin-Zinn} and we use
some of their results. We first represent the unit vector field as
\begin{equation}
\vec{n} = \vec{\sigma}_{0} + \vec{\pi},
\label{B0}
\end{equation}
 where $ \langle\vec{\sigma}_{0}\rangle$ is finite and $ \vec{\sigma}_{0}
\cdot \vec{\pi} = 0$. The sublattice magnetization $N_{0}$ 
in spin-S antiferromagnet is expressed as 
$N_{0} = \tilde{S}\langle \sigma_{0} \rangle $, where $\tilde{S} =
S Z_S$. The renormalization factor $Z_{S}$ 
accounts for the order parameter fluctuations  at
short (lattice) scales which have to be integrated out in the mapping to
sigma model 
from the original spin Hamiltonian on the lattice.
 Upon substituting (\ref{B0}) into (\ref{nlsz}),
 the functional integral becomes
\widetext
\begin{displaymath}
Z   =  \int {\cal D} \sigma_{0} \, {\cal D} \pi_{l} \,\, 
\delta({\sigma}^{2}_{0}  +  \pi^{2}_{l} -
1)~~~~~~~~~~~~~~~~~~~~~~~~~~~~~~~~~~~~~~~
\end{displaymath}
\begin{equation}
~~~~~~\exp \left( - \frac{\rho^{0}_{s}}{2\hbar} \int d^2 r \int^{\infty}_{0}
d \tau \left [ (\nabla_{r} \sigma_{0})^{2}  + (\nabla_{r} 
\pi_{l})^{2} + \frac{1}{c^{2}_{0}}\left(
(\partial_{\tau} \sigma_{0})^2 + (\partial_{\tau} \pi_{l})^2 \right)
\right] \right),
\label{B1}
\end{equation}
\narrowtext
where the index $l$ now runs from 1 to $N-1$, $\rho^{0}_{s}$ is the bare
spin-stiffness and $c_{0}$ is the bare spin-wave velocity.
All of the  discussion in this Appendix will use the 
relativistic cutoff scheme 
when the momenta and frequency satisfy $k^2 + \omega^2 < \Lambda^2$. In 
this situation,
 the full relativistic invariance is preserved at each
order in the perturbation theory and we will not have to consider
explicitly the renormalization of the spin-wave velocity.
To simplify the presentation, 
 below we use the units where $\hbar  = c_{0} = 1$.
As in Sec \ref{NLsmodel},
 we introduce a Lagrange multiplier $\lambda$ into the functional 
integral to impose the constraint, and integrate over $\pi_{l}$. This gives
\widetext
\begin{equation}
Z = \int {\cal D} \vec{\sigma}_{0} {\cal D}
 \lambda \exp \left( - \frac{\rho^{0}_{s}}{2}
 \int d^2 r \int^{\infty}_{0}
d \tau \left [ (\partial_{\mu} \sigma_{0})^{2} + \lambda ( \sigma^{2}_{0} - 1)
\right ] - \frac{N-1}{2} \log || - \partial_{\mu}^{2} + \lambda || \right ).
\label{B2}
\end{equation}
\narrowtext
Here $\partial_{\mu}^{2} = (\nabla_{r})^2 + (\partial_{\tau})^2$.
The saddle-point point equation is easily obtained by taking a variational
derivative  over
$\lambda$ and neglecting fluctuations in $\sigma_{0}$. This yields
\begin{equation}
\frac{(N-1)}{N} \,  g \int \frac{d^2 k d\omega}{(2\pi)^3}
 G_{0}(\vec{k}, i\omega) 
 = 1 - \langle \sigma_{0} \rangle^2 ,  
\label{B3}
\end{equation}
Here $g = N/\rho^{0}_{s}$ is the coupling constant and
$G_{0}(\vec{k}, i\omega)  = (\vec{k}^2 +
\omega^{2})^{-1}$ is the zero-temperature 
propagator of the $\pi_{l}$ field at $N=\infty$.
The $1/N$ corrections to (\ref{B3}) are calculated as described
earlier\cite{Polyakov}, with the modification that we have to consider
the fluctuations of $\sigma_{0}$ around its mean value. We 
find the $T=0$ analog to the
polarization operator 
\begin{equation} 
\Pi^{*}({k},i\omega) = \Pi({k},i\omega) \, + 
\frac{2}{g} \, \langle\sigma_{0}\rangle^2 \, G_{0}({k},i\omega) ,    
\label{B4}
\end{equation}                                    
and the correlator of the $\sigma_{0}$ field:
\begin{eqnarray}
\langle(\sigma_{0}(k,i\omega))^{2}\rangle\, && = 
\langle\sigma_{0} \rangle^2 \delta(\omega) \delta^2 (k) \nonumber\\
+ &&
\frac{G_{0}({k}, i\omega)}{\rho_s^0}  \left(1 -
\frac{2}{g} \, \langle\sigma_{0}\rangle^2 \,
\frac{G_{0}({k},i\omega)}{\Pi^{*}({k},i\omega)} \right). 
\label{B5}
\end{eqnarray}                                        
The condition  $\sigma^{2}_{0} + \pi^{2}_{l} =1$,  then yields
\begin{eqnarray}
1 - \langle\sigma_{0}\rangle^2   = && 
 g \int \frac{d^2 k d\omega}{(2\pi)^3} \,
G({k}, i\omega)  \,\nonumber\\
&&~~ - \frac{2 \langle\sigma_{0}\rangle^2}{N} \,  \int
 \frac{d^2 k d\omega}{(2\pi)^3} \, \frac{G^{2}_{0}({k},
i\omega)}{\Pi^{*}({k},i\omega)}.   
\label{B6}
\end{eqnarray}               
Here $G$ is related to $G_{0}$ in the usual way:
\begin{equation}
 G^{-1}({k},i\omega) = G^{-1}_{0}({k},i\omega) +
 \Sigma({k},i\omega) ,
\end{equation}
\begin{equation}
\Sigma({k},i\omega) = \frac{2}{N}\,
 \int \frac{d^2 q d\epsilon}{(2\pi)^3} ~
\frac{G_{0}({q},i\epsilon)}{\Pi^{*}(\vec{q},i\epsilon)}.
\end{equation}
Below, we will express $N_{0}$ directly in terms of the fully renormalized
spin-stiffness.
 It is instructive, however, to compute the critical exponent for
$N_{0}$ directly from (\ref{B6}). For this we observe that the first
term in the r.h.s. in (\ref{B6}) is simply a constant so that  with the accuracy
 to $1/N$,
\begin{equation}
\langle\sigma_{0}\rangle^2  = \left (\frac{g_{c} - g}{g_{c}} \right)\, \left(1 + 
 \frac{2}{N} \,  \int \frac{d^2 k ~d\omega}{(2\pi)^3} 
 \, \frac{G^{2}_{0}(\vec{k},
i\omega)}{\Pi^{*}(\vec{k},i\omega)}\right),   
\label{B7}
\end{equation}                                            
where $g_{c}$ is the non-universal critical coupling. 
The value of the integral depends on the precise form of the
polarization operator at $k, \omega \sim \Lambda$, which can be very
complicated. However, for the logarithmic contribution in (\ref{B7}),
 we only need to know the form of $\Pi^{*}(\vec{k},i\omega)$ for momentum and 
frequency well below the upper cutoff. For such $k$ and $\omega$ the 
evaluation of $\Pi (k, i\omega)$ at $T=0$  is straightforward and we obtain
\begin{equation}
\Pi(\vec{k},i\omega) = \frac{1}{8 \sqrt{k^2 + \omega^2}} .
\label{BB}
\end{equation}
We then use (\ref{B4}) for $\Pi^{*}$ and evaluate the integral in
(\ref{B7}) with the logarithmic accuracy. We obtain 
\begin{equation}
\langle\sigma_{0}\rangle^2  = \left (\frac{g_{c} - g}{g_{c}} \right)\, \left(1 + 
\frac{8}{N \pi^2} \, \log \frac{g_{c}}{g_{c} - g} \right) .
\label{B8}
\end{equation}
The correction can be exponentiated in the usual way and we get
\begin{equation}
\langle\sigma_{0}\rangle^2 = \left(\frac{g_{c} - g}{g_{c}}\right)^{2\beta} .
\label{B9}
\end{equation}
where 
\begin{equation}
2\beta = 1 -  \frac{8}{N \pi^2} .
\label{B10}
\end{equation}

Our next goal is to express $N_{0}$  in terms of the fully
renormalized spin stiffness at $T=0$. At $N=\infty$, we have from
(\ref{rhosNinfty})
\begin{equation}
\rho_s =  N \left( \frac{1}{g} - \frac{1}{g_c} \right) .
\label{BB1}
\end{equation}
Now we have to
compute $1/N$ corrections to (\ref{BB1}).  In principle, it is possible
to evaluate $\rho_{s}$ directly by calculating the response to the twist of the
order parameter in the momentum space. In practice, however, it is more
convenient to calculate the static susceptibility $\chi_{\perp}$, which measures
the response to the twist in $\tau$-direction; the value of 
 $\rho_{s}$ then can be obtained using Lorentz invariance: $\rho_{s}
 \equiv c^{2}_{0} \cdot \chi_{\perp}$.

The calculations of $\chi_{\perp}$ to order $1/N$ were described in Sec
\ref{NLsmodel1N}.
As before,  they
have to be modified to account for nonzero 
order parameter. For $N=\infty$,  we  use 
 (\ref{B0}) for the $n-$field,  substitute it into the bubble diagram 
in Fig.~\ref{figfeyn},
and after simple manipulations obtain at $T=0$,  
\begin{equation}
\chi^{N=\infty}_{\perp} = \frac{\rho^{0}_{s}}{c^{2}_{0}} \, 
\langle\sigma_{0}\rangle^2 . 
\label{B9a}
\end{equation}
The calculation of $1/N$ terms proceeds along the same lines
as in Sec \ref{NLsmodel1N}.
 We substitute (\ref{B0}),  into the $1/N$ diagrams in Fig.~\ref{figfeyn}
  and using (\ref{B5}) obtain after 
some algebra,
\begin{equation}
\chi_{\perp} = \chi^{N=\infty}_{\perp}\, \left (1 - \frac{2}{N} \, 
I_{T=0}\right ) ,              
\label{B10a}           
\end{equation}
where
\begin{equation}
I_{T=0} =  \int \frac{d^2 k d\omega}{(2\pi)^3}  \,
 \frac{G^{2}_{0}({k},
i\omega)}{\Pi^{*}({k},i\omega)} \, \frac{(k^2 - 3 \omega^2)}{(k^2 + 
\omega^2)}.                       
\label{B11}
\end{equation}
We emphasize that the calculations 
at $T=0$ are much simpler than 
that at finite $T$ because in fact we have to keep only the terms $\sim 
~\langle\sigma_{0}\rangle$; 
all other contributions give zero after {\it{integration}} 
over intermediate frequency. This indeed  is clearly seen from the expression 
for $\chi_{\perp}$ at finite $T$ (Eqn (\ref{chist1n1}))
 where each term  contains derivatives of the Bose functions.

Finally, from  (\ref{B9}) and
(\ref{B10}), and the definition of $N_0$, we obtain:
\begin{equation}
\frac{N_{0}^2}{\rho_{s}} =  \frac{g \tilde{S}^2}{N}\, \left (1 - \frac{2}{N} \, 
I_{T=0}\right ) .              
\label{B12}
\end{equation}

Clearly, the value of the integral in (\ref{B11}) depends on the form of
$\Pi^{*}({k},i\omega)$ near the upper cutoff and the result for 
$N_{0}^2 /\rho_{s}$ is therefore model dependent. 
However, we explicitly showed in Sec~\ref{secqc}
that the universal functions for 
observables are insensitive to the behavior of $\Pi^{*}$ at $k, \omega
\sim \Lambda$. In view of this, we use in Sec \ref{secqc} and Sec \ref{secrc}
the result for $N_{0}^2 /\rho_{s}$ obtained from (\ref{B12}) in the case when
the polarization operator is computed without a cutoff in the momentum
and frequency integration. Then $\Pi (k, \omega)$ is given by
(\ref{BB}) and performing the integration in (\ref{B11}), we find
\begin{equation}
\frac{N_0^2}{\rho_s} = \frac{g \tilde{S}^2}{N}
 \left[ 1 - \frac{8}{3\pi^2 N}~ \log \left(
\frac{N \Lambda}{16 \rho_s} \right) \right] .
\label{BB2}
\end{equation}                    
This equation we use in   (\ref{n02rs}).

We will also need the result for $\rho_s$ expressed in terms of the
spin-stiffness at $N=\infty$. From (\ref{B10}) (\ref{B7}) and (\ref{BB1}), we
obtain
\begin{equation}
\rho_s = \rho^{N=\infty}_s ~ \left(1 + 
 \frac{2}{N} \,  \int \frac{d^2 k ~d\omega}{(2\pi)^3} 
 \, \frac{G^{2}_{0}(\vec{k},
i\omega)}{\Pi^{*}(\vec{k},i\omega)}~\frac{4 \omega^2}{(k^2 + 
\omega^2)} \right).                         
\label{BB3}
\end{equation}
					    
Finally,  we  deduce from (\ref{BB3}) the critical exponent for $\rho_s$.
For this, we  perform the integration in   (\ref{BB3}) 
with the logarithmic accuracy using  (\ref{BB}) and (\ref{B4}), and
exponentiate the result. We then obtain
\begin{equation}
\rho_{s} \sim (\frac{g_{c} - g}{g_{c}})^{\nu} ,
\label{B14}
\end{equation}
where
\begin{equation}
\nu = \frac{2 \beta}{1 + \eta} = 1 - \frac{32}{3\pi^2 N} ,
\label{B15}
\end{equation}
and
 $\eta = {8}/({3 \pi^2 N})$ is the critical exponent for
spin correlations at $g_{c}$. 

\narrowtext

\begin{figure}
\caption{Phase diagram of ${\cal H}$ (Eqn. (\protect\ref{calH})) as a
function of $g$ and temperature $T$ (after Ref.~\protect\cite{CHN}). 
The coupling $g$ measures the
strength of the quantum fluctuations.  
 It is inversely proportional to $S$ for large spin and  its value also
depends on the  ratios of the $J_{ij}$.
The parameters $x_1 = N k_B T / (2 \pi \rho_s)$, and $x_2 = k_B T /
\Delta$ control the scaling properties of the antiferromagnet (here
$\rho_s$ is the spin-stiffness of the N\'{e}el-ordered ground state,
and $\Delta$ is the spin 1 gap in the quantum-disordered ground state).
}
\label{phasediag}
\end{figure}
\begin{figure}
\caption{Properties of the nearly-critical antiferromagnet as a
function of the observation wavevector $k$, or frequency $\omega$ in
the three different regions of Fig.~\protect\ref{phasediag}. The
appropriate regime is determined by the larger of $\hbar c k / (k_B
T)$ or $\hbar \omega / k_B T$. In the renormalized-classical regime,
$\xi$ is the actual correlation length, while $\xi_{J}$ is a Josephson
correlation length related to the spin-stiffness by $\hbar c/ \xi_{J} =
\rho_s /\Upsilon$, with $\Upsilon$ a universal number. In the quantum
disordered region, $\Delta$ is the gap for $S=1$ excitations  at $T=0$.
The thermodynamic behavior in the various regions is discussed in the text.} 
\label{komdep}
\end{figure}
\begin{figure}
\caption{The scaling function $\mbox{Im} \Phi_{1s} ( \overline{k},
\overline{\omega}, \infty )$ for the staggered susceptibility in the 
quantum-critical region. The results have been computed in a $1/N$
expansion to order $1/N$ and evaluated for $N=3$. The shoulder on the
peaks is due to a threshold to three spin-wave decay.}
\label{figphi1s}
\end{figure}
\begin{figure}
\caption{Scaling function $\Xi_1 ( \overline{k} , \infty )$ for the 
structure factor in the quantum-critical region.}
\label{figsfac}
\end{figure}
\begin{figure}
\caption{Scaling function $F_1 ( \overline{\omega} , \infty)$ for the
imaginary part of the local susceptibility in the quantum-critical
region. The oscillations at large $\overline{\omega}$ are due to a
finite step-size in the momentum integration.}
\label{figchil}
\end{figure}
\begin{figure}
\caption{Feynman graph for the staggered and uniform 
susceptibilities to order $1/N$. The solid line represents the
$n_{\ell}$ field and the dashed line represents the propagator of the
$\lambda$ field (polarization operator),
 where $\lambda$ is the  Lagrange multiplier imposing the
constraint.} 
\label{figfeyn}
\end{figure}
\begin{figure}
\caption{Quantum Monte-Carlo \protect\cite{Makivic} (squares) 
and our theoretical (line)
results
for the uniform susceptibility $\overline{\chi}_u = (3 J (a\hbar/g \mu_B)^2)
\chi_u^{{\rm st}}$ of a square lattice 
spin-1/2 Heisenberg antiferromagnet ($a$ is the 
lattice spacing). The experimental results for weakly doped $La_{2}CuO_4$ are
very close to the Monte-Carlo data~\protect\cite{Johnson}.
There are {\em no} adjustable parameters in the theoretical 
result (\protect\ref{chistdef}). 
Over the range of $T$ plotted, the function 
$\Omega_1 (x_1)$ (recall $x_1 = N k_B T / ( 2 \pi \rho_s)$ )
is very close
to its large $x_1$ behavior given in Eqn. 
(\protect\ref{Omega1subleading},\protect\ref{Omega1}).
We used the theoretical result at $N=3$.
The theoretical and experimental slopes 
agree remarkably well. The good agreement
in the intercept is somewhat surprising as
 its theoretical value is known only at $N=\infty$.}    
\label{figunisus}
\end{figure}
\begin{figure}
\caption{Theoretical scaling function at $N=\infty$, Eqn. 
(\protect\ref{E6}) (line) and Quantum Monte-Carlo data 
\protect\cite{Makivic} (squares) 
for the equal-time structure factor in the $S=1/2$ Heisenberg antiferromagnet
at $k_B T = 0.42 J$. The correlation length is taken from the Monte-Carlo data
($\xi \sim 5.7 a$). We expect that for most of the values of
$k\xi$ plotted, the
antiferromagnet is in the quantum-critical regime where $1/N$ corrections
to (\protect\ref{E6}) are small. 
The only adjustable parameter in the theoretical
curve is the $k-$independent overall factor $\bar{S}$ in
(\protect\ref{E6}). The best fit value of $\bar{S}$ was found to be
$\sim
1.45$ times larger than our $N=\infty$ result.}
\label{figsk}
\end{figure}
\begin{figure}
\caption{Schematic of the renormalization group flows of the action
$S$ (Eqn. (\protect\ref{XYmodel})) of a classical $XY$ model on a
cubic lattice. The coupling $h_4$ is a cubic anisotropy perturbation
which is irrelevant at the critical fixed point $T=T_c$, $h_4 = 0$.
Its neglect is however dangerous in the low temperature phase because
$h_4$ is relevant at the $T=0$, $h_4 = 0$ fixed point.}
\label{figberry}
\end{figure}

\begin{table}
\caption{Results of the Monte Carlo simulation 
of the classical statistical-mechanics
model (\protect\ref{appmontecarlo}) at $K=K_c = 0.6930$. We used a box of
$L \times L \times L_{\tau}$ sites with periodic boundary conditions. 
The stiffness $\rho_{\tau}$ is defined in (\protect\ref{apprhodef}). The last column
is obtained by a polynomial extrapolation in inverse powers of $1/L$.
The three runs had 70000, 70000, 210000 flips per spin respectively.
A weighted average of the three runs was used in the extrapolation.}
\label{tabmontecarlo}
\begin{tabular}{dddddd}
$L_{\tau}$ & $L$ & \multicolumn{3}{c}{$L_{\tau} \rho_{\tau}$} &
$\lim_{L \rightarrow \infty} L_{\tau} 
\rho_{\tau}$ \\
 & & Run 1 & Run 2 & Run 3 & \\
\tableline
\tableline
   & 10  & 0.3983 & 0.3866 & 0.3898 & \\
   & 15  & 0.3693 & 0.3701 & 0.3663 & \\
5  & 20  & 0.3596 & 0.3515 & 0.3527 & 0.3257\\
   & 25  & 0.3511 & 0.3574 & 0.3478 & \\
   & 30  & 0.3483 & 0.3529 & 0.3511 & \\
\tableline
   & 10  & 0.4138 & 0.4087 & 0.4099 & \\
   & 15  & 0.3718 & 0.3766 & 0.3747 & \\
7  & 20  & 0.3650 & 0.3625 & 0.3506 & 0.3037\\
   & 25  & 0.3529 & 0.3419 & 0.3442 & \\
   & 30  & 0.3424 & 0.3405 & 0.3402 & \\
\tableline
   & 15  & 0.4079 & 0.4049 & 0.4020 & \\
10 & 20  & 0.3861 & 0.3743 & 0.3713 & 0.2890\\
   & 25  & 0.3650 & 0.3671 & 0.3511 & \\
   & 30  & 0.3548 & 0.3521 & 0.3425 &   
\end{tabular}
\end{table}

\begin{references}                                                 
\bibitem{theory}  S. Chakravarty in {\it{High-Temperature 
Superconductivity}}, eds. K. Bedell, D. Coffey, D.E. Meltzer, D. 
Pines and J.R. Schrieffer, Addison-Wesley, p.136 (1990).

\bibitem{numerics} E.Manousakis, Rev. Mod. Phys, {\bf{63}}, 1 (1991).

\bibitem{trieste} S. Sachdev, in {\em Low Dimensional Quantum Field Theories
for Condensed Matter Physicists\/}, Proceedings of the Trieste Summer School
1992, World Scientific, to be published, and references therein.
Available as paper 9303014  on cond-mat@babbage.sissa.it.

\bibitem{ramirez} A.P. Ramirez, G.P. Espinosa, and A.S. Cooper, Phys. Rev. Lett.
{\bf 64}, 2070 (1990).
		     
\bibitem{broholm} C. Broholm, G. Aeppli, G.P. Espinosa, and A.S.
Cooper, Phys. Rev. Lett. { \bf 65}, 3173
(1991).

\bibitem{otherafm} G. Aeppli, C. Broholm, and A. Ramirez in Proceedings of the
Kagom\'{e} Workshop, NEC Research Institute, Princeton, NJ (unpublished).

\bibitem{elser} V. Elser, Phys. Rev. Lett., {\bf 62}, 2405 (1990).

\bibitem{clarke} S.J. Clarke, A. Harrison, T.E. Mason, G.J. McIntyre and D.
Visser, J. Phys. Cond. Matter, {\bf 4}, L71 (1992).

\bibitem{harrison} For a review on recent experiments see e.g. A. Harrison,
in "Annual Reports on the Progress of Chemistry (Royal Soc. of Chemistry, UK),
{\bf 87A}, 211, (1992); {\bf 88A}, 447, (1992).

\bibitem{Pokr} V.L. Pokrovskii, Adv. Phys. {\bf 28}, 595 (1979).

\bibitem{Andreev} A.F. Andreev and V. I. Marchenko, Sov. Phys. Usp. {\bf 23},
21, (1980). 

\bibitem{CHN} S. Chakravarty, B.I. Halperin and D.R. Nelson, Phys. Rev. Lett.
{\bf 60}, 1057 (1988); 
Phys. Rev. B {\bf{39}}, 2344 (1989).

\bibitem{Tyc} S. Tyc, B.I. Halperin, S. Chakravarty, Phys. Rev. Lett.
{\bf 62}, 835 (1989).

\bibitem{Kalm_Laugh} V. Kalmeyer and R.B. Laughlin, Phys. Rev. Lett., {\bf 59},
2095 (1987); X.G. Wen, F. Wilczek and A. Zee, Phys. REv. B {\bf 39}, 11413
(1989). 

\bibitem{sun} N. Read and S. Sachdev, Phys. Rev. Lett. {\bf 62}, 1694 (1989);
\prb {\bf 42}, 4568 (1990).

\bibitem{spn} N. Read and S. Sachdev Phys. Rev. Lett. {\bf 66}, 1773 (1991); 
S. Sachdev
and N. Read, Int. J. Mod. Phys. {\bf B5}, 219 (1991).

\bibitem{CCL} P.Chandra, P. Coleman and A.I. Larkin, J. Phys. Cond. Matter {\bf
2}, 7933 (1990).

\bibitem{Affl_Marst} I. Affleck and J. Brad Marston, Phys. Rev. B {\bf 37},
3774, (1988).

\bibitem{Wieg-Khv} P.B. Wiegmann, Phys. Rev. Lett. {\bf 60}, 821, (1988); D.V.
Khveshchenko and P.B. Wiegmann, Mod. Phys. Lett. B {\bf 4}, 17 (1990). 

\bibitem{Anders} P.W. Anderson,  Science {\bf 235}, 1196 (1987).

\bibitem{jinwu} S. Sachdev and Jinwu Ye, Phys. Rev. Lett. {\bf 69}, 2411 (1992).

\bibitem{andrey} A.V. Chubukov and S. Sachdev, Phys. Rev. Lett. 
{\bf 71}, 169 (1993); erratum, to be published.

\bibitem{tsvelik} B. Andraka and A.M. Tsvelik, Phys. Rev. Lett.
{\bf 67}, 2886 (1991).

\bibitem{millis} A.J. Millis, preprint; 
J.A. Hertz, Phys. Rev. B {\bf 14}, 525 (1976).

\bibitem{hasen1} P. Hasenfratz and F. Niedermayer, Phys. Lett. B {\bf 268},
231 (1991); University of Bern preprint BUTP-92/46.

\bibitem{hasen2}  P. Hasenfratz, M. Maggiore and F. Niedermayer,
 Phys. Lett. B {\bf 245}, 
522 (1990); P. Hasenfratz and F. Niedermayer, Phys. Lett. B {\bf 245},
529 (1990).

\bibitem{Chak-Orbach} S. Chakravarty and R. Orbach, 
Phys. Rev. Lett {\bf{64}}, 224
(1990).

\bibitem{Yamada} Y. Endoh {\em et. al.\/}, Phys. Rev. B {\bf{37}}, 
7443 (1988); K. Yamada {\em et. al.\/}, Phys. Rev. B {\bf{40}}, 4557 (1989).

\bibitem{gabe} S.M. Hayden, G. Aeppli, H. Mook, D. Rytz, M.F.
Hundley, and Z. Fisk, Phys. Rev. Lett. {\bf 66}, 821 (1991);
S.M. Hayden, G. Aeppli, R. Osborn, A.D. Taylor, T.G.
Perring, S.-W. Cheong, and Z. Fisk, \prl, {\bf 67}, 3622 (1991).

\bibitem {Slichter} T. Imai, C.P. Slichter, K. Yoshimura and K. Kosuge,
Phys. Rev. Lett {\bf 70}, 10002, (1993).

\bibitem{Johnson} D.C. Johnson, Phys. Rev. Lett. {\bf{62}}, 957 (1989).

\bibitem{Singh-Gelfand} R.R.P. Singh and M. Gelfand, Phys. 
Rev. B {\bf{42}}, 966 (1990).

\bibitem{Makivic} H.Q. Ding and M. Makivic, Phys. Rev. Lett, 
{\bf{64}}, 1449 (1990); Phys. Rev. B {\bf{43}}, 3662 (1990).
		  
\bibitem{Sokol} A. Sokol, S. Bacci and E. Gagliano, Phys. Rev. B {\bf 47}, 14646
(1993).

\bibitem{birg} B. Keimer, R.J. Birgeneau, A. Cassanho, Y. Endoh,
R.W. Erwin, M.A. Kastner, and G. Shirane, Phys. Rev. Lett. {\bf 67}, 1930 (1991).

\bibitem{keimer2} B. Keimer, N. Belk, R.J. Birgeneau, A. Cassanho,
C.Y. Chen, M. Greven, M.A. Kastner, A. Aharony, Y. Endoh, R.W. Erwin,
and G. Shirane, Phys. Rev. B {\bf 46}, 14034 (1992).

\bibitem{varma} C.M. Varma P.B. Littlewood, S. Schmitt-Rink,
E. Abrahams, and A.E. Ruckenstein, Phys. Rev. Lett. {\bf 63}, 1996 (1989). 

\bibitem{Ioffe_Lar} L.B. Ioffe and A.I. Larkin, Int. J. Mod. Phys. B {\bf 2},
203 (1988).

\bibitem{Chub1} A.V. Chubukov, Phys. Rev. B {\bf 44}, 392 (1991). 

\bibitem{halpsas} B.I. Halperin and W.M. Saslow, Phys. Rev. B {\bf 16},
2154 (1977).

\bibitem{dombre} T. Dombre and N. Read, Phys. Rev. B {\bf 39}, 6797
(1989).

\bibitem{delamotte} P. Azaria, B. Delamotte, and T. Jolicoeur,
Phys. Rev. Lett. {\bf 64}, 3175 (1990).

\bibitem{kagtotal} A.B. Harris, C. Kallin and A.J. Berlinsky, Phys. Rev B {\bf
45}, 2889 (1992); J.T. Chalker, P.S. Holdsworth and E.F. Shender, Phys. Rev.
Lett. {\bf 68}, 855 (1992); P. Chandra, P. Coleman and I. Ritchey, Phys. Rev. B,to appear;  A.V. Chubukov, Phys. Rev. Lett. {\bf 69}, 832
(1992); J. von Delft and C.L. Henley, Phys. Rev. Lett. {\bf 69}, 3236 (1992); 
J.N. Reimers, A.J. Berlinsky and A.-C Shi, Phys. Rev. B {\bf 43}, 865 (1991);
R.R.P. Singh and D. Huse, Phys. Rev. Lett. {\bf 68}, 1706 (1992); D.L. Huber
and W.Y. Cheong, Phys. Rev. B {\bf 47}, 3220 (1993).  

\bibitem{kagome} S. Sachdev, Phys. Rev. B {\bf 45}, 12377 (1992).

\bibitem{Ma} S-k. Ma, {\em Modern Theory of Critical
Phenomena}, Benjamin/Cummings, Reading (1976). 

\bibitem{Park}  J.B. Parkinson, J. Phys. C {\bf 2}, 2012 (1969); 
B.S. Shastry and B. Shraiman, Phys. Rev. Lett. {\bf 65}, 1068 (1990);
R.R.P. Singh, Comments Cond. Matt. Phys. {\bf 15}, 241 (1991).

\bibitem{helicity} M.E. Fisher, M.N. Barber, and D. Jasnow, 
Phys. Rev. A {\bf 8}, 1111 (1973).

\bibitem{josephson} B.D. Josephson, Phys. Lett. {\bf 21}, 608 (1966).

\bibitem{eduardo} A.H. Castro Neto and E. Fradkin, paper 9301009 on
cond-mat@babbage.sissa.it

\bibitem{abe} A.N. Vasil'ev, Yu.M. Pis'mak and Yu.R. Honkonen,
Teor. Mat. Fiz. {\bf 46}, 157 (1981).

\bibitem{holm} C. Holm and W. Janke, preprint; P. Peczak, A.M. Ferrenberg,
and D.P. Landau, Phys. Rev. B {\bf 43}, 6087 (1991).

\bibitem{erratum} In an earlier version of the manuscript we had an incorrect
conjecture of the form of the subleading term in the large $x_{1,2}$ expansion.
S. Chakravarty pointed out to us, and we determined by explicit computation
the present correct form.

\bibitem{stauffer} D. Stauffer, M. Ferer, and 
M. Wortis, Phys. Rev. Lett. {\bf 29}, 345 (1972).

\bibitem{two-scale} P.C. Hohenberg, A. Aharony, B.I. Halperin,
and E.D. Siggia, Phys. Rev. B {\bf 13}, 2986 (1976);
C. Bervillier, {\em ibid.} {\bf 14}, 4964 (1976).

\bibitem{matt} M.P.A. Fisher, G. Grinstein, and S.M. Girvin, Phys. Rev. Lett.
{\bf 64}, 587 (1990); K. Kim and P.B. Weichmann, Phys. Rev. B {\bf 43},
13583 (1991); 
M.-C. Cha {\em et.al.\/}, Phys. Rev. B {\bf 44}, 
6883 (1991). 

\bibitem{binder} M.N. Barber in {\em Phase Transitions and Critical Phenomena},
ed. C. Domb and J. Lebowitz, v. 8, pg. 145, Acad. Press (New York).

\bibitem{privman} V. Privman and M.E. Fisher, Phys. Rev. B {\bf 30},
322 (1984).

\bibitem{Fisher} D.S. Fisher, Phys. Rev. B {\bf 39}, 11783 (1989).

\bibitem{degennes} M.E. Fisher and 
P.-G. de Gennes, C.R. Acad. Sci. Ser. B {\bf 287},
207 (1978).

\bibitem{zama} A.B. Zamalodchikov, Pis`ma Zh. Eksp. Teor. Fiz {\bf 43},
565 (1986); [JETP Lett. {\bf 43}, 730 (1986).

\bibitem{Blote} H.W.J. Blote, J.L. Cardy, and M.P. Nightingale,
Phys. Rev. Lett. {\bf 56}, 742 (1986); I. Affleck Phys. Rev. Lett.
{\bf 56}, 746 (1986).

\bibitem{AA} D.P. Arovas and D. Auerbach, Phys. Rev. B
{\bf 38}, 316 (1988); Phys. Rev.
Lett. {\bf 61}, 617 (1988).

\bibitem{Rosen} B. Rosenstein, B.J. Warr and S.H. Park, Nucl. Phys. B {\bf
336}, 435 (1990).

\bibitem{spherical} E. Brezin, J. de Physique {\bf 43}, 15 (1982);
M. Henkel and C. Hoeger, Z. Phys. B {\bf 55}, 67 (1984);
S. Singh and R.K. Pathria, Phys. Rev. B {\bf 31}, 4483 (1985)
and references therein.

\bibitem{Dashen} R.F. Dashen, S-K. Ma and R. Rajaraman, Phys. Rev. D {\bf 11},
1499 (1975).                                           

\bibitem{polylog} S. Sachdev, Phys. Lett. B {\bf 309}, 285 (1993).

\bibitem{Polyakov} A.M. Polyakov, 
{\em Gauge Fields and Strings}, Harwood, New York (1987).

\bibitem{Colem} S. Coleman, R. Jackiw and D. Politzer, Phys. Rev. D {\bf 10},
2491 (1974).                                           

\bibitem{Aref} I.A. Aref'eva, Ann. Phys. (N.Y.) {\bf 117}, 393 (1979).

\bibitem{cardy} J.L. Cardy, J. Phys. A. {\bf 17}, L385 (1984);
R. Shankar and S. Sachdev, unpublished.

\bibitem{Pol} A.M. Polyakov, Phys. Lett. B {\bf 59}, 79 (1975).

\bibitem{Wiegmann} P.B. Wiegmann, Pis'ma Zh. Eksp. Teor. Fiz. {\bf 41}, 79 (1985)
[JETP Lett. {\bf 41}, 95 (1985).

\bibitem{Brezin-Zinn} E. Brezin and J. Zinn-Justin, 
Phys. Rev. B {\bf 14}, 3110 (1976).

\bibitem{Pol_Wiegmann} A.M. Polyakov and P.B. Wiegmann, Phys. Lett. B {\bf 131},
121 (1983).

\bibitem{Drusha} A.V. Chubukov, Phys. Rev. B {\bf 44}, 12318 (1992). 

\bibitem{Kag_Chub} M.I. Kaganov and A.V. Chubukov, Usp. Fiz. Nauk {\bf 153},
537 (1987) [Sov. Phys. Usp. {\bf 30}, 1015 (1987)]; in "Spin Waves and Magnetic
Dielectrics" eds. A.S. Borovik-Romanov and S.K. Sinha, Elsevier Science 
Publ. (1988).

\bibitem{TycHalp} S. Tyc and B.I. Halperin, Phys. Rev. B {\bf 42}, 2096 (1990).

\bibitem{Hohen}    B.I. Halperin and P.C. Hohenberg, Phys. Rev. {\bf{177}}, 952
(1969).

\bibitem{extra} In the recent paper by one of us~\protect\cite{Drusha}, it was suggested
that the damping  may become comparable to the real part
of the quasiparticle energy at the spatial scales which parametrically exceed
 $\xi$. The more sophisticated analysis presented here shows that
it is more likely that 2D antiferromagnet has the same typical spatial scale
(correlation length) for both static and dynamic phenomenon, as is 
predicted by the dynamical scaling hypothesis.

\bibitem{Igarashi} J. Igarashi, Phys. Rev. B {\bf 46}, 10763 (1992).

\bibitem{Singh} R.R.P. Singh, Phys. Rev. B {\bf 39}, 9760 (1989); see also R.R.P.
Singh and D. Huse, Phys. Rev. B {\bf 40}, 7247 (1989).

\bibitem{MFHub} A. Singh and Z. Tesanovic, Phys. Rev. B {\bf 41}, 614 (1990); A. Singh, Z.
Tesanovic and J.H. Kim, Phys. Rev. B {\bf 44}, 7757 (1991); A. Auerbach and B.E.
Larson, Phys. Rev. B {\bf 43}, 7800 (1991). 

\bibitem{boris} B.I. Shraiman and E.D. Siggia, Phys. Rev. Lett. {\bf 61}, 467
(1988); Phys. Rev. B {\bf 42}, 2485 (1990).

\bibitem{borislast} B.I. Shraiman and E.D. Siggia, 
 Phys. Rev. B {\bf 46}, 8305 (1992). 

\bibitem{Frenk} A.V. Chubukov and D. Frenkel, Phys. Rev. B {\bf 46},
 11884 (1992).

\bibitem{Col_Gan_Andrey} J. Gan, N. Andrey and P. Coleman, J. Phys. Condens.
Matter {\bf 3}, 3537 (1991).

\bibitem{jinwuunpub} S. Sachdev, unpublished.

\bibitem{Singh_chi} R.R.P. Singh, Phys. Rev. B {\bf 39}, 9760 (1989).

\bibitem{Andy} A. Millis and H. Monien, Phys. Rev. Lett. {\bf 70}, 2810 (1993).

\bibitem{chiexpdop}  D.C. Johnson, S.K. Sinha, A.J. Jacobson and
J.M. Newsam, Physica (Amsterdam),  {\bf 153-155 C}, 572 (1988).

\bibitem{Mila-Rice} F. Mila and T.M. Rice, Physica C {\bf 157}, 561 (1989);
S. Shastry Phys. Rev. Lett., {\bf 63}, 1288 (1989).

\bibitem{MMP} A.J. Millis, H. Monien and D. Pines, Phys. Rev. B {\bf 42}, 167 (1990).


\bibitem{Manous-MC} E. Manousakis and R. Salvador, 
Phys. Rev. B {\bf{39}}, 575 (1989);
E. Manousakis, Phys. Rev. B {\bf {45}}, 7570, (1992).

\bibitem{Man} E. Manousakis and R. Salvador, Phys. Rev B {\bf{40}}, 
2205 (1989). 

\bibitem{Sasha} A. Sokol, private communication.

\bibitem{t2paper} A. Chubukov, S. Sachdev and A. Sokol, Phys. Rev. B, submitted.

\bibitem{Jarrell} M. Makivic and M. Jarrell, Phys. Rev. Lett., {\bf 68},
1770 (1992).

\bibitem{weichmann} M.P.A. Fisher, P.B. Weichmann, G. Grinstein, and
D.S. Fisher, Phys. Rev. B {\bf 40}, 546 (1989).

\bibitem{bhatt_lee} R.N. Bhatt and P.A. Lee, \prl {\bf 48}, 344 (1982).

\bibitem{ludwig} I. Affleck and A.W.W. Ludwig, Nucl. Phys. 
{\bf B360}, 641 (1991).

\bibitem{young_binder} K. Binder and A.P. Young, Rev. Mod. Phys. 
{\bf 58}, 801 (1986).  

\bibitem{murthy} G. Murthy and S. Sachdev, Nucl. Phys. {\bf B344},
557 (1990).

\bibitem{daniel} D.S. Fisher, private communication.

%
%
%
\bibitem{wolffalgo} U. Wolff, Phys. Rev. Lett. {\bf 62}, 361 (1989).

%
%
%
%
%
%
\end{references}
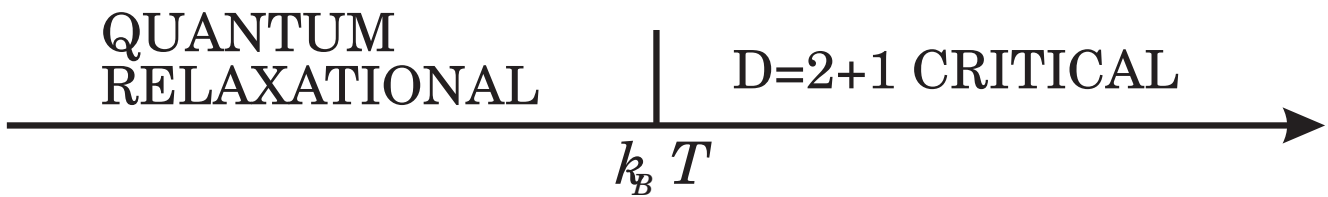
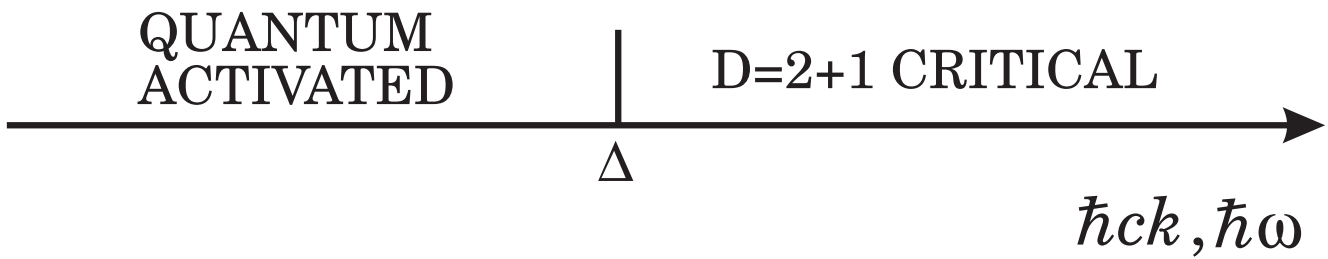
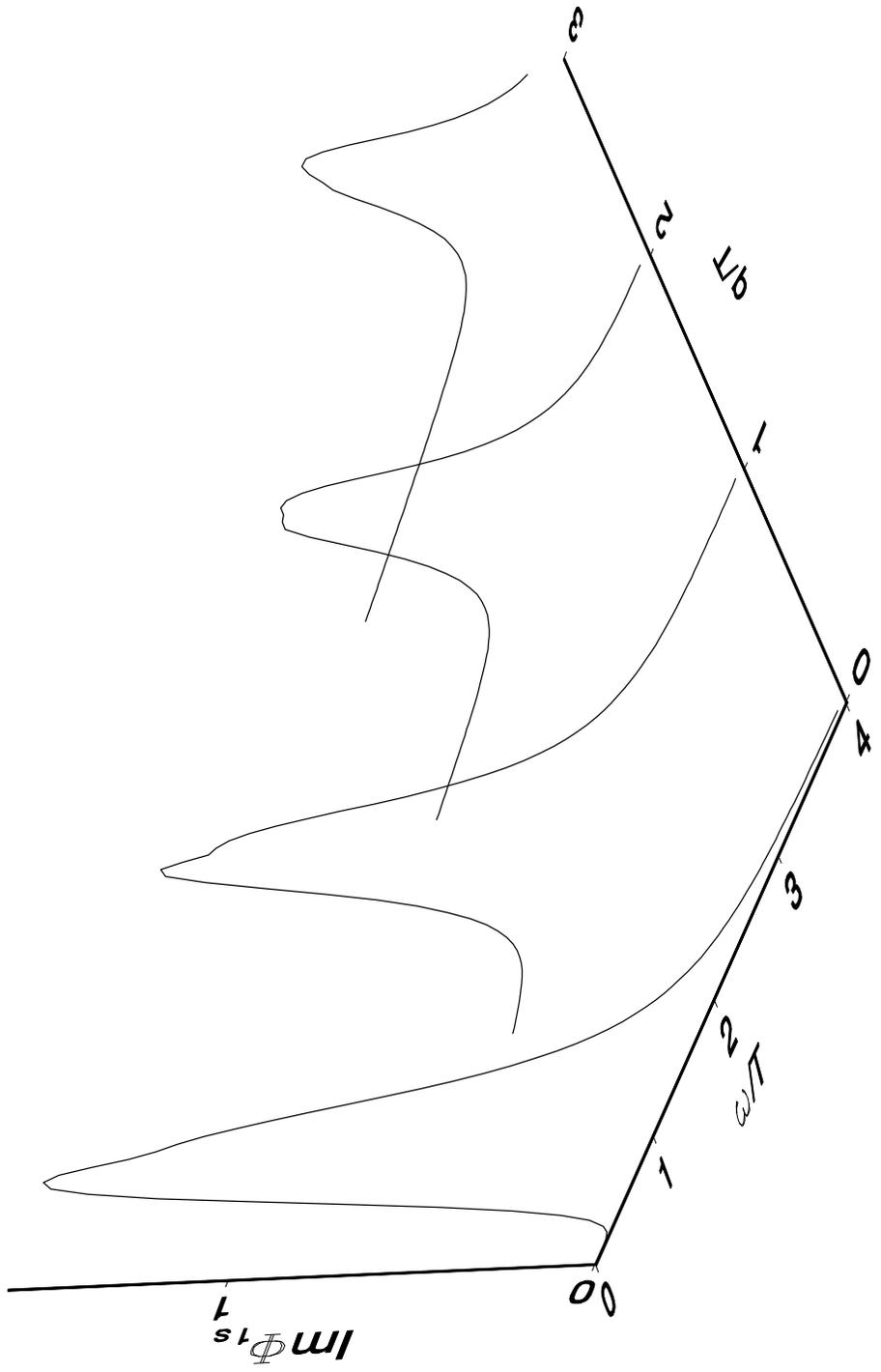
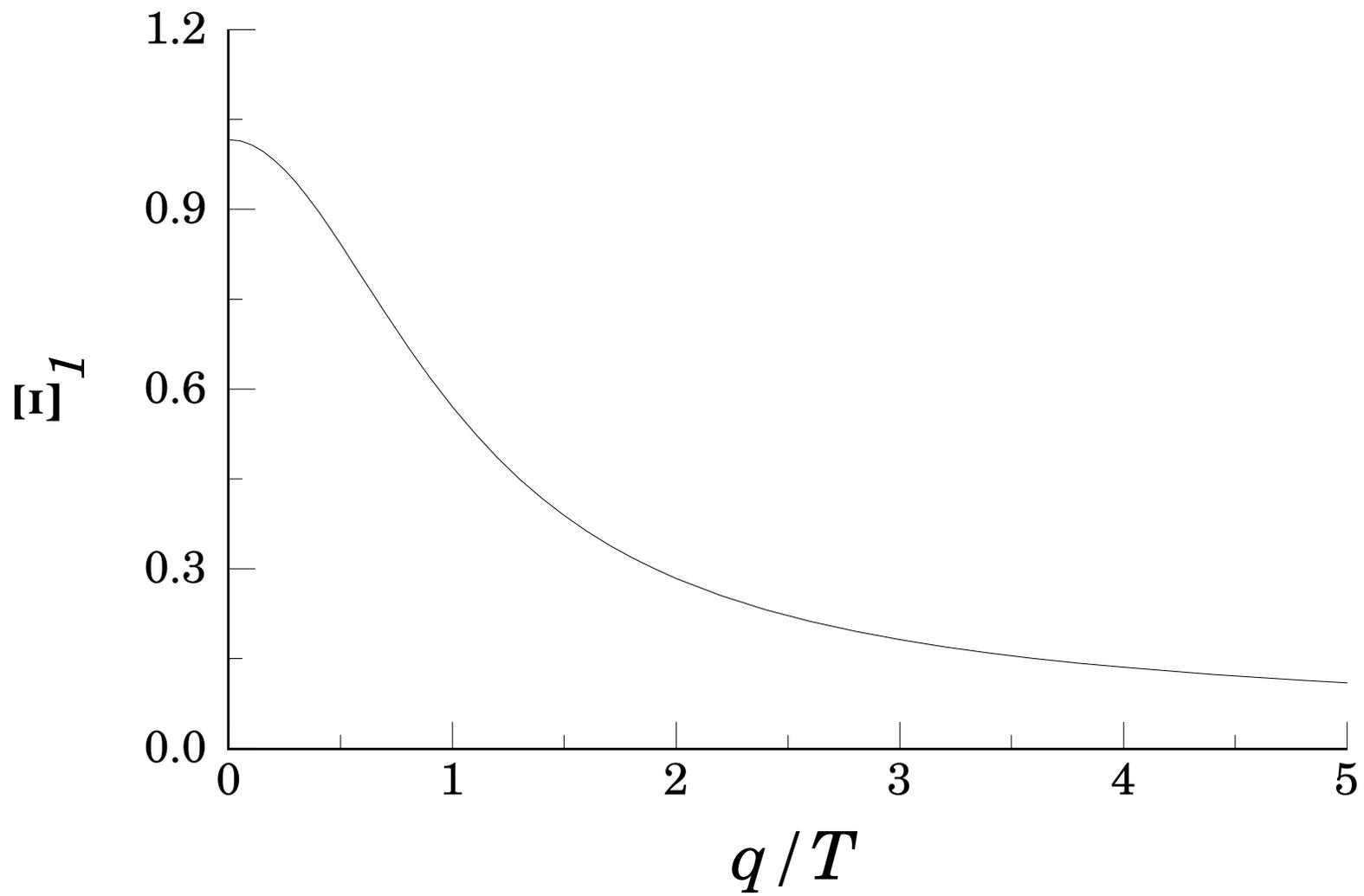
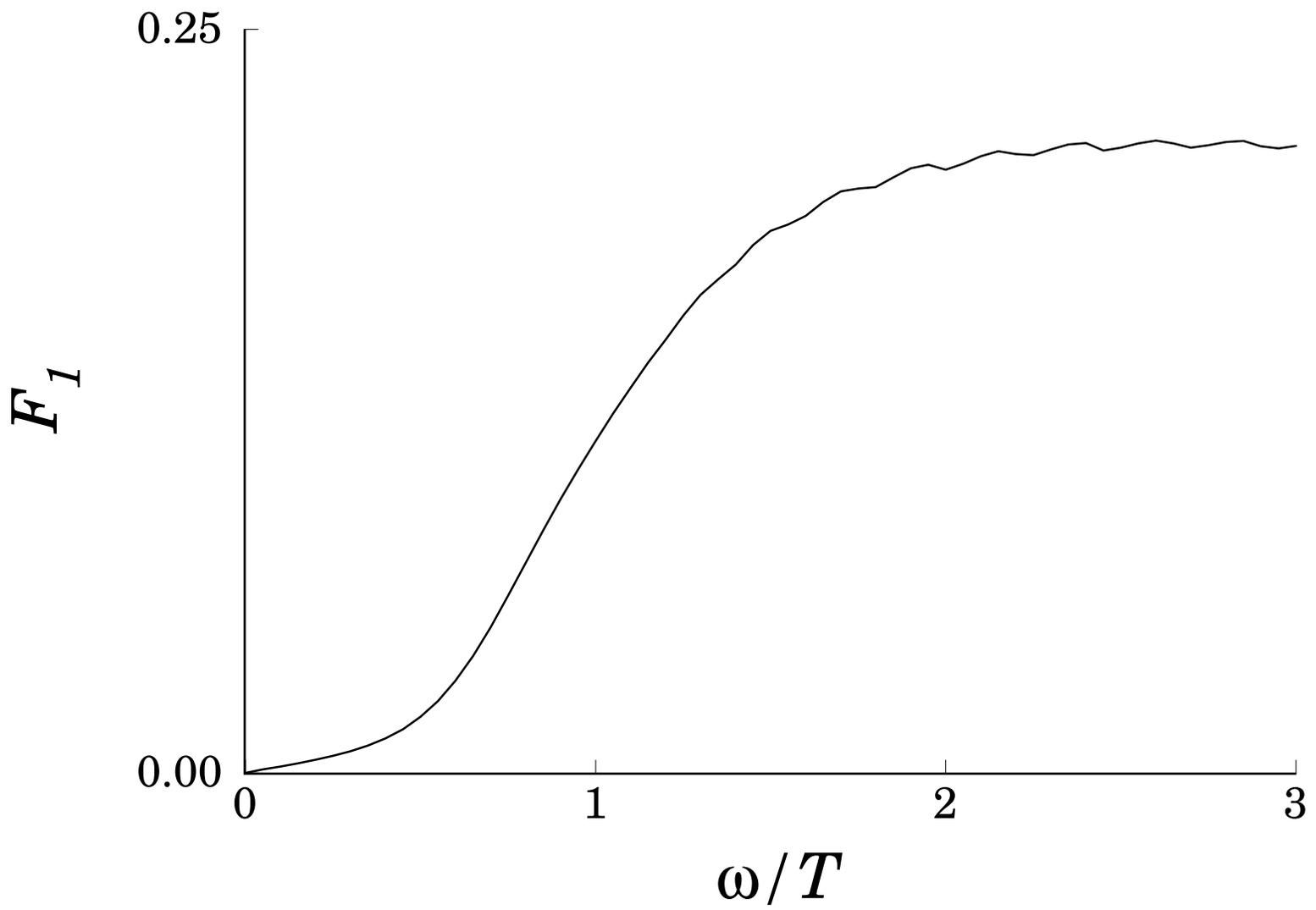
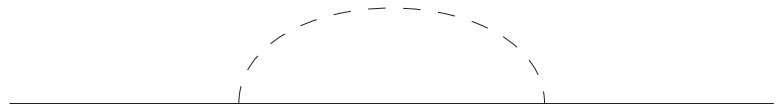
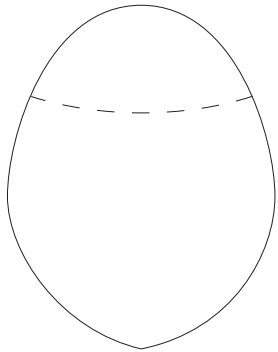
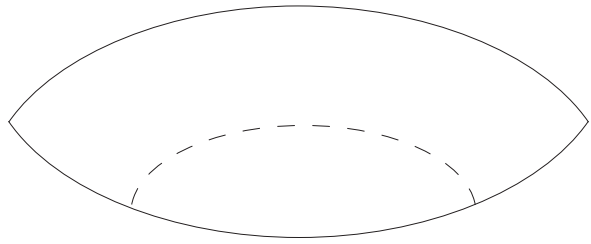
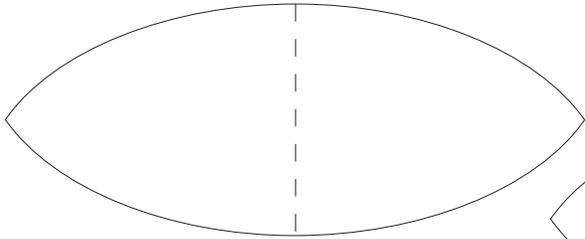
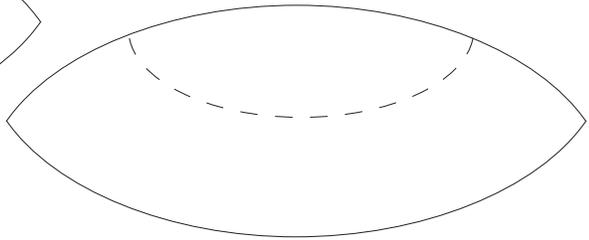
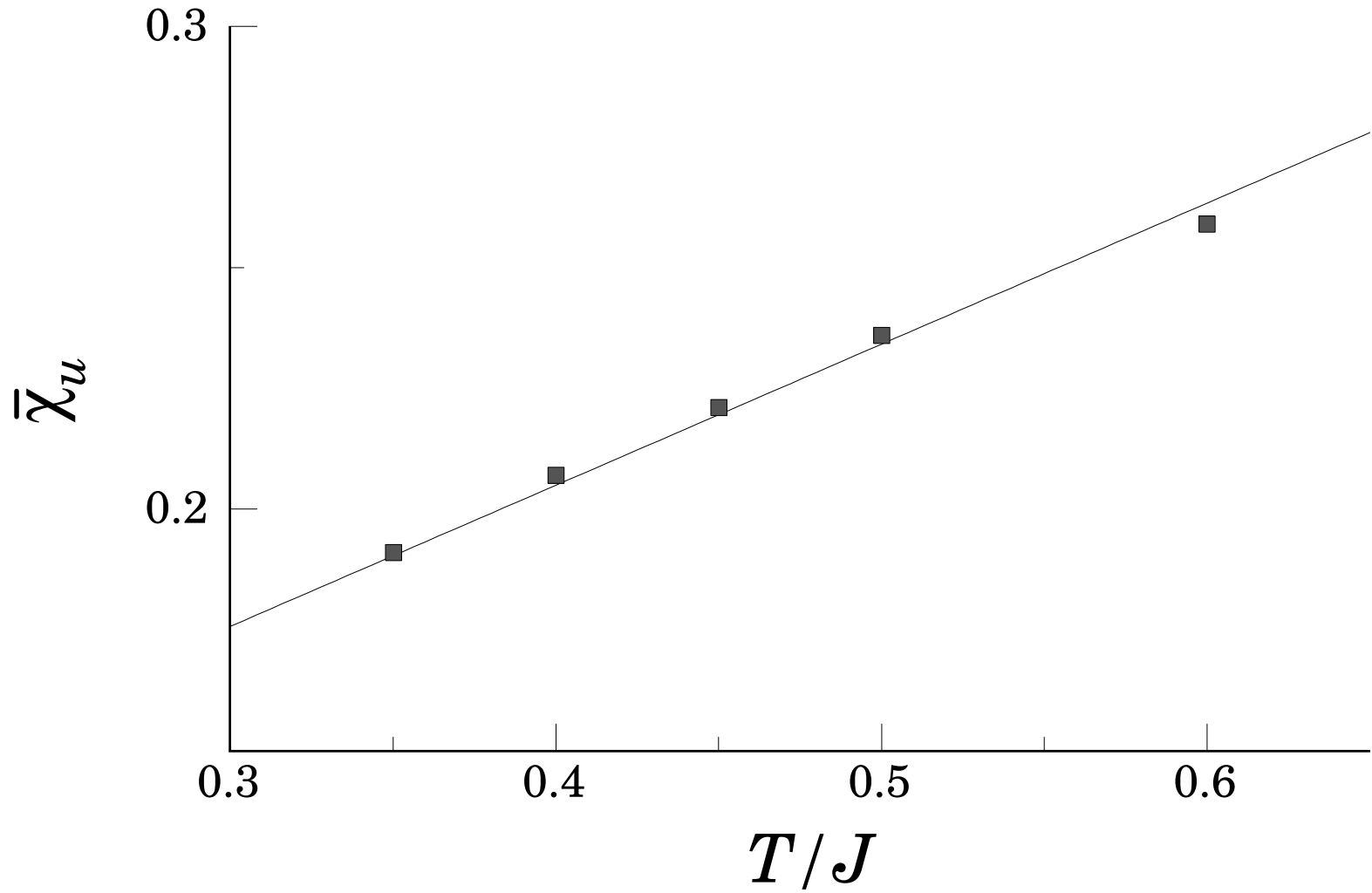
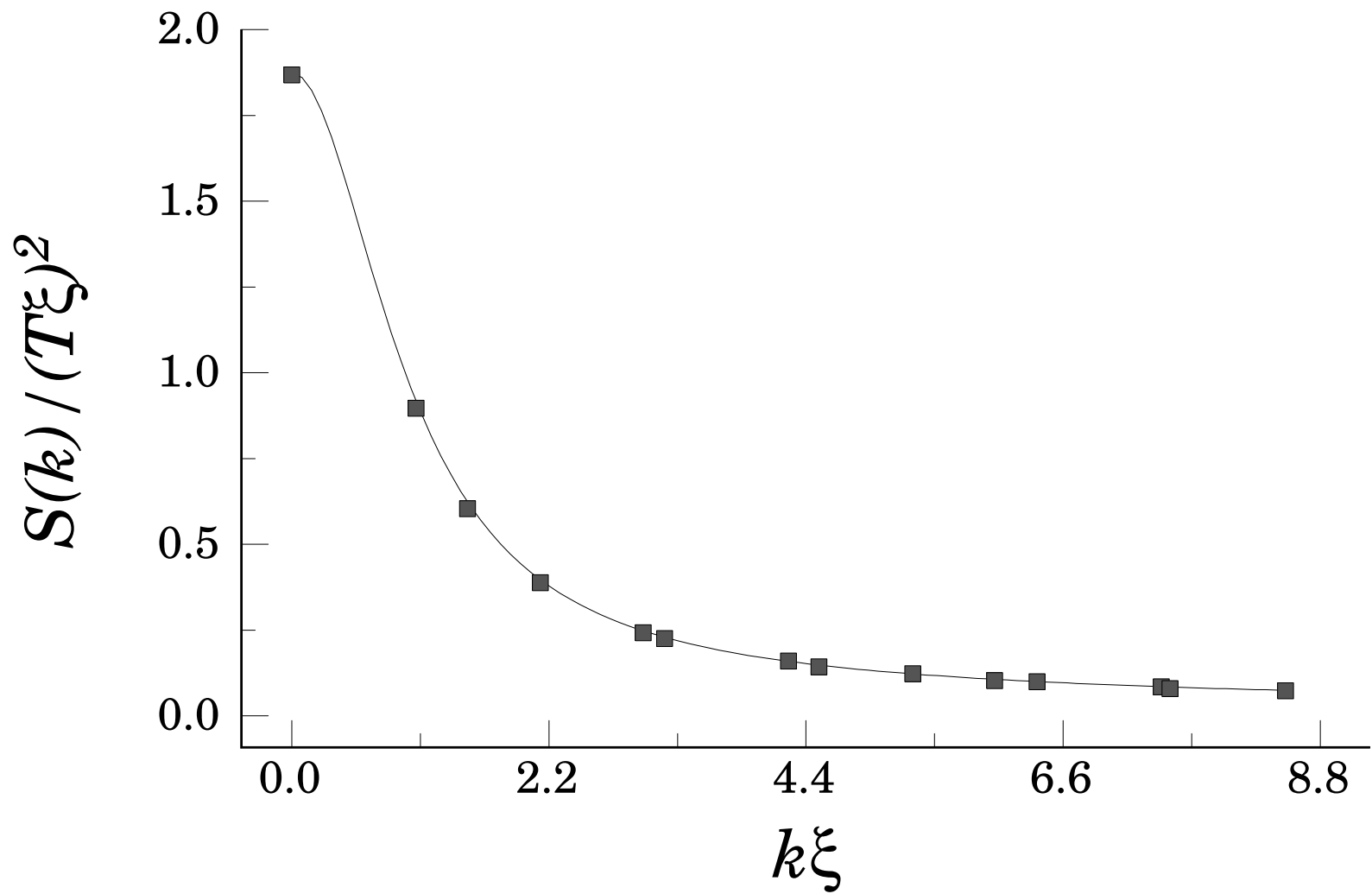
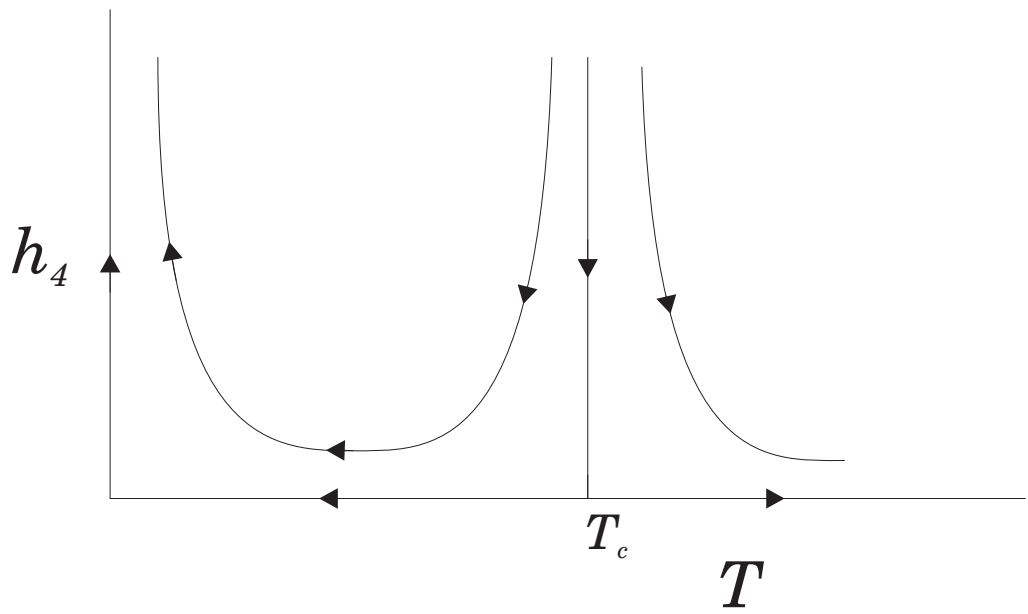
\end{document}